\documentclass[12pt,preprint]{aastex}

\usepackage{amsmath}
\usepackage{amstext}
\usepackage{booktabs}

\usepackage{natbib}
\usepackage{subfigure}
\usepackage{pdflscape}

\newcommand{\epsAur}{$\epsilon$ Aurigae}
\newcommand{\SIMTOI}{\textit{SIMTOI}}
\newcommand{\LIBOI}{\textit{liboi}}
\newcommand{\MSolar}{$M_\odot$}
\newcommand{\RSolar}{$R_\odot$}
\newcommand{\mas}{mas}
\newcommand{\UDD}{$\theta_{\text{UDD}}$}
\newcommand{\LDD}{$\theta_{\text{LDD}}$}

\shorttitle{Interferometry of $\epsilon$ Aurigae: Stellar and disk models}
\shortauthors{Kloppenborg et al.}

\begin{document}


\title{Interferometry of $\epsilon$ Aurigae: Characterization of the asymmetric
eclipsing disk}


% Who knows where:
\author{Kloppenborg, B.K. \altaffilmark{1,2,3}}
    \affil{CHARA/Georgia State University}
    \email{bkloppenborg@chara.gsu.edu}
\author{Stencel, R.E.}
    \affil{University of Denver}
    \email{Robert.Stencel@du.edu}
\author{Monnier, J.D.}
    \affil{University of Michigan}
    \email{monnier@umich.edu}
\author{Schaefer, G. H.}
    \affil{CHARA/Georgia State University}
    \email{schaefer@chara-array.org}
\author{Baron, F.}
    \affil{Georgia State University}
    \email{baron@phy-astr.gsu.edu}
\author{Tycner, C.}
    \affil{NPOI / Central Michigan University}
    \email{tycne1c@cmich.edu}
\author{Zavala, R. T.}
    \affil{U.S. Naval Observatory Flagstaff Station}
    \email{bzavala@nofs.navy.mil}
\author{Hutter, D.}
    \affil{U.S. Naval Observatory Flagstaff Station}
    \email{djh@nofs.navy.mil}
\author{Zhao, M.}
    \affil{Penn. State}
    \email{astro.mingzhao@gmail.com}
\author{Che, X.}
    \affil{University of Michigan}
    \email{xche@umich.edu}
\author{ten Brummelaar, T. A.}
    \affil{CHARA/Georgia State University}
    \email{theo@chara-array.org}
\author{Farrington, C.D.}
   \affil{CHARA/Georgia State University}
   \email{farrington@chara.gsu.edu}
\author{Parks, R.}
    \affil{Georgia State University}
    \email{parksj@chara.gsu.edu}
\author{McAlister, H. A.}
    \affil{CHARA/Georgia State University}
    \email{hmca@gsu.edu}
\author{Sturmann, J.}
    \affil{CHARA/Georgia State University}
    \email{judit@chara-array.org}
\author{Sturmann, L.}
    \affil{CHARA/Georgia State University}
    \email{sturmann@chara-array.org}
\author{Sallave-Goldfinger, P.J.}
    \affil{CHARA/Georgia State University}
    \email{pj@chara-array.org}
\author{Turner, N.}
    \affil{CHARA/Georgia State University}
    \email{nils@chara-array.org}
\author{Pedretti, E.}
    \affil{School of Engineering and Physical Sciences, Heriot-Watt University, Edinburgh, Scotland, UK}
    \email{e.pedretti@hw.ac.uk}
\author{Thureau, N.}
    \affil{University of St. Andrews, Scotland, UK}
    \email{nt15@st-andrews.ac.uk}

\altaffiltext{1}{Georgia State University}
\altaffiltext{2}{Max-Planck-Institut F\"ur Radioastronomie}
\altaffiltext{3}{University of Denver}


\begin{abstract}
We report on a total of 106 nights of optical interferometric observations of the
$\epsilon$ Aurigae system taken during the last 14 years by four beam combiners
at three different interferometric facilities. This long sequence of data
provides an ideal assessment of the system prior to, during, and after the
recent 2009-2011 eclipse. We have reconstructed model-independent
images from the 10 in-eclipse epochs which show that a disk-like object is
indeed responsible for the eclipse. Using new 3D, time-dependent modeling
software, we derive the properties of the F-star (diameter, limb darkening),
determine previously unknown orbital elements ($\Omega$, $i$), and access the
global structures of the optically thick portion of the eclipsing disk using
both geometric models and approximations of astrophysically relevant density
distributions. These models may be useful in future hydrodynamical modeling of the
system. Lastly, we address several outstanding research questions including
mid-eclipse brightening, possible shrinking of the F-type primary, and any
warps or sub-features within the disk.
\end{abstract}





\section{Introduction}

Epsilon ($\epsilon$) Aurigae (FKV 0183, \object{HD 31964}) is an intrinsically bright
(V $\sim$ 3.0) star system that has confounded generations of astronomers since its
two-year long, $\Delta V \sim 0.75$ mag fading was first discovered nearly 200
years ago \citep{Fritsch1824}. In the early 1900s, it was established that
\epsAur{} was a single-line spectroscopic binary with a 27.1 year period
\citep{Ludendorff1903}. Application of the then recently developed eclipsing
binary theory \citep{Russell1912d,Russell1912} to \epsAur{} came to the
perplexing result: the companion must be nearly equal in mass to the primary,
but nearly invisible \citep[see note 18 of][]{Shapley1915}. Although many theories
have been proposed to explain this conundrum \citep[see][for comprehensive
summaries of the literature]{Kloppenborg2012, Guinan2002, Carroll1991} it was
\cite{Kopal1954} who first proposed and \cite{Huang1965} who later
(independently) developed the theory that the system is a nearly edge-on
eclipsing binary composed of an F0Ia supergiant (primary, hereafter the “F-star”)
and an unseen companion that was believed to be enshrouded in a disk of opaque
material. It was theorized that the passage of the disk in front of the F-star
that caused the anomalously long fading of the system. This theory received
significant support during the 1984 eclipse when \cite{Backman1984} detected
the presence of a $500 \pm 150$ K blackbody that remained detectable when
the F-star was eclipsed.

The 2009-2011 eclipse of \epsAur{} caused a resurgence of interest in this
enigmatic binary, yielding a wealth of new scientific results. Historic data
played a pivotal role in several studies. \cite{Stefanik2010} and
\cite{Chadima2010} published two new, nearly identical, spectroscopic orbital
solutions using spectroscopic and photometric data going back to the mid-1800s.
Likewise, \cite{Griffin2013} conducted a meticulous study of historic and new
spectroscopic observations covering the three eclipses. They discovered that the
precision by which the spectroscopic evolution of the eclipse unfolds indicates
that the structure of the disk has not appreciably changed on a time scale of at
least the last 100 years. Similarly, infrared studies reported by
\cite{Stencel2011} have confirmed persistent behavior of disk features
over at least the last two eclipse cycles.

New photometric and spectroscopic observations have revealed the progression of
the eclipse in hitherto unprecedented detail. Photometric monitoring, largely by
amateur astronomers, captured nearly 3,700 photometric points in a myriad of
filters from the UV to near-IR \citep{Hopkins2012}. Spectroscopic monitoring has
shown that the equivalent width of some spectral lines follow a stair step-like
evolution pattern \citep{Leadbeater2012}, suggesting some substructure exists
within the eclipsing disk. For the first time, a comprehensive SED of the system
was assembled from the UV to radio wavelengths. These data clearly show the
presence of not only the F-star and a 550 K disk, but also a slight amount
of far-UV flux thought to originate from a B-type companion at the center of the
disk \citep{Hoard2010}. Subsequent work has shown the disk has an asymmetric
temperature structure ranging from 550 - 1150 K due to external heating from the
F-star \citep{D.W.HoardandD.LadjalandR.E.StencelandS.B.Howell2012}.

In addition to the classic observational methods, the comparatively new
technique of optical interferometry was recently applied to the \epsAur{}
eclipse for the first time. The model-independent H-band images prove, beyond
any reasonable doubt, that an eclipsing disk is indeed responsible for the
observed fadings \citep{Kloppenborg2010}. Complementary spectro-interferometric
observations of the H$\alpha$ line have shown that the F-star has an extensive
P-Cygni-like wind region and that the disk contains a substantial gaseous
atmosphere that can eclipse a greater fraction of the F-star than the dark disk
itself \citep{Mourard2012}.

Yet with as much information we have about this system, we have yet to unravel
many of its fundamental properties. Foremost, the distance to the system is
extremely uncertain, ranging from 0.4 - 4.0 kpc as estimated by HIPPARCOS
\citep{PerrymanM.A.C.1997, VanLeeuwen2007}. Dynamical parallaxes from the Yerkes
\citep{Strand1959} and Sproul \citep{Heintz1994, VandeKamp1978} observatories
narrow this range to 0.5 - 0.7 kpc, yet there is in significant disagreement with
the $1.5 \pm 0.5$ kpc suggested via interstellar absorption and reddening
\citep{Guinan2012}. Second, the evolutionary state of the system has been called
into doubt. Some suggest that the F-star is a massive ($\sim 16$ \MSolar{})
horizontal branch supergiant, whereas others consider the F-star to be a less
massive ($\sim 2-3$ \MSolar{}) post-AGB star \citep[see][for a review and recent
developments concerning this topic]{Guinan2002, Hoard2010,
SadakaneD;KambeE.SatoB;HondaS;Hashimoto2010}. Lastly, although several models
have been proposed for the eclipsing disk \citep[e.g.][]{Huang1965, Wilson1971,
Huang1974a, Takeuti1986a, Ferluga1990, Lissauer1996, Takeuchi2011, Budaj2011},
these efforts were primarily based on light curve modeling. Reproducing the
photometry is a necessary condition of every model, but is not sufficient for
proving the validity of the hypothesis. In particular, such work cannot separate
out the degeneracies between radial and height-dependent optical profiles in
the disk.

We begin to address several of these issues by presenting newly determined
orbital elements and new models for the disk based upon simultaneous
photometric and interferometric modeling of the 2009-2011 eclipse during the
ingress (RJD $=$ JD $-$ 2,400,000 $\sim$ 55,062 - 55,193), totality
(RJD $\sim$ 55,193 - 55,631), and egress (RJD $\sim$ 55,631 - 55,693) phases.
In Section
\ref{sec_data}, we present our observations and data reduction methods.
Section \ref{sec_analysis} and \ref{sec_image_reconstruction_methods} include a
discussion of our modeling process, image reconstruction methods, and
statistical analysis of  the resulting data. Results are presented in
Section \ref{sec_results}, and finally, we draw conclusions in Section
\ref{sec_discussion}.

\section{Observations and data reduction}
\label{sec_data}

This work summarizes a total of 106 nights of interferometric observations
taken by four beam combiners at three different interferometric arrays. An
account of the observations including the array, baselines, combiner, spectral
configuration, and calibrators is in Table \ref{tbl_observations}. Calibrator
identifiers, positions, proper motions, uniform disk diameters (\UDD{}), and the
array at which each calibrator was used is listed in Table \ref{tbl_calibrators}.
The online supplementary
material for this paper contain both calibrated and uncalibrated data saved
in the Optical Interferometry Exchange format \citep[OIFITS; ][]{Pauls2005}.
A sample plot of the UV coverage, squared visibilities, and bispectra for the
2009-11 CHARA-MIRC epoch is provided in Figure \ref{fig_data_example}.
Equivalent plots for all other epochs can be found in the online appendix,
Figures \ref{data_plots_start} - \ref{data_plots_end}.
In the next several paragraphs we describe
the interferometric arrays, combiners, and reduction methods in detail.

\subsection{Palomar Testbed Interferometer (PTI)}

Our first interferometric data set on \epsAur{} were acquired using the Palomar
Testbed Interferometer \citep[PTI;][]{Colavita1999} located on Mount Palomar in
California. The facility consisted of three $40$ cm siderostats, each located at
the termination of one of the interferometer's three arms. Pairwise combination
provided baselines between $85$ and $110$ m. The beam combiner at PTI operated in
several low resolution spectral modes providing up to 11 spectral channels
across the K-band ($2.2 \mu$m).

Much of the wide-band visibility data has been discussed previously in
\cite{Stencel2008a}, therefore, we will not consider it further. Here we have
re-reduced the spectrally dispersed (i.e., narrow-band) data subject to the
calibrator diameters in Table \ref{tbl_calibrators}. We have used the
narrow- and wide-band calibration routines \textit{nbCalib} and
\textit{wbCalib}, respectively, from the \textit{V2calib} software package.
These programs are available as a web service, \textit{webCalib}, on the
NASA Exoplanet Science Institute (formerly Michelson Science Center)
website\footnote{http://nexsciweb.ipac.caltech.edu/cgi-bin/webCalib/webCalib.cgi}.
We have selected the PTI defaults with the following exceptions: (1) the
calibration window was extended to four hours; (2) no ratio correction was
applied, and (3) no minimum uncertainty was enforced for reasons discussed in
\cite{Stencel2008a}. Output from this pipeline were saved in OIFITS format.

\subsection{Navy Precision Optical Interferometer (NPOI)}

The Navy Precision Optical Interferometer \citep[NPOI,][]{Armstrong1998a} is a
six telescope optical interferometer that started operation in 1994. The array
may be configured in either an astrometric or imaging mode. Data from the
imaging subarray comes from six movable $50$ cm siderostats with baselines
between 16 m and 79 m. The NPOI beam combiner operates at visible wavelengths (0.5 -
0.85 $\mu$m) in 16 spectral channels. The NPOI observational setup and data
recording procedure can be found in \cite{Hummel2003} and
\cite{BensonJamesA.2003}. Post-processing and data reduction were performed
using C. Hummel's OYSTER software package.

The NPOI observed \epsAur{} on a total of 29 nights between 2006 Feb. and 2010
Apr. as shown in Table \ref{tbl_observations}. The data were initially calibrated
with respect to HD 32630 assuming a uniform disk diameter of 0.507 $\pm$ 0.025
milliarcseconds (\mas{}) and saved as OIFITS files. Prior to modeling these data,
we recalibrated the OIFITS files by multiplying the visibilities and closure
amplitudes by the ratio of the uniform disk function of the former and new
diameter as listed in Table \ref{tbl_calibrators}.

\subsection{Center for High Angular Resolution Astronomy (CHARA)}

Georgia State University's Center for High Angular Resolution Astronomy
\citep[CHARA;][]{TenBrummelaar2005} is an interferometric array located on Mount
Wilson, CA. The array consists of six 1-m telescopes that can be combined to
form up to 15 baselines ranging in length from 34 to 331 meters. Using the
longest baselines, a resolution of down to $0.5$ \mas{} in the H-band
($0.7$ \mas{} in the K-band) can be realized.

Initial calibration observations of \epsAur{} were taken far in advance of the
eclipse, in 2008-Oct./Nov./Dec.. Semi-regular observations were scheduled around
the photometric eclipse, beginning in 2009 Oct./Nov. and ending in 2011 Nov.
In total, \epsAur{} was observed on 38 nights,
yielding 19 individual epochs after consecutive nights were merged. The first
two eclipse ingress epochs were previously discussed in \cite{Kloppenborg2010}
in which reconstructed images and a preliminary model (comprised of an infinitely
thin, but optically thick disk seen in projection) for the eclipsing disk were
presented.

\subsubsection{Michigan InfraRed Combiner (MIRC)}

The Michigan InfraRed Combiner \citep[MIRC; described in][]{Monnier2006,
MonnierJohnD.2004} went through several revisions during our observing
program. As it was first used, the combiner was configured for four-telescope
beam combination using a camera sensitive in the H- and K-bands with three
spectral resolution options ($R \sim 44$, $150$, or $400$). We followed the
standard observing procedures and reduced the data using the data reduction
pipeline described in \cite{Monnier2007}. After co-adding frames, background
subtraction, and a Fourier transform of the raw data, fringe amplitudes and
phases are used to form the squared visibilities and triple products. These data
are written to OIFITS files for further analysis.

In the earliest data, photometric calibration was achieved using choppers that
temporally encoded the flux coming from each telescope by periodically
blocking each beam at a unique frequency. In August 2009,
the choppers were replaced by dedicated photometric channels \citep{Che2010}
that utilize a fraction of light from each science beam for calibrated,
spectrally dispersed photometry. This led to a dramatic improvement in
uncertainties. In the spring of 2011, MIRC was
upgraded again to combine light from all six of CHARA's telescopes \citep[see
initial report in][]{Monnier2010}, permitting measurement of 15 non-redundant
visibilities and all 10 independent closure phases available at CHARA.

\subsubsection{CLassic Interferometry with Multiple Baselines (CLIMB)}

During the egress phase of the 2009-2011 eclipse, when MIRC was unavailable for
use, we employed the CLassic Interferometry with Multiple Baselines
\cite[CLIMB;][]{Sturmann2010} combiner. CLIMB is a three-telescope beam combiner
which operates in one of five broadband spectral modes.
For our observations we configured CLIMB in with the K’ band
($\bar{\lambda} = 2.133$ $\mu$m, FWHM $0.349$ $\mu$m) filter installed and
destructive readout mode. Because of the large hour angle during our
observations, the UV coverage afforded by these data is quite limited.

We used the standard observing technique and reduced our data with the CLIMB
version 2.1 reduction pipeline \citep{TenBrummelaar2012}. Data were calibrated
subject to the calibrators listed in Table \ref{tbl_calibrators} and written
to OIFITS format for further analysis.

\section{Modeling and analysis}
\label{sec_analysis}

When our interferometric observations at CHARA began, no publicly available
software was capable of modeling both interferometric and photometric data in a
fully 3D, time dependent fashion. We wrote two software packages to solve this
problem: The SImulation and Modeling Tool for Optical Interferometry, \SIMTOI{},
and the OpenCL Interferometry Library \LIBOI{}. Because there are no prior
publications discussing these software, we present a brief overview of their
capabilities here.

\SIMTOI{}\footnote{SIMTOI: \url{https://github.com/bkloppenborg/simtoi}}
\citep{Kloppenborg2012b} is an open source (GPL) C/C++ program for fitting
time-dependent, 3D models to large data sets. Instead of using analytic models,
\SIMTOI{} uses the Open Graphics Library (OpenGL) to tessellate the ``surfaces''
of the astronomical objects being simulated. Geometrical effects, such as limb
darkening, are applied using programs written in the OpenGL Shading Language (GLSL).
The models may be positioned statically in their 3D environment using fixed
Cartesian coordinates or be determined dynamically from a Keplerian orbit. All
models also have an intrinsic inclination, $i$, position angle, $\Omega$, and
rotational zero point, $\omega$. The geometrical primitives are rotated,
translated, and orthographically rendered to a multisample anti-aliasing buffer.
OpenGL computes the total flux in each pixel using a method akin to ray tracing,
creating 2D images against which observed data is compared. At present,
\SIMTOI{} supports both interferometric and photometric data. Additional data
types can be supported by subclassing and registering the new data type with
\SIMTOI{}'s plugin API. \SIMTOI{} provides a similar API for registering
minimization engines. Presently implemented minimizers include a  recursive
grid search, \textit{Levmar} \citep{lourakis04LM}, \textit{MultiNest}
\citep{Feroz2008a,Feroz2009}, and a bootstrapping minimizer based on
\textit{Levmar}. The validity of models produced by \SIMTOI{} has been tested
against iota Peg, the CHARA-MIRC closure phase calibrator \citep{Monnier2007}
and LitPro \citep{Tallon-Bosc2008} model results.

\SIMTOI{} uses \LIBOI{}\footnote{\LIBOI{}: \url{https://github.com/bkloppenborg/liboi}}
\citep{Kloppenborg2012c} to generate interferometric observables. \LIBOI{}
is an open-source (LGPL) C/C++ library which implements the backend from the GPu
Accelerated Image Reconstruction \citep[GPAIR;][]{Baron2010a} program. \LIBOI{}
aims to provide software developers with convenient access to fast routines for
common interferometric tasks. The software heavily relies on the heterogeneous
computing environment of the Open Compute Language (OpenCL) to target a wide
range of traditional and multi-core CPUs; servers, hand-held/embedded devices,
specialized hardware, and Graphical Processing Units (GPUs). The computational
correctness of \LIBOI{} is provided by a series of built-in unit tests to
analytical functions.

\subsection{Bayesian model selection}
\label{sec_bayesian_model_selection}

In this work, we use Bayesian statistics to assess the relative goodness of fit
between our proposed models rather than traditional chi-squared methods that
are ill-adapted to make such inferences \citep{Marshall2006}. Bayesian
statistics provides a consistent approach to estimate a set of parameters,
$\Theta$, in a hypothesis (e.g. an image or model), $H$, given some observed
data, $D$. Bayes' theorem states that

\begin{equation}
P(\Theta | D,H) = \frac{P(D|\Theta,H) P(\Theta|H)}{P(D|H)},
\end{equation}

\noindent
where $P(\Theta | D,H)$ $\equiv$ $P(\Theta)$ is the posterior probability
distribution of the parameters, $P(D|\Theta,H) \equiv L(\Theta)$ is the
likelihood, $P(\Theta|H) = \pi(\Theta)$ is the prior, and $P(D|H) \equiv Z$ is
the Bayesian Evidence. In parameter estimation problems, where the model remains
the same, the normalization factor, $Z$, is often ignored as it is independent
of the parameters $\Theta$. When selecting between various models, the evidence
plays a central role through the Bayes Factor:

\begin{equation}
R = \frac{P(H_1|D)}{P(H_0|D)} = \frac{P(D|H_1)P(H_1)}{P(D|H_0)P(H_0)} = \frac{Z_1}{Z_0}\frac{P(H_1)}{P(H_0)}.
\end{equation}

The evidence is the average of the likelihood over the prior. Thus a simple
model with greater likelihood over the parameter range will be favored over
a more complex model with lower likelihood over the parameter range, unless
the latter is significantly better at explaining the data. Therefore the
Bayes factor automatically implements the Occam razor principle.

A principal difficulty in using Bayesian evidence for model selection is that
the multidimensional integral,

\begin{equation}
P(D|H) \equiv Z = \int L(\Theta) \pi(\Theta) d^D \Theta,
\end{equation}

\noindent
must be evaluated. For this work, we decided to use the \textit{MultiNest}
library \citep{Feroz2008a,Feroz2009,Feroz2013} that numerically estimates this
integral by intelligently exploring the parameter range using Markov chain
methods and ellipsoidal bounding conditions. In the model fitting process
we have assumed non-informative (flat) priors and used the standard likelihood
function:

\begin{equation}
L(\Theta) = \prod_i \frac{1}{\sqrt{2 \pi \sigma_i^2}}
    \exp \left [
        \frac{ ( M(\Theta)_i - D_i )^2 }{2 \sigma_i^2}
    \right ],
\end{equation}

\noindent where $M(\theta)_i$ is the model prediction, $D_i$ is the data,
and $\sigma_i$ is the uncertainty associated for the $i^\text{th}$ data point.

\subsection{F-star model}

Within the \SIMTOI{} framework, the F-star is modeled as a single uniformly
illuminated sphere, located at the origin, to which GLSL shaders (simulating
limb darkening) were applied. The PTI observations are very high up on the
visibility curve, therefore, they provide no reliable measurement of limb
darkening. Consequently, we fit all PTI data with a uniform disk model. The
CHARA and NPOI observations frequently resolve the F-star near or beyond the
first visibility null. These data show clear departures from uniform disk
behavior. To account for this, we implemented several limb darkening laws in
GLSL. These include linear, logarithmic, square root, power law
\citep[via.][]{Hestroffer1997}, and a few multi-parameter laws
\citep[e.g.][]{ClaretA.2000a, Claret2003, Fields2003}.

\subsection{Disk models}
\label{sec_disk_models}

The previous models for the \epsAur{} disk were purely geometric representations
with hard edges \citep[e.g.][and references therein]{Kloppenborg2010, Kloppenborg2012}.
In this work we have created geometric and astrophysical density distribution
models with position-dependent optical properties.
Because of the edge-on nature
of the eclipse, we elected to represent all disks as a series of concentric
rings of infinitesimal thickness and uniform total height, $h$, that are
equally spaced between an inner, $r_\text{in}$, and outer, $r_\text{out}$,
radius. The rings are connected at the midplane by another surface.

Inspired by the appearance of proplyds in the Orion Nebula
\citep[e.g.][]{Ricci2008}, the geometrical models have opacity (via. OpenGL
source transparency, $\text{src}_\alpha$) that is controlled by a double power
law that is a function of radius and height:

\begin{equation}
\text{src}_\alpha(r, z) = \left( \frac{r}{r_\text{out}} \right)^{-\alpha}
    \left( \frac{z}{h/2} \right)^{-\beta}.
\end{equation}

For the two astrophysical disk models we keep the concentric ring representation
of the disk, but modify the opacity of each ring according to a real density
distribution. The first model from \cite{Pascucci2004} consists of a power-law
in the radial direction and a scale height exponential taper in the vertical
direction:

\begin{equation}
\rho(r, z) = \rho_0 \left ( \frac{r}{r_c} \right)^{-\alpha}
    e^{-\frac{1}{2} \left (
        \frac{z}{h}
        \right)^2 }
\end{equation}

\noindent where $\rho_0$ is the density, $r_c$ is the scale radius,
$h = h_c \left( \frac{r}{r_c} \right)^{-\beta}$,
$h_c$ is the scale height and units are in angular quantities (e.g. mas or
mas$^{-3}$ when appropriate).
The second density distribution is characterized by a power-law in the inner
disk and an exponential taper at large radii \citep[cf.][]{Andrews2009}:

\begin{equation}
\rho(r, z) = \rho_0 \left ( \frac{r}{r_c} \right)^{-\gamma}
    e^{-\frac{1}{2} \left ( \frac{z}{h} \right)^2 }
    e^{-\left( \frac{r}{r_c} \right)^{2 - \gamma} }.
\end{equation}

\noindent For these models, the OpenGL transparency is calculated as
$\text{src}_\alpha = 1 - e^{- \kappa(\lambda) \rho(r,z)}$ to ensure the pixel intensity
is computed following radiative transfer conventions. Because the composition
and precise optical properties of the disk have yet to be ascertained, we have
treated the product $\kappa_0 \rho_0$ as a single quantity in our minimization.

In this work we also introduce time-dependence by positioning the disk according
to a Keplerian orbit. The spectroscopic
orbital solutions from \cite{Stefanik2010} and \cite{Chadima2010} are in
excellent agreement with each other; however, the astrometric orbital parameters
are not well constrained. Therefore, our model adopts the spectroscopic orbital
parameters (longitude of periastron, $\omega$, eccentricity, $e$, period, $P$,
and time of periastron, $T$) from \cite{Stefanik2010} and we derive
the astrometric parameters (inclination, $i$, longitude of the ascending node,
$\Omega$, and total orbital semi-major axis, $\alpha_T$) via. minimization to our
data.

\subsection{Modeling process}
\label{sec_modeling_process}

Within the framework of the aforementioned models the number of parameters needed
ranges from six in the case of a uniformly illuminated F-star and cylindrical
disk (e.g. $i$, $\Omega$, $\alpha_T$, \UDD{}, $r_\text{disk}$,
and $h_\text{disk}$) to ten with a limb darkened disk and more complex opacity
model for the disk (see Table \ref{tbl_parameters} for a summary of all parameters
used and their permitted ranges).
To reduce the number of degrees of
freedom we elected to establish bounds by solving a series of subproblems first,
then lift restrictions to generalize the results.

\begin{enumerate}

\item First use Bayesian evidence and the post-eclipse CHARA-MIRC data to
establish bounds for the diameter and limb-darkening of the F-star.

\item Once determined, the diameter was used to approximate the total orbital
semi-major axis, $\alpha_T$, for the system using two methods. To first order, we
may assume the orbit is circular and the eclipse transects the equator of the
F-star. Then $\alpha_T$ may be found, to first order, by equating the fraction of
the orbit spent in ingress, with the equivalent sector of the orbit via,

\begin{equation}
\frac{T}{t_\text{ingress}} = \frac{p}{s} \approx \frac{2 \pi \alpha_T}{\theta_\text{star}},
\end{equation}

\noindent
where $p$ is the perimeter of the orbit, $s$ is the orbital sector, $\alpha_T$ is
the separation of the components, and $\theta_\text{star}$ is the diameter of
the F-star in radians. If $s << p$, we may perform a second-order approximation
for the sector using the Ramanujan approximation for the perimeter of an ellipse.
In this case, we find:

\begin{equation}
\frac{T}{t_\text{ingress}} \approx \pi \alpha_T \left[ 3(1+\sqrt{1-e^2}) - \sqrt{10 \sqrt{1-e^2} + 3 (2 - e^2)} \right] \frac{\cos\omega}{\theta_\text{star}},
\end{equation}

\noindent
where $\omega$ and $e$ are the aforementioned orbital quantities.

\item Approximate the position angle of the ascending node, $\Omega$,
by computing the average of the position angles determined from single-epoch
in-eclipse minimizations.

\item Determine the best-fit disk model by performing a simultaneous
fit to the photometric and interferometric data. The F-star's diameter and limb
darkening are held constant. $\alpha_T$, and $i$ are free, whereas $\Omega$,
$\omega$, $e$, $P$, $T$ are constant.

\item Derive the best-fit F-star's diameter, limb darkening coefficient, disk
height, and disk transparency at each epoch while holding the remaining
parameters constant.

\item Lastly, lift the constraints on our models insofar as possible to
derive statistical information of the aforementioned parameters via
bootstrapping individual epochs.

\end{enumerate}

\section{Image reconstruction}
\label{sec_image_reconstruction_methods}

We performed image reconstruction for the figures presented in this publication
using both SQUEEZE \citep{Baron2010}, a logical successor to the Markov Chain
Imager \citep[MACIM][]{Ireland2006}, and the BiSpectrum Maximum Entropy Method
\citep[BSMEM;][]{BuscherD.F.1994, Baron2008}.
The theory behind image reconstruction, namely the minimization of the $\chi^2$
datum plus a regularization function, is common to these packages; however, the
programs use different approaches when solving the minimization problem.
SQUEEZE/MACIM perform global stochastic minimization by simulated annealing
whereas BSMEM uses a local gradient-based approach. Despite the differences in
implementation, the images produced by these packages are in remarkable
agreement. This is proof that given the same data, the software frequently
converge to the same solution. Because of this we present the SQUEEZE total
variation regularization images here and other SQUEEZE and BSMEM images in
the online appendix.

To access the presence of artifacts in our image reconstruction method, we
adopted a pragmatic approach in which (1) we find the best-fit model from our
minimization process, (2) sample the model using the same UV coverage as the
original to create synthetic data, (3) redistribute the nominal values in the
synthetic data using the uncertainties from the real data, and (4) reconstruct
the simulated model image using the same methods as the real data.
Steps two and three were performed using
\textit{oifits-sim}\footnote{oifits-sim: \url{https://github.com/bkloppenborg/oifits-sim}}.
The three
images (real, model, and simulated model) are then compared qualitatively.
Features present in the model that are also present in the real and simulated
model images are likely true. Conversely, features seen in the real or simulated
model image that are not present in the model are likely artifacts of the
reconstruction process. Lastly, features present in the real image but not
explained otherwise may be real, but warrant further investigation.

\section{Results}
\label{sec_results}

\subsection{Model selection for the F-star's limb darkening law}

We have used the four post-eclipse CHARA-MIRC observations to compare seven
different limb darkening laws for the F-star. As seen in Table
\ref{tbl_ld_selection}, the uniform disk model is always a poor fit to the data
compared to the limb darkening models. In all but one epoch, the quadratic limb
darkening law was found to be a better fit to the interferometric data than
other models. On 2011-10-10, the two-parameter law described in
\cite{Fields2003} was a better fit.

It would appear that the quadratic limb darkening law is most
appropriate; however, an inspection of the fits reveals that all models predict
essentially the same visibility function. Interestingly, all models also predict
significantly lower flux at the limb than is implied by either plane parallel or
spherical stellar atmosphere codes.
We verified that the \SIMTOI{} results matched LitPro's analytical models,
therefore, it is unlikely that this effect is fictitious.
We have noticed small, few degree, non-zero closure phases and two epochs
where the location of the first visibility null differs between baselines. Thus
it is possible that the F-star may be slightly oblate or have surface features.
We will explore these possibilities in greater detail in a future publication.
Because it is simple and can reproduce the data, we pragmatically adopted the
power-law limb darkening law to represent the F-star in this work.
The mean out-of-eclipse diameter and limb darkening coefficient are
$2.22 \pm 0.09$ \mas{} and $0.50 \pm 0.26$, respectively. This range of values
compares favorably with published diameters from the NPOI
\citep[\UDD{} $2.18 \pm 0.05$ \mas{} at $0.5 - 0.85 \mu$m,][]{Nordgren2001}, the
Mark III \citep[limb darkened diameter, \LDD{}, $1.888 - 2.136$ \mas{}
at $0.4 - 0.8 \mu$m,][]{Mozurkewich2003}, and PTI
\citep[\UDD{} $2.27 \pm 0.11$ \mas{} at 2.2 $\mu$m,][]{Stencel2008a} interferometers.

\subsection{Initial estimation of the orbital parameters}

Using the equations in \ref{sec_modeling_process}, we have estimated the orbital
semi-major axis. Assuming $T = 9896 \pm 1.6$ days \citep{Stefanik2010}, the
H-band ingress time $145 \pm 15$ days \citep{Hopkins2012}, and a circular orbit,
we estimate $\alpha_T = \alpha_1 + \alpha_2 \sim 24 \pm 2.6$ \mas{} where
$\alpha_1$ and $\alpha_2$ are the semi-major axes of the F-star and disk with
respect to the system's center of mass. In the elliptical
case, the spectroscopic elements from \cite{Stefanik2010} and \cite{Chadima2010}
orbital elements yield nearly identical results of
$\alpha_T \sim 31 \pm 3.7$ and $33 \pm 4.5$ \mas{}, respectively. The dominant source
of uncertainty in these values comes from the $\sim 10$\% errors in $\omega$ and
$t_\text{ingress}$. Recognizing estimates for the F-star’s contribution to
$\alpha_T$ are some $13-24$ \mas{} \citep{Strand1959,VandeKamp1978,Heintz1994},
we establish bounds of $13 < \alpha_T < 38$ \mas{} for our minimization. A
non-equatorial intersection, as seen in our previous work
\citep{Kloppenborg2010}, will strictly decrease the upper bound on $\alpha_T$.

Next, we estimated $\Omega$ by fitting the in-eclipse CHARA-MIRC data with a
model consisting of a circular F-star with power-law limb darkening and
cylindrical disk. We set $\alpha_T =
31$ \mas{} and $r_\text{disk} = 10$ \mas{} while permitting the stellar and
orbital inclination parameters to remain free. An average of $\Omega = 296 \pm
3$ deg was obtained from the six totality epochs.

\subsection{Disk model selection via. multi-epoch minimization}
\label{results_disk_selection}

From the aforementioned disk models we created eight variants. The geometric
models are (1) a hard-edged cylinder, and three variants with power-law
transparency in (2) both radius and height, (3) height only, and (4) radius only.
The models from astrophysical density distributions are (5) a
\citeauthor{Pascucci2004} disk, (6) an \citeauthor{Andrews2009} disk.
Although not supported by the interferometric images, we decided to test for
the central clearing hypothesis \citep[c.f.][and references therein]{Ferluga1990}
we created (7) a \citeauthor{Pascucci2004} disk with a variable inner radius.
Lastly we test whether the \cite{Kemp1986} polarization model by creating
(8) a \citeauthor{Pascucci2004} disk that may be tilted out of the orbital plane.

Using the \textit{MultiNest} minimizer, we derived $\log Z$ estimates by
simultaneously fitting each model to a subset of the H-band photometric data
(consisting of 200 observations spaced at approximately equal intervals throughout
the eclipse) and six interferometric epochs from CHARA-MIRC (2009-11, 2009-12,
2010-08, 2010-11, 2011-01, and 2011-09-18). The best-fit parameters,
posterior odds ratio ($\Delta \log R$),
and average $\chi^2$ values for each model are shown in Table
\ref{tbl_disk_model_selection}. The $\Delta \log R$ values indicate that model 8
(the tilted Pascucci disk) provides the best simultaneous fit to the data and
was therefore adopted for the remainder of our work.

We have rendered the best-fit version of each disk model in Figure
\ref{fig_disk_models}. With the exception of the cylindrical disk, all disk
models have a characteristic size of $\sim 6$ \mas{} radius and $\sim 0.75$ \mas{}
height before becoming optically thin. Plots of the corresponding H-band
photometry in Figure \ref{fig_photometry} show that although the disk
models appear physically different, they all do a reasonable job reproducing
the global properties of the light curve. This clearly demonstrates that
although reproducing the photometry is a necessary but not sufficient
condition to prove the validity of any particular disk model.
We note that none of these symmetric disk models are capable of reproducing all
of the photometric features seen during the eclipse.

Using model 7 (a Pascucci disk with a variable inner radius) we test the notion
that the disk's central clearing is responsible for the alleged mid-eclipse
brightening. We find that the disk remains edge-on, but has an inner radius
of some $3.8$ \mas{}. This implies that the inner 60\% of the disk could be
devoid of any opaque material. Despite this fact, the impact on the light curve
is minimal. Given the geometry of the eclipse, light penetrating the central
clearing cannot be responsible for any mid-eclipse brightening. We will discuss
this result in greater detail in Section \ref{sec_discussion}.

\subsection{Bootstrapping and aggregate statistics}

The photometry predicted by the symmetric disk models creates an interesting
corollary: if one assumes the disk is symmetric, one may immediately conclude,
to the contrary, that the disk must be asymmetric because
the residuals between the observed and predicted photometry far exceed
the F-star's $\Delta H \sim 0.05$ mag variations seen outside of eclipse.
We tested this asymmetric conjecture by performing several single-epoch
MultiNest minimizations to the in-eclipse MIRC data.
We created three additional disk models which (a) forced the disk to reside in
the orbital plane, (b) tilted the disk out of the orbital plane at a fixed angle,
and (c) permitted both the position and inclination of the disk with respect to
the orbital plane to vary.
We find that the disk is tilted out of the orbital plane by less than $4$ degrees
($1.33 \pm 0.67$ deg with rejection of one outlier) and has significant variations
in structure (see Table \ref{tbl_eclipse_multinest}).

To derive statistical uncertainties and simulated photometry we used the
best-fit values from the aforementioned minimizations as starting points and
bootstrapped each interferometric epoch 10,000 times. During each bootstrap
we dynamically recalibrated the data using the uncertainty distribution of
the calibrator. Then we created a new realization of the data using the measured
uncertainties, taking into account any known correlations (i.e. as seen in
spectrally dispersed visibilities) in the data when required.
The results of this effort are shown in Table \ref{tbl_bootstrap_results} with
a subset of the results plotted in Figure \ref{fig_bootstrapped_interferometric_results}.

By inspection of Table \ref{tbl_bootstrap_results}, one can see that the angular
diameter and limb darkening profile of the F-star are largely consistent within
one sigma. This implies that there are no egregious systematic calibration
errors between our data sets. Furthermore, these results show that there has
been no secular change to the F-star's diameter over the last 14 years.
However, our observations are not sufficiently precise to definitively exclude
the 0.6\% yr$^{-1}$ contraction rate suggested by \cite{Saito1986}.

In the top panel of Figure \ref{fig_bootstrapped_interferometric_results} we
plot the H-band photometry as observed, simulated from the symmetric model, and
predicted from the bootstrap process described above. Not all of the photometric
values are in perfect agreement, but this is expected as the photometry was not
used as a constraint in the bootstrapping process. We note that a small change
in disk structure can have a substantial impact on the observed photometry
(e.g. a 0.1 \mas{} difference in thickness, five times smaller than our
resolution limit, results in $\Delta H \sim 0.1$ mag), thus the predicted
photometry is in reasonable agreement.

In the bottom three panels of Figure \ref{fig_bootstrapped_interferometric_results}
we plot the angular diameter, limb darkening coefficient, disk scale height, $h_c$,
and height power-law exponent, $\beta$, as a function of time. The 1-sigma
estimates for $h_c$ and $\beta$ from model 8 do not overlap well with the
single-epoch bootstraps. This is likely due to the
$\sim 1.6$ degree difference in disk inclinations between the models.
The changes in $h_c$ and $\beta$ (greater $h_c$, smaller $\beta$) indicate that
the disk is more spatially extended in height before mid-eclipse than it is
after mid-eclipse.
Combined with the asymmetric evolution of several neutral
absorption lines during eclipse \citep[c.f.][]{Leadbeater2012, Lambert1986},
the evidence suggests that the disk is not purely symmetric. We suspect that
these features could be explained by asymmetric heating \citep[c.f.][]{Takeuchi2011}
and sublimation of the disk on the side facing the F-star.
Verifying this claim by hydrodynamical radiative transfer simulations is
beyond the scope of this work.

Because of the limited UV coverage, the egress CHARA-CLIMB data was fit in
conjunction with an interpolated photometric point. The result reveals that
the F-star is similar in size, but the disk is more extended in the
vertical direction than the CHARA-MIRC observations three months earlier would
imply. This interpretation is supported by the appearance of the model-independent
images (discussed below); however, we caution the reader that this may simply
be an artifact of the limited UV coverage of this data set.

In Table \ref{tbl_aggregate_stats}, we show the aggregate statistics derived
from single-epoch bootstrapping for each beam combiner. We believe these values
represent the general characteristics of the system.
In Table
\ref{tbl_physical_values}, we summarize these quantities for a variety of distance
estimates for the system. Due to the large scatter of possible distances ($0.5 -
4$ kpc), we use only the nominal values and do not propagate any uncertainties
from the distance measurements. We caution the reader that the aggregation of
data to create Table \ref{tbl_physical_values} was performed without regard
to either the wavelength of observation or any asymmetries we advocate exist.
Hence, the outliers have biased and skewed the resulting value. We provide this
last table to assist with the creation of a full radiative transfer model of the
system including dust physics rather than provide a definitive measurement of
the properties of the system.

\subsection{Reconstructed images and artifacts}
\label{sec_imaging_results}

In Figure \ref{fig_images} we present the best-fit model and model-independent
SQUEEZE reconstructions using the total variation regularizer.
By following the qualitative comparison method discussed in Section
\ref{sec_image_reconstruction_methods}, we may distinguish whether the features
seen in these images are real signals or artifacts of the reconstruction
process. A detailed discussion of this process, as well as BSMEM and other SQUEEZE
reconstructions, can be found in Appendix \ref{sec_reconstruction_comments}
and Figures \ref{2008-11_image_grid} - \ref{post_eclipse_image_grid}.
Summarizing this account, the true features are as follows: (1) the dark lane in
the F-star's southern hemisphere, interpreted to be the disk; (2) flux that
appears on the far South-West (or South-East) edge of the F-star during ingress
(egress); and (3) the presence or absence of the F-star's southern pole.
Although the post-eclipse interferometric observations do feature small non-zero
closure phases and photometry of the \epsAur{} system does show an intrinsic
$\Delta V \sim 0.1$ mag photometric variation outside of eclipse, we presently
regard any flux variations on the surface F-star as artifacts.

\section{Conclusions and Discussion}
\label{sec_discussion}

We have analyzed 106 nights of interferometric observations provided by four beam
combiners at three different interferometric facilities to derive properties of
the F-star, determine previously unknown orbital elements, and access the global
structures of the optically thick portion of the eclipsing disk. We have
reconstructed a series of model-independent images using SQUEEZE and BSMEM
programs with a variety of regularization functions. The images show that the
F-star appears circular (round) and free of egregious asymmetries prior to and
after the 2009-2011 eclipse. During the eclipse, most of the southern hemisphere
on the F-star is obscured. The appearance and persistence of this feature lead
us to confirm the Huang-Lissauer disk hypothesis for the eclipse. During most of
the epochs, the southern pole of the F-star is visible, thereby providing an
opportunity to measure the thickness of the disk at sub-milliarcsecond
resolution.

Our interferometric modeling efforts were complex: under a Bayesian framework,
we tested the observations of the F-star against seven different analytic limb
darkening prescriptions and differentiated between eight proposed disk models.
The Bayes factors listed in Table \ref{tbl_disk_model_selection} are
all exceptionally large. After conducting a comprehensive overview of \SIMTOI{}'s
rendering pipeline, \LIBOI{}'s unit test framework, and our use of MultiNest,
we found no mistakes in our implementation.
Hence these values are either true, or there is some unforeseen systematic
error for which or modeling process did not account.
Nevertheless, we are confident that our best-fit model is indeed the most
probable as it achieves the lowest reduced chi-squared estimate.
The pre- and post-eclipse observations indicate that the F-star has remained at
a more-or-less constant diameter for the last 14 years;
however, our observations are not sufficiently precise to definitively exclude
the 0.6\% yr$^{-1}$ contraction rate suggested by \cite{Saito1986}.
The average power-law limb
darkening coefficient, $\sim 0.5$, is much higher than predicted for an F-type
supergiant ($\sim 0.1 - 0.2$). The presence of small, $<5$ degree, non-zero
closure phases on the longest baselines in the post-eclipse observations,
coupled with variations in radius
 and limb darkening seen during the eclipse,
suggest that there may be convective cells or some other feature (e.g. spots)
on the surface of the F-star. These features, if they exist, have only a
minimal impact on our results.

We have simultaneously fit the disk models to a subset of the interferometric
and photometric data. We find that the data can be adequately fit by a variety
of models; however, the most consistent model was that of tilted disk derived
from an astrophysical density distribution.
The opaque region of the disk is seen nearly edge on and is remarkably uniform.
These conclusions are supported by spectroscopic work by \cite{Griffin2013},
who have shown the precise repetition of disk-related spectral features have
not changed appreciably over the last century.
Therefore, it is unlikely that the disk is significantly twisted or warped
\citep{Kumar1987}.
We do, however, find evidence that the disk may be slightly tilted out of the
orbital plane. If this is true, the difference between our model and the
\cite{Kemp1986} polarization result could be attributed to precession.

The thickness and inclination of the disk exclude the possibility that a the
purported mid-eclipse brightening is caused by light penetrating a central
clearing in the disk \citep[cf.][]{Wilson1971} or the notion that a series of
semi-transparent rings are responsible for the photometric variations seen
during totality \citep[cf.][]{Ferluga1990,Ferluga1989}.
We suggest that these light curve features, if true, have other physical causes.
For example, a mid-eclipse brightening could be due to scattering above and
below the plane of the disk \citep[e.g.][]{Budaj2011, Muthumariappan2012},
perhaps in the same region responsible for the increase in He 10830 \AA{} absorption
\citep{Stencel2011}. Likewise, the manifestation of $\Delta V \sim 0.1$ mag
variations during totality are probably orbitally-excited non-radial pulsation
of the F-star \citep{KloppenborgB.K.2012}, rather than substructure in the disk.

Lastly we predict that the secondary eclipse will occur
between $\sim$ JD 2,461,030 - 2,461,860 (2025 Dec. 20 - 2028 Mar. 29). We encourage
a comprehensive photometric campaign during this time focusing on NIR, mid-IR,
and far-IR observations to confirm this prediction.

\acknowledgments
{
\textbf{Acknowledgements:}
The CHARA Array, operated by Georgia State University, was built with funding
provided by the National Science Foundation grant AST-0606958, Georgia State
University, the W. M. Keck Foundation, and the David and Lucile Packard Foundation.
This research is
supported by the National Science Foundation as well as by funding from the
office of the Dean of the College of Arts and Science at Georgia State
University. MIRC was supported by the National Science Foundation. The Navy
Precision Optical Interferometer is a joint project of the Naval Research
Laboratory and the US Naval Observatory, in cooperation with Lowell Observatory,
and is funded by the Office of Naval Research and the Oceanographer of the Navy.
Participants from the University of Denver are grateful for the bequest of
William Hershel Womble in support of astronomy at the University of Denver. They
acknowledge support from National Science Foundation through ISE grant
DRL-0840188 to the American Association of Variable Star Observers and AST grant
10-16678 to the University of Denver.
}

{\it Facilities:} \facility{CHARA}, \facility{NPOI}, \facility{PTI}.

% Begin the appendix:
\appendix
\noindent
\setcounter{figure}{0} \renewcommand{\thefigure}{A.\arabic{figure}}

% First section discussing BSMEM images.
\section{Image reconstruction and artifact discussion}
\label{sec_reconstruction_comments}

In the following section we elaborate on the image reconstruction and artifact
detection process for all epochs. We present the best-fit SIMTOI model and
reconstructed with SQUEEZE and BSMEM. All images are rendered with North up and
East to the left. The $0.5$ \mas{} H-band or $0.7$ \mas{} K-band resolution
limit of the data is indicated by the circle in the lower-left hand corner of
the model image.

The SQUEEZE reconstructions were conducted using the Laplacian (abbreviated LA),
Total Variation (abbreviated TV), and L0-norm (abbreviated L0) regularizers.
Images were reconstructed in a grid with regularizer weights ranging from
$0.1 - 100,000$ in semi-logarithmic steps. The images with the lowest reduced
chi-squared were selected for presentation. Most of the epochs were reconstructed
from a Dirac starting image. However, the sparse UV coverage in the 2008-11,
2010-02, and 2011-04 mandated we use the best-fit model images to initialize
the flux distribution. SQUEEZE was executed with 50 threads with 500
realizations each.

All BSMEM reconstructions were performed using flat priors and a $1.4$ \mas{}
diameter Gaussian for entropy estimation. We have used the ``full'' elliptical
approximation for the bispectra uncertainties (see BSMEM documentation for
details) which we found dramatically suppress reconstruction artifacts.

\subsection{2008-11}

Figure \ref{2008-11_image_grid} shows the resulting model and image from this
set of two four-telescope MIRC observations that were taken at nearly the
same hour angle. The image reconstruction in this epoch is quite poor when
compared with later epochs. By inspection of the SED
\citep{D.W.HoardandD.LadjalandR.E.StencelandS.B.Howell2012,Hoard2010}
we know that the F-star is the dominant source of flux in the H-band, hence
we interpret this to be an image of the F-star. All of the images show that the
flux is mostly constrained within the bounds of the best-fit model.
The model-independent image shows the star as approximately
round with some surface features. These features are mostly replicated in the
model-independent reconstruction from the synthetic data, hence the non-circular
structure and small photometric variations seen on the F-star are artifacts of
the reconstruction process. Therefore the F-star before the eclipse does not appear
to have any egregious asymmetries which might interfere with later observations.

\subsection{2009-11}

In Figure \ref{2009-11_image_grid} we show the best-fit model and model-independent
images reconstructed from this eclipse ingress phase epoch. For reasons discussed
above \citep[and in greater detail in][]{Kloppenborg2010}, we again interpret
the bright source to be the F-star and the dark region occuring in the southern
half of the image as the disk intruding into the line of sight. We are not aware of any
evidence which suggests the F-star’s rotation is misaligned to the binary's
orbit, hence we shall call the un-eclipsed portion of the F-star the ``northern
hemisphere.'' Likewise, we will refer to the (mostly) eclipsed portion as the
``southern hemisphere.'' The North pole of the F-star would be located at a position
angle of $\sim 26$ degrees.

By comparing the best-fit model from SIMTOI to the sampled model reconstruction,
we may qualitatively access the presence of artifacts in the image. The dark
spot in the northern hemisphere and two bright spots in the East/West near the
limb appear to be artifacts. The straight edges along the perimeter of the F-star
are a common artifact caused by the UV coverage of the data set.

Despite the large number of artifacts, several real features may be discerned.
For example, the southern pole is seen in the real image, model, and synthetic
reconstruction; hence we feel this feature is real. Likewise, the small amount
of flux seen on the western edge of the disk intrusion is also real.

\subsection{2009-12}

The best-fit model and reconstructed images of this second ingress epcoh are
shown in Figure \ref{2009-12_image_grid}.
This data has excellent UV coverage and appears similar, in many regards, to
the 2009-11 epoch. Like previous observations, the straight-edge appearance of
the F-star is an artifact of the UV coverage. It is probable that the spots
seen in the northern hemisphere of the F-star are also artifacts.

The southern pole again appears quite strong in the real image, model, and
synthetic reconstruction, implying this feature is likely real. For the same
reason, we regard the small quantity of flux at the west edge of the disk
intrusion to be a real feature rather than an artifact of the reconstruction
process.

\subsection{2010-02}

The UV coverage at this epoch is extremely poor, consisting of two
four-telescope observation with MIRC. Hence the model-independent images shown
in Figure \ref{2010-02_image_grid} are difficult to interpret without information
garnered from the model and H-band photometry. Both the model and images imply
that the entire southern hemisphere and a small fraction of the northern
hemisphere are covered. This conclusion is supported, at least circumstantially,
by the H-band photometry being at its faintest at this time.
The appearance of bright spots in the northern hemisphere is most likely caused
by limited UV coverage or the reconstruction process, rather than any real
surface flux variations on the F-star.
The model independent and synthetic images agree quite well about the over-all
appearance of the F-star during this epoch.

\subsection{2010-08 / 2010-09 / 2010-10 / 2010-11 / 2010-12 / 2011-01}

The qualitative appearance of these epochs is quite similar (see
Figure \ref{totality_image_grid}), hence they will be
discussed in aggregate. The obscuration by the disk remains remarkably
consistent across five months of observations. The occasional spot in the
F-star’s northern hemisphere, scalloped edge of the disk along the F-star’s
equator, and flux variations along the F-star’s equator are frequently seen in
the real data and synthetic reconstructions, hence these are likely artifacts.
The southern pole has re-appeared. It appears in the real image, model, and
synthetic image therefore we regard this as a true feature in the image.

It is important to note that the mid-eclipse observation (2010-08) shows the
disk as entirely opaque. Hence mid-eclipse brightening hypotheses, which rely
on a large opening in the disk, are unlikely. Likewise, although photometric
variations were seen during this phase of the eclipse, no significant flux
variations were seen within the disk plane, hence the photometric variations
are not likely a result of flux penetrating semi-transparent gaps in the disk's
midplane.

\subsection{2011-04}

In Figure \ref{2011-04_image_grid} we present the only interferometric images
of the \epsAur{} during the egress phase and the first image created with
data taken by the CLIMB beam combiner. Unlike all other data sets, the L0-norm
images also required the use of the uniform disk regularizer in SQUEEZE.
These model-independent K-band images show that a portion of the F-star's
South-East edge is no longer obscured by the disk.
This notion is in excellent agreement with the SIMTOI model and observed
photometry (e.g. see Figure \ref{fig_bootstrapped_interferometric_results}).
A bulk comparison of the reconstructed model vs. real data implies that the
large concentration of flux in the F-star's northern hemisphere is probably an
artifact, whereas the small crescent of the F-star seen in the East is real.

\subsection{2011-09-18 and later}

All of the post-eclipse images are displayed in Figure \ref{post_eclipse_image_grid}.
Much like the ingress phase images, these reconstructions show that the F-star
has no egregious asymmetries that cannot be explained by UV coverage or
reconstruction artifacts. We note that two of the post-eclipse epochs
(2011-10-10 and 2011-11-03) do show non-zero closure phase and different locations
of the first visibility null, indicating that the F-star may harbor spots or be
slightly oblate. Both of these effects, if real, are insignificant compared to
the variations that the disk imparts upon the interferometric data during the
eclipse. We will attempt to quantify the presence of spots or asymmetry in
greater detail in a future publication.




\bibliography{epsAur-interferometry}

\clearpage

\setcounter{figure}{0} \renewcommand{\thefigure}{\arabic{figure}}

%% the file, uncomment the \plotone, \plottwo, and
%% \includegraphics command.gifs
%%
%% If you need a layout that cannot be achieved with \plotone or
%% \plottwo, you can invoke the graphicx package directly with the
%% \includegraphics command or use \plotfiddle. For more information.gif,
%% please see the tutorial on "Using Electronic Art with AASTeX" in the
%% documentation section at the AASTeX Web site, http://aastex.aas.org/
%%
%% The examples below also include sample markup for submission of
%% supplemental electronic materials. As always, be sure to check
%% the instructions to authors for the journal you are submitting to
%% for specific submissions guidelines as they vary from

%% This example uses \plotone to include an EPS file scaled to
%% 80% of its natural size with \epsscale. Its caption
%% has been written to indicate that additional figure parts will be
%% available in the electronic journal.

\clearpage
\begin{figure}
\includegraphics[height=\linewidth]{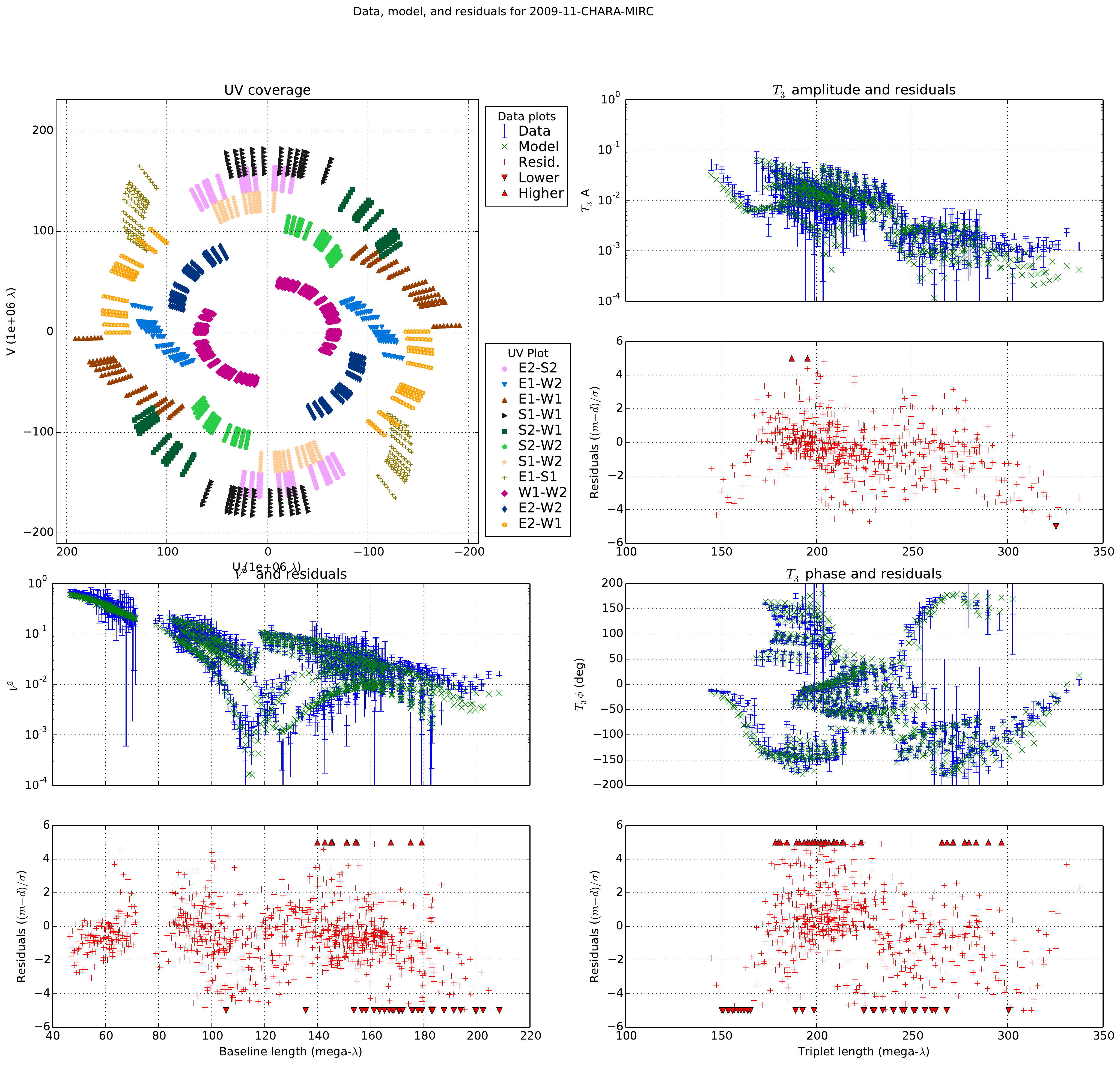}
\caption{UV coverage, data, best-fit bootstrapped model, and residual plot for
the 2009-11 CHARA-MIRC epoch. This data, taken during the ingress phase of
the eclipse, shows clear departure from circular symmetry in all measured
quantities as evidenced by the visibilities, triple amplitudes and closure
phases being significantly different at similar baseline/triplet lengths
(baseline lengths summed in quadrature).
Equivalent figures for other epochs can be found in
the online appendix, Figures \ref{data_plots_start} - \ref{data_plots_end}.
\label{fig_data_example}}
\end{figure}

\clearpage
\begin{figure}
\plotone{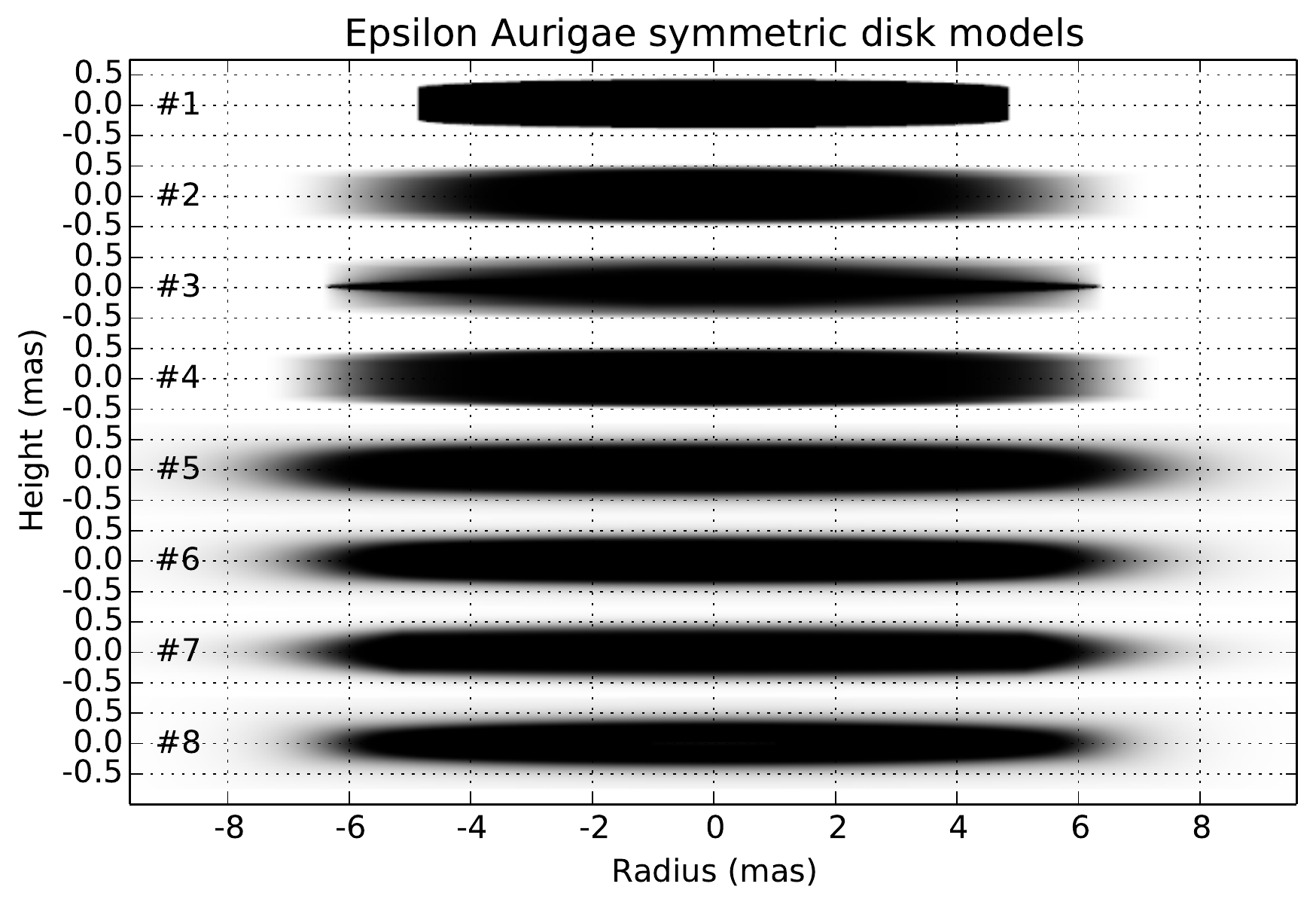}
\caption{The eight best-fit symmetric disk models resulting from a simultaneous
minimization to both the photometric and interferometric data. The models are
as described in Section \ref{sec_disk_models}. All models have been
rotated in position angle, but otherwise appear as they would when occulting
the F-star's photosphere. Model \#8 has the greatest evidence value and was
adopted for the remainder of this work.}
\label{fig_disk_models}
\end{figure}

\clearpage

\begin{figure}
\plotone{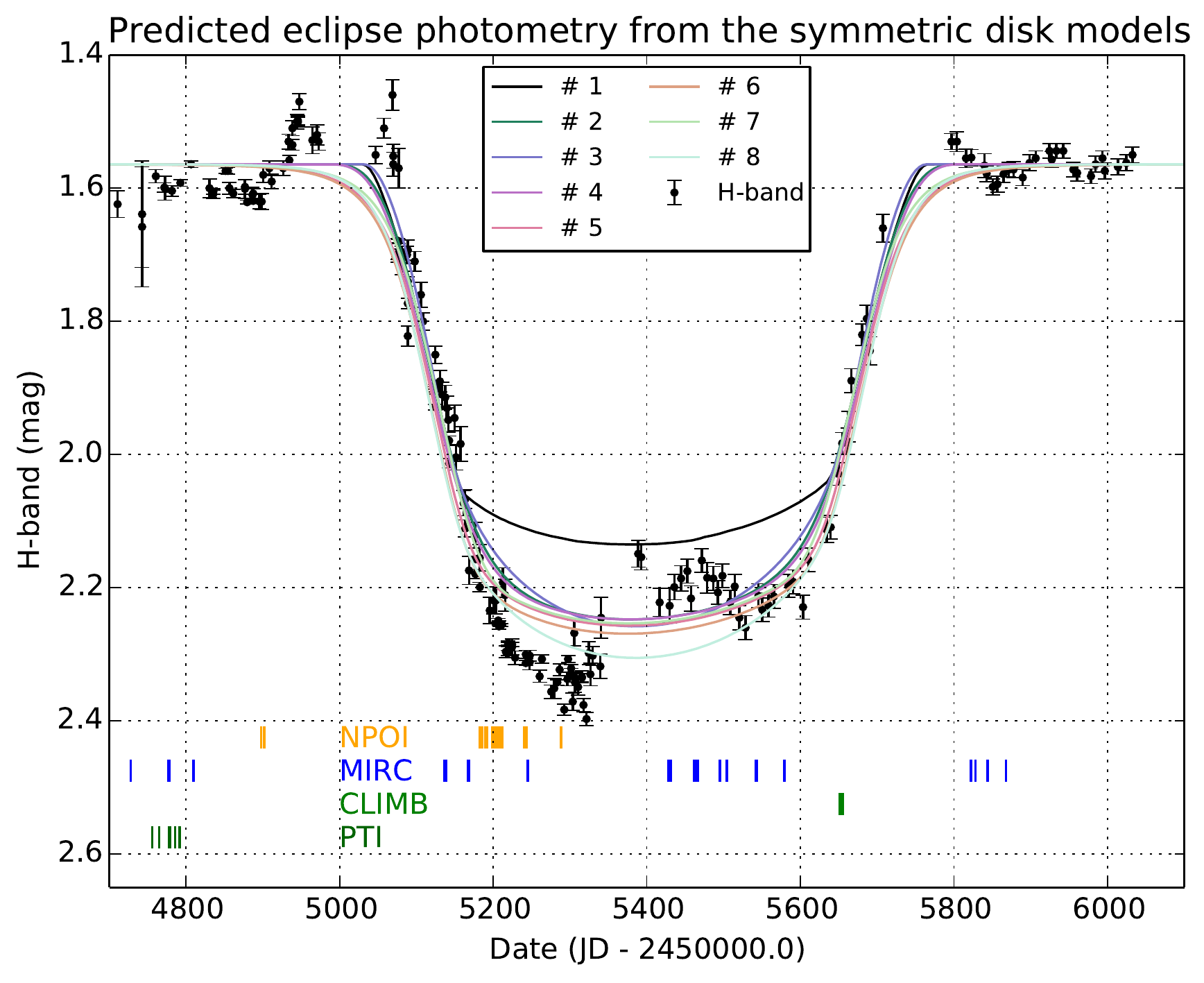}
\caption{
Observed and simulated H-band photometry from the eight best-fit symmetric disk
models as a function of time. The four rows of tick marks at the bottom of the
plot indicate the time of observation by the corresponding beam combiner. Tick
marks appear thicker when successive nights of observations occurred. None of
the symmetric disk models entirely reproduce all features of the light curve,
implying that the disk is not symmetric. The F-star's $\Delta H \sim 0.05$ mag
out-of-eclipse variations are clearly seen at all phases of the eclipse.}
\label{fig_photometry}
\end{figure}

\clearpage

\begin{figure}
\includegraphics[width=0.7\linewidth]{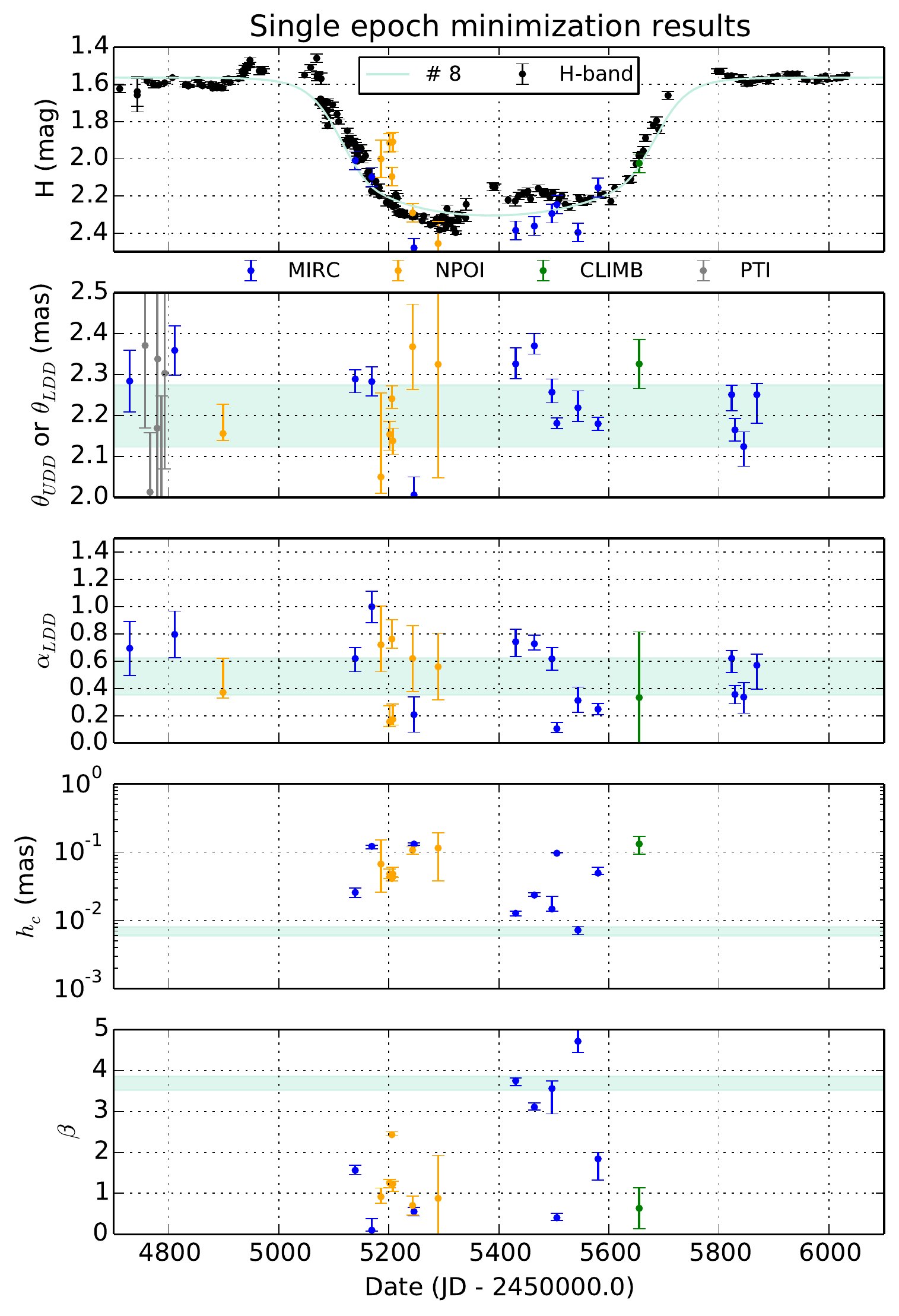}
\caption{\label{fig_bootstrapped_interferometric_results}
\small
Results of single-epoch minimizations to the interferometric (and in two cases
photometric) data. Top: The observed and predicted H-band photometry as a function
of time. Superimposed is the best-fit symmetric disk model (\#8) and predicted
H-band magnitudes from the single-epoch fits. Bottom: the best-fit F-star
diameter (limb darkened or uniform); limb darkening coefficient; disk scale
height; and disk radial exponent, $\beta$, as a function of time. The horizontal
bands denote the 1-$\sigma$ limits on the parameters from model eight.
The diameter of the F-star is largely consistent, regardless of the wavelength
of observation. This indicates there is not an egregious systematic calibration
error between the different interferometric facilities.
}
\end{figure}

%
% Reconstructed image array. This is broken over several pages
%
\clearpage
\begin{landscape}
\setcounter{figure}{4}
\begin{figure}
\includegraphics[width=0.50\linewidth]{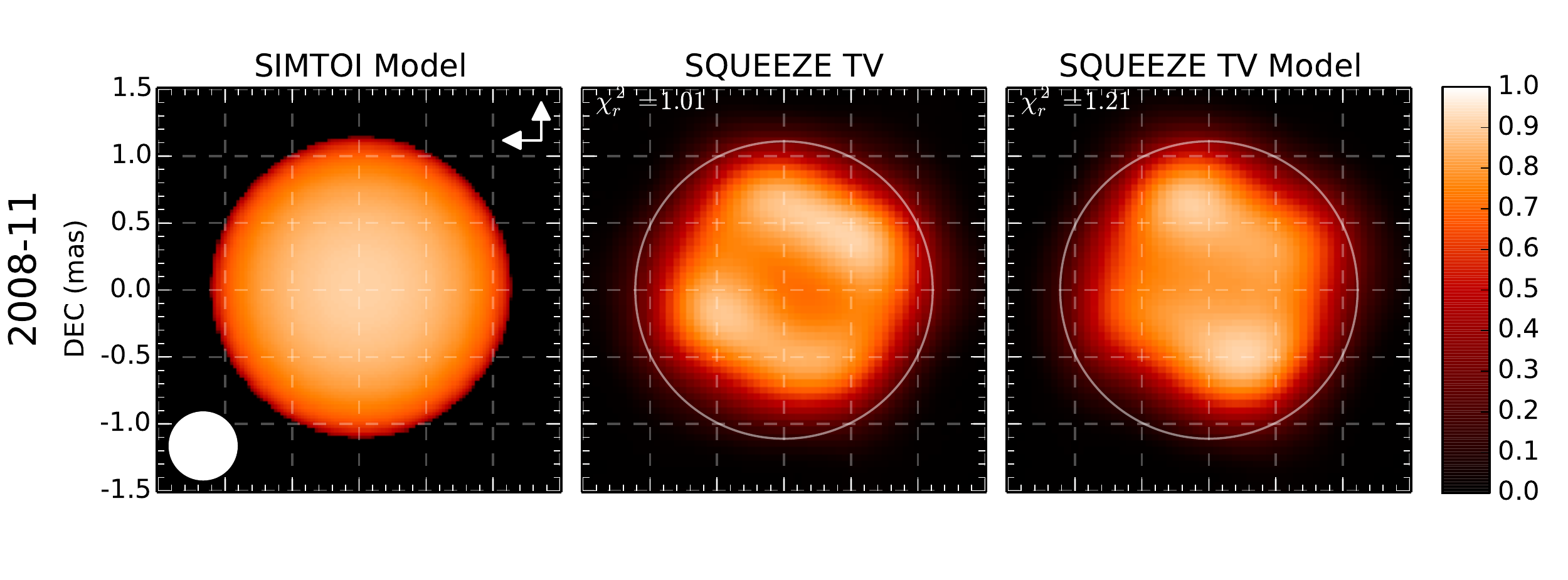}
\includegraphics[width=0.50\linewidth]{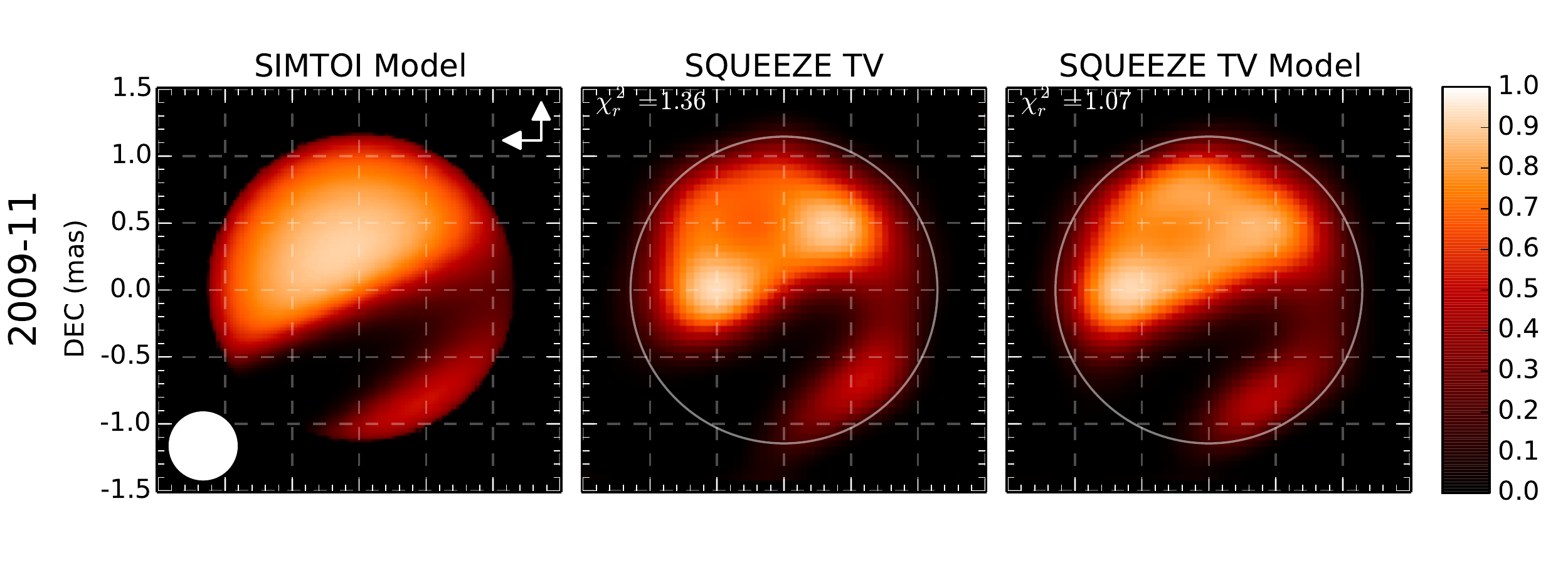}
\includegraphics[width=0.50\linewidth]{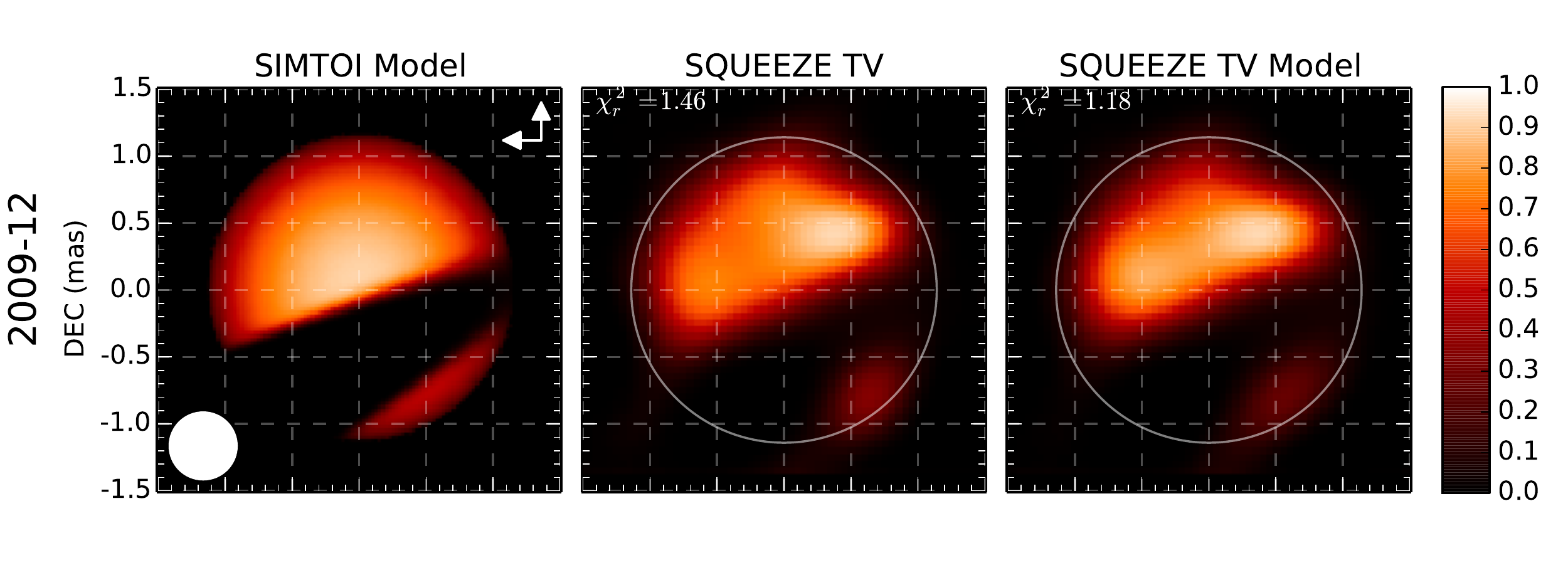}
\includegraphics[width=0.50\linewidth]{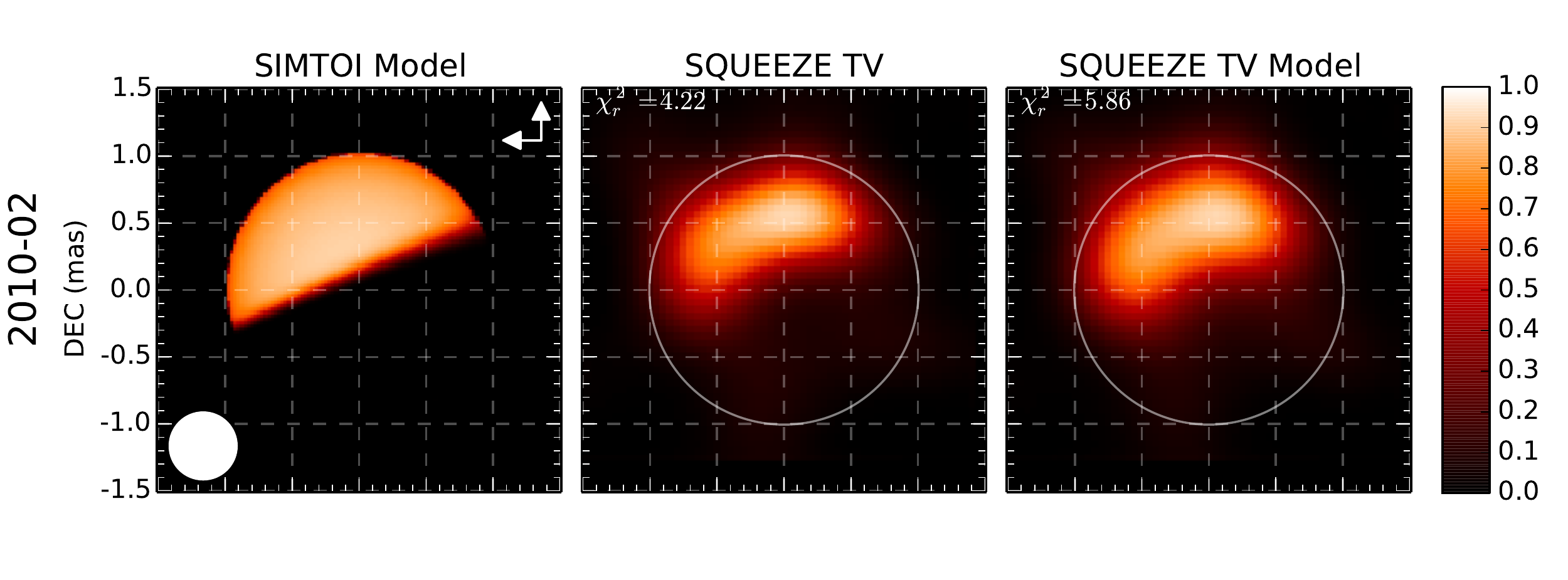}
\includegraphics[width=0.50\linewidth]{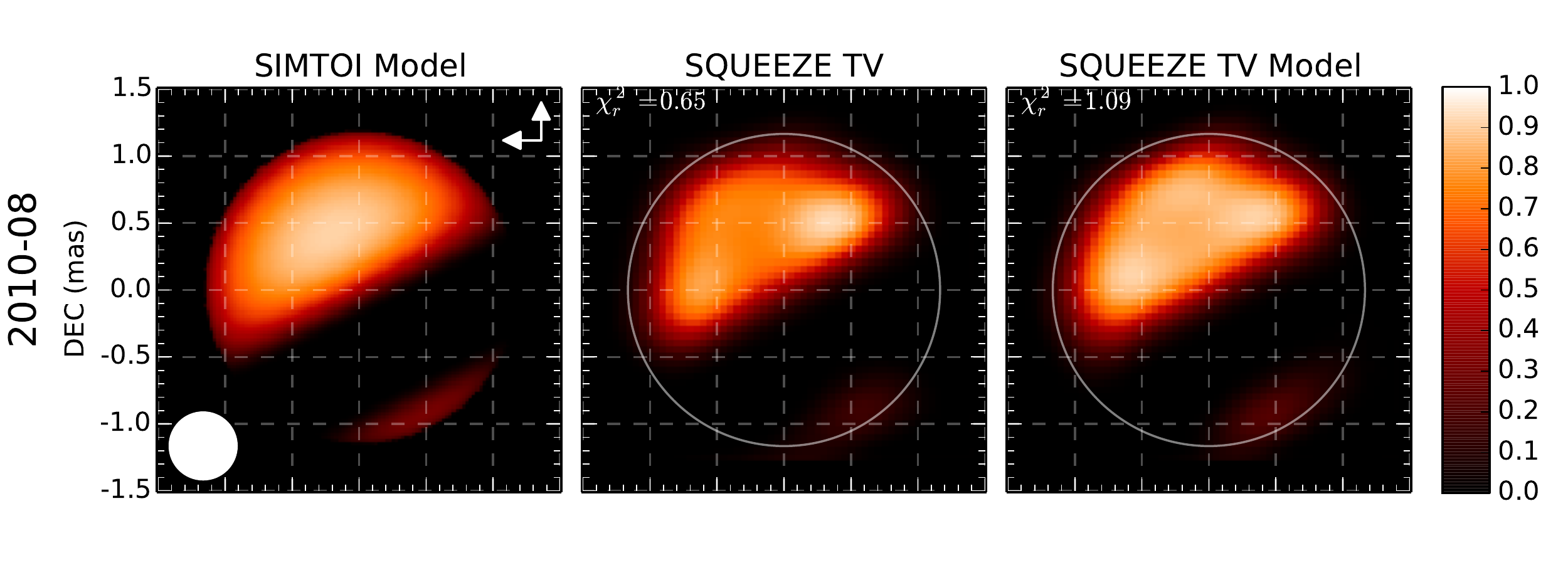}
\includegraphics[width=0.50\linewidth]{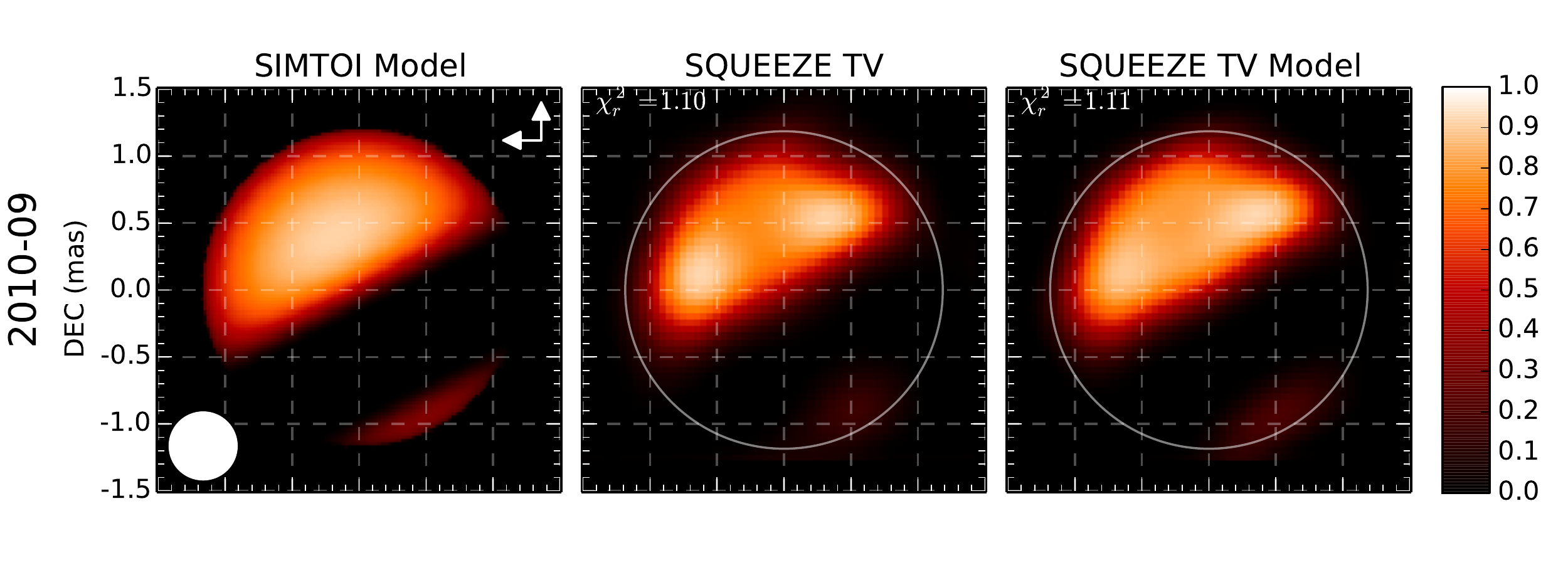}
\caption{Best-fit models and SQUEEZE total variation (TV) regularizer reconstructions
for each CHARA epoch. All images are displayed with North up and East to the left.
The $0.5$ \mas{} resolution limit of MIRC and $0.7$ \mas{} limit of CLIMB are
indicated by the solid white circles in the lower-left of the first image.
The remaining white circle indicates the best-fit SIMTOI angular diameter.
In each epoch the columns contain the following images (left) the best-fit
model from SIMTOI, (center) the model-independent reconstruction of the
interferometric data, and (right) a reconstruction of synthetic data generated
from the model as described in Section \ref{sec_image_reconstruction_methods}.
For a detailed discussion of each epoch please see Appendix \ref{sec_reconstruction_comments}.
}
\label{fig_images}
\end{figure}

\setcounter{figure}{4}
\begin{figure}
\includegraphics[width=0.50\linewidth]{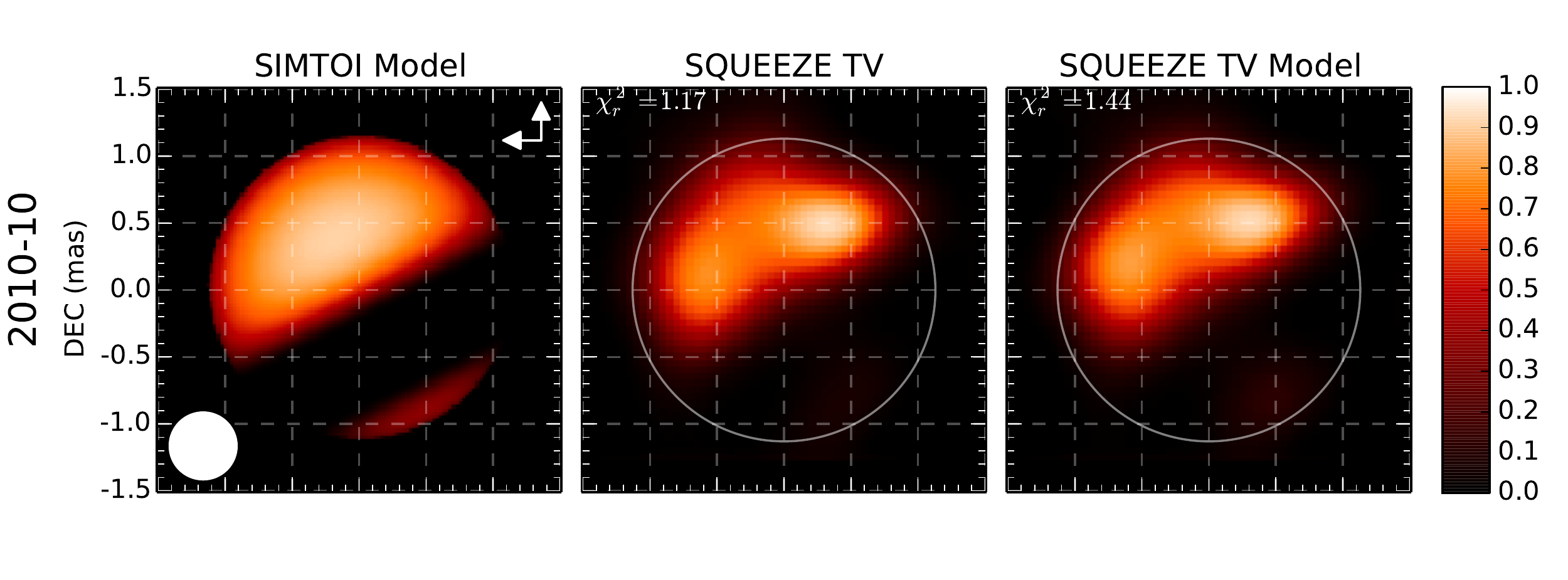}
\includegraphics[width=0.50\linewidth]{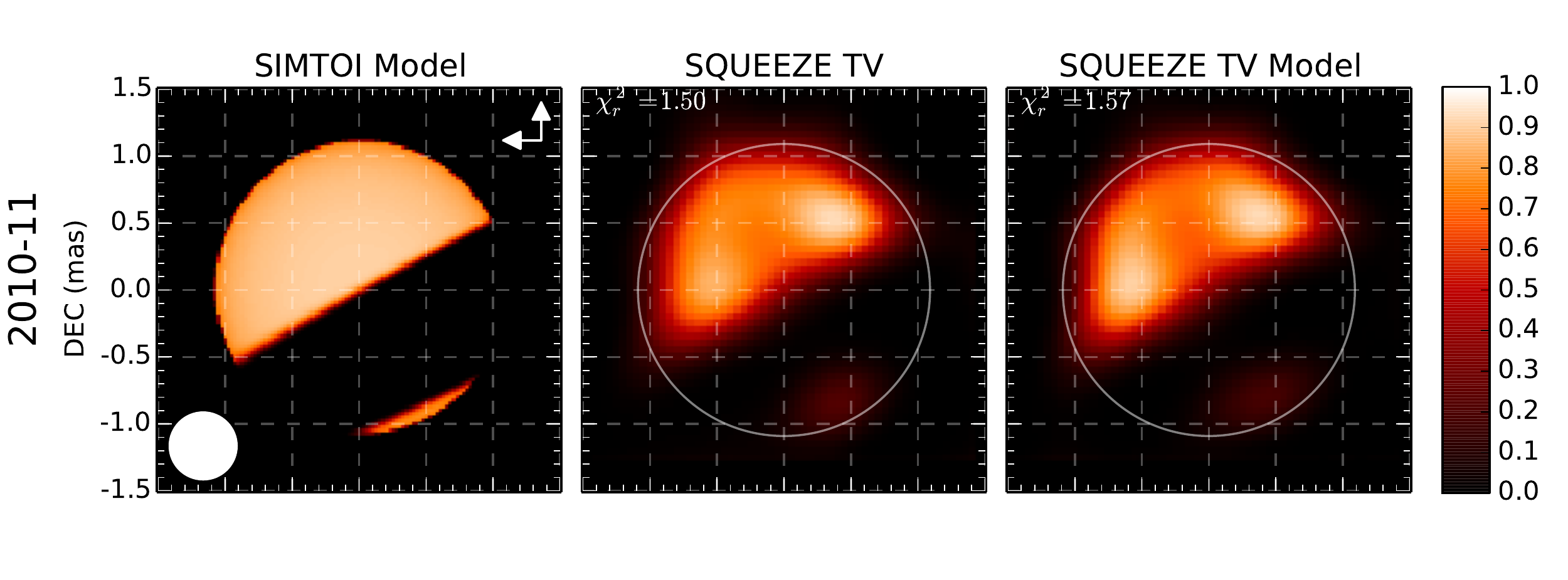}
\includegraphics[width=0.50\linewidth]{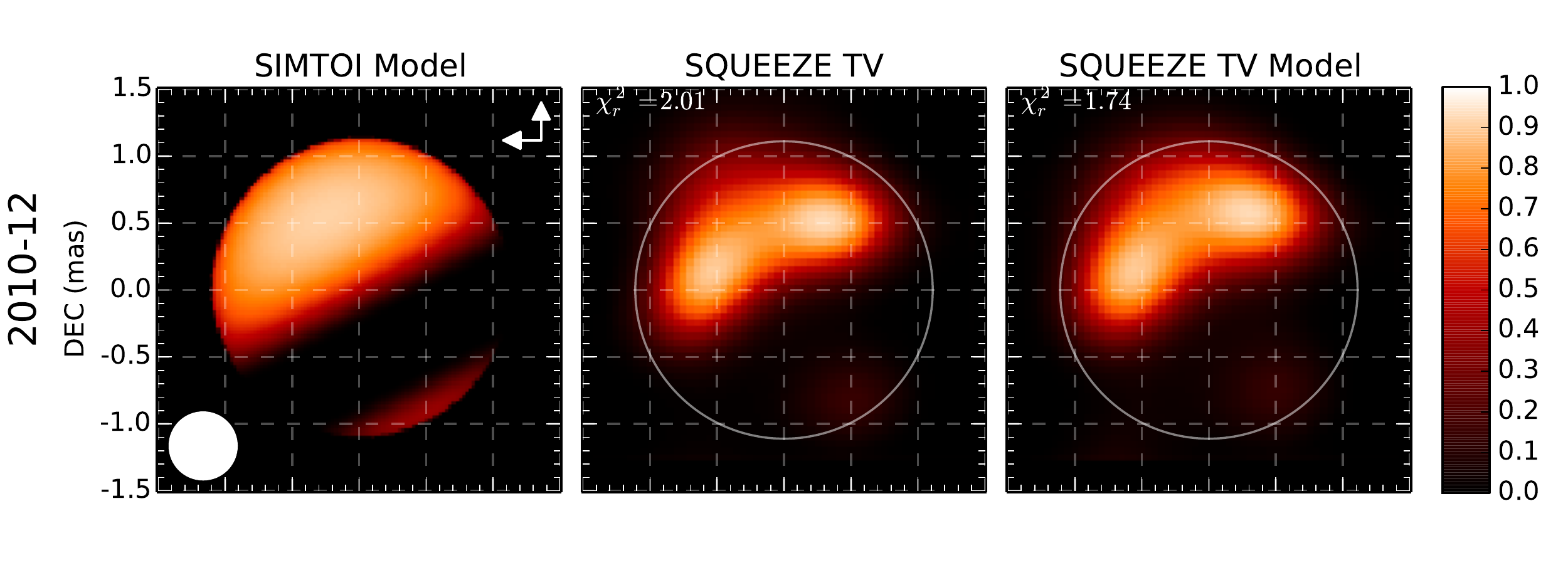}
\includegraphics[width=0.50\linewidth]{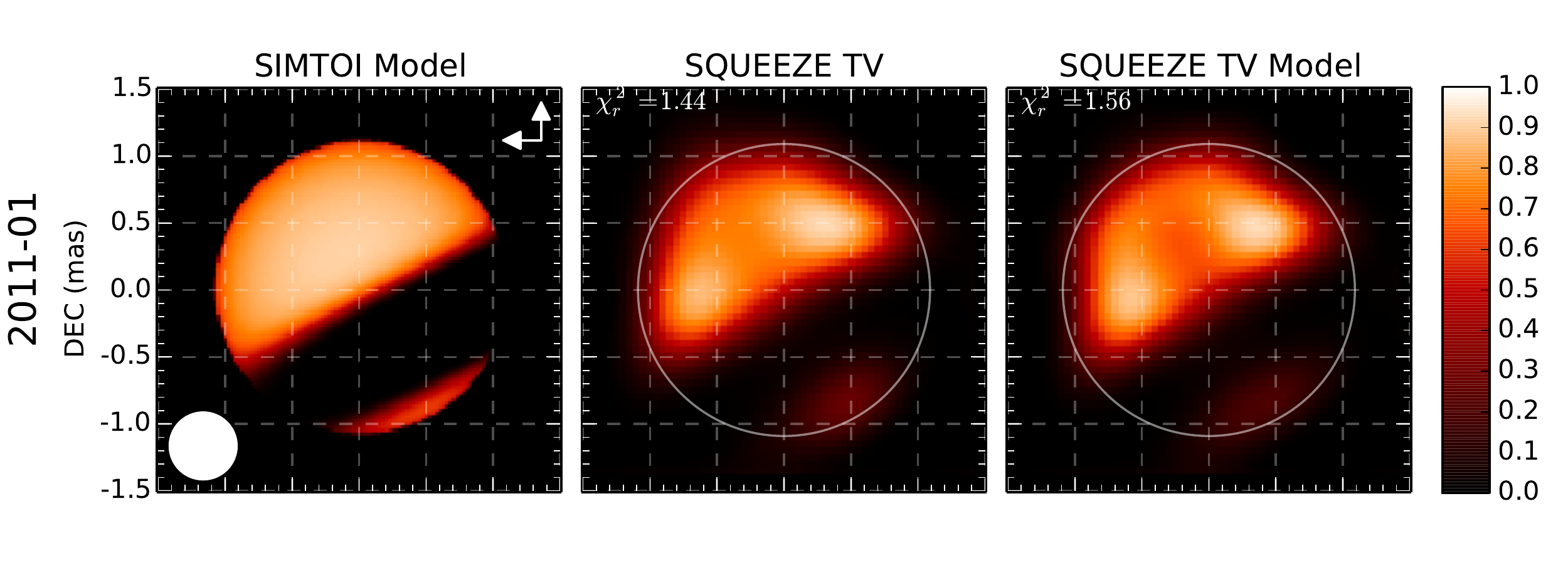}
\includegraphics[width=0.50\linewidth]{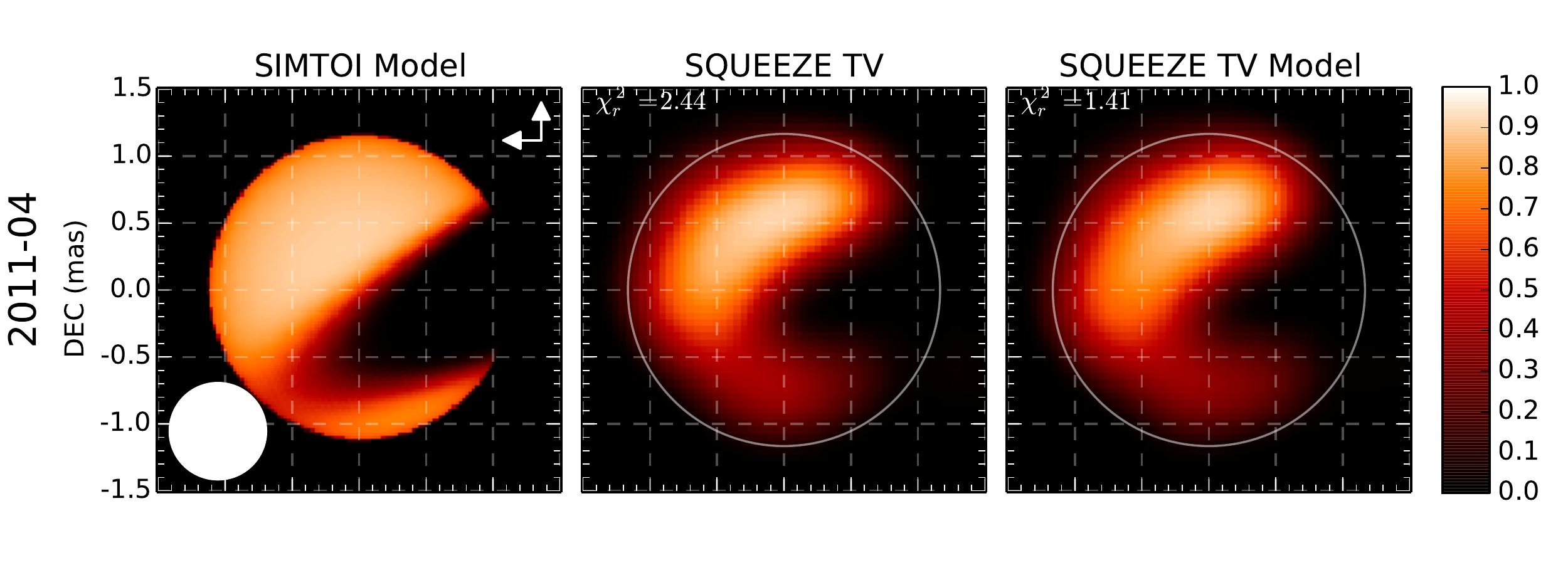}
\caption{(continued)}
\end{figure}

\setcounter{figure}{4}
\begin{figure}
\includegraphics[width=0.50\linewidth]{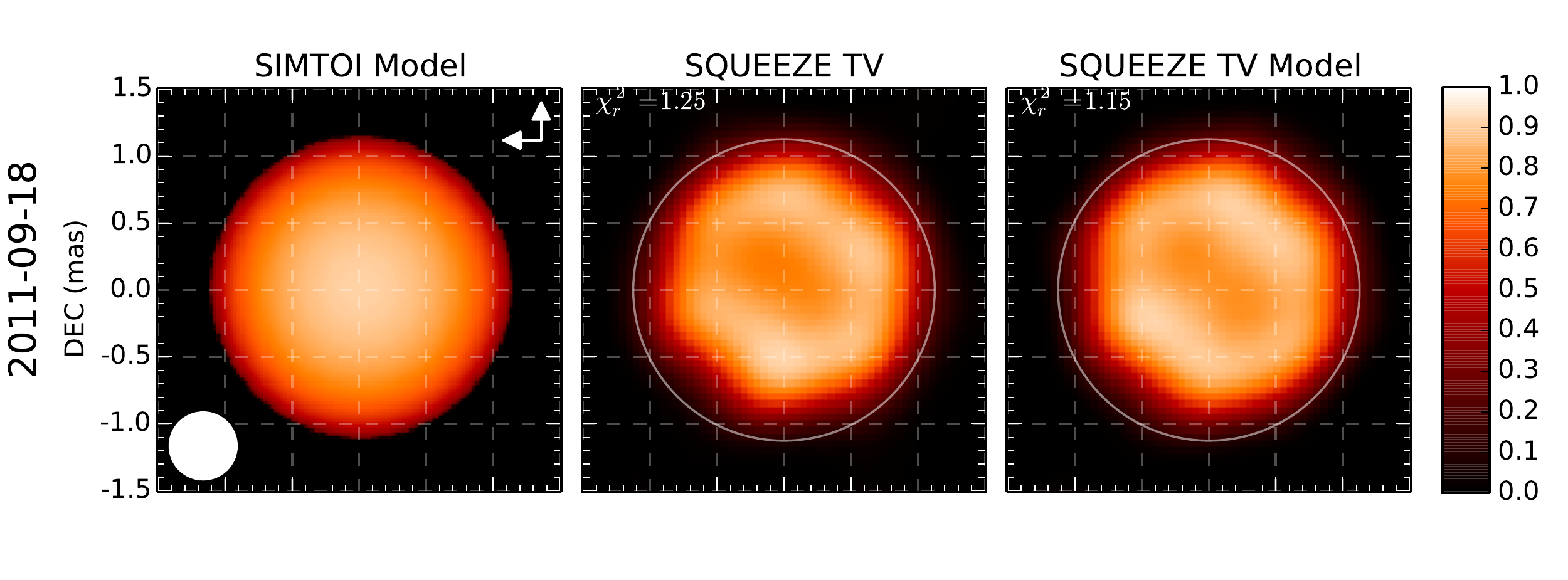}
\includegraphics[width=0.50\linewidth]{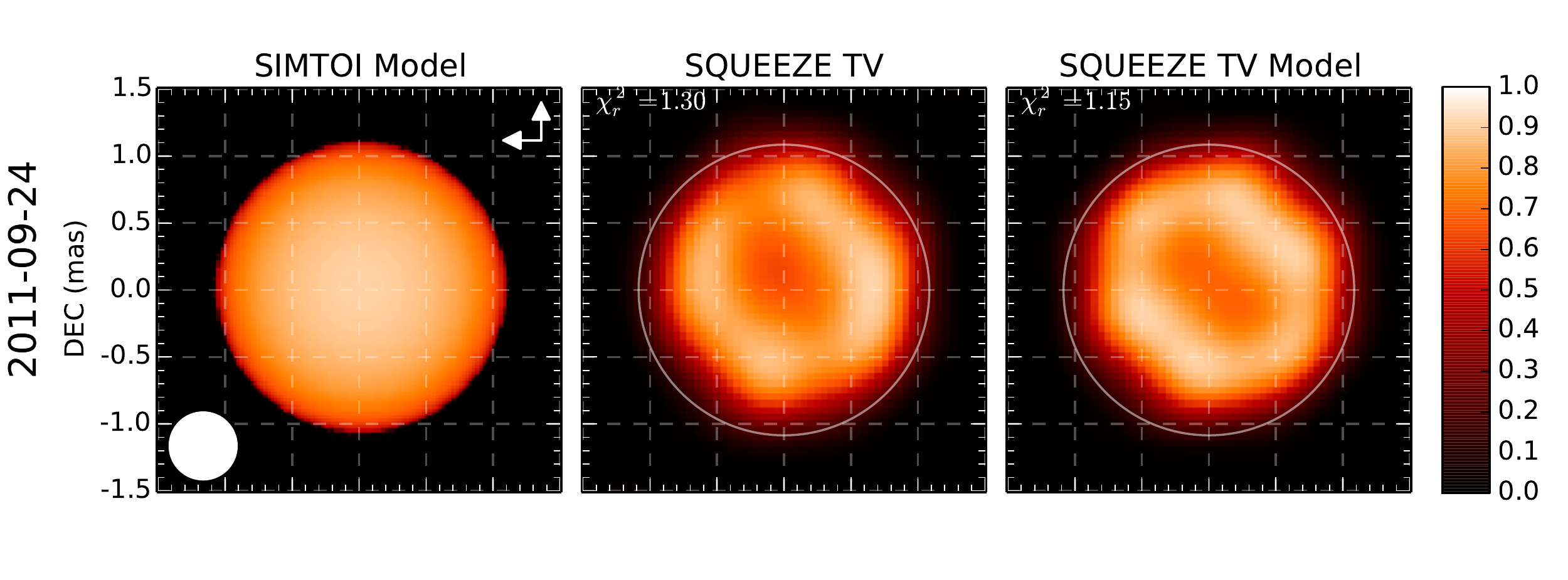}
\includegraphics[width=0.50\linewidth]{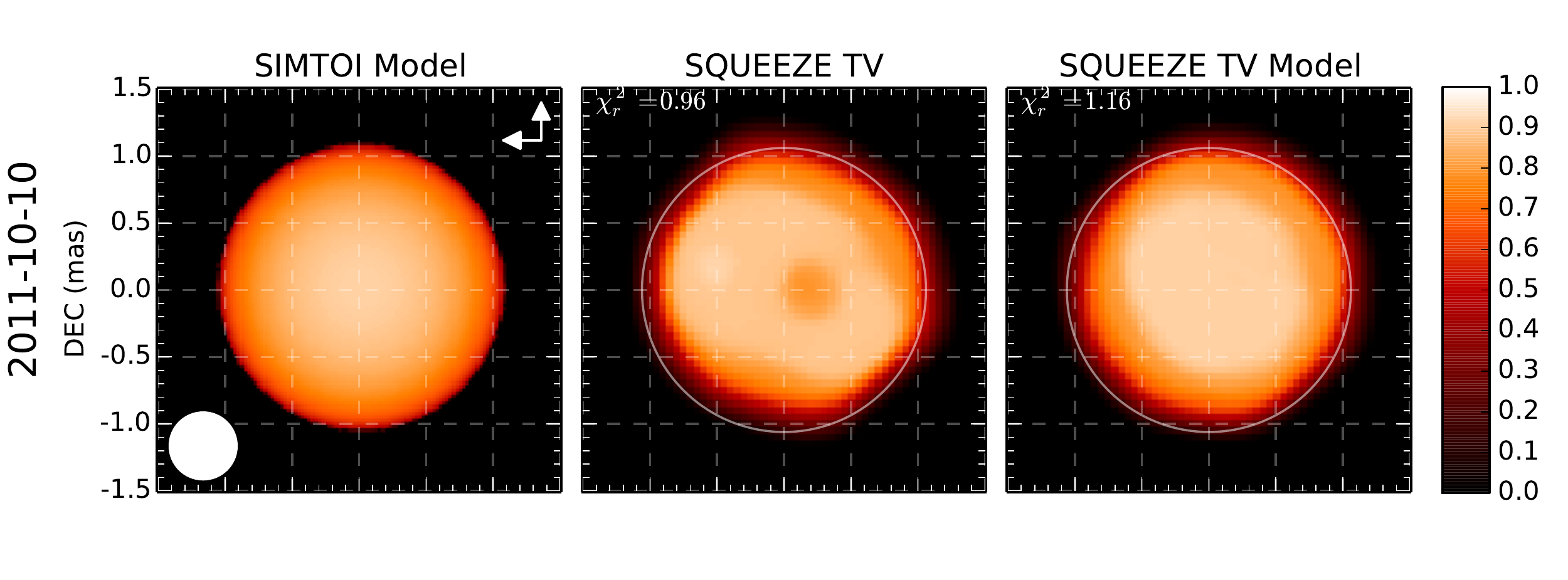}
\includegraphics[width=0.50\linewidth]{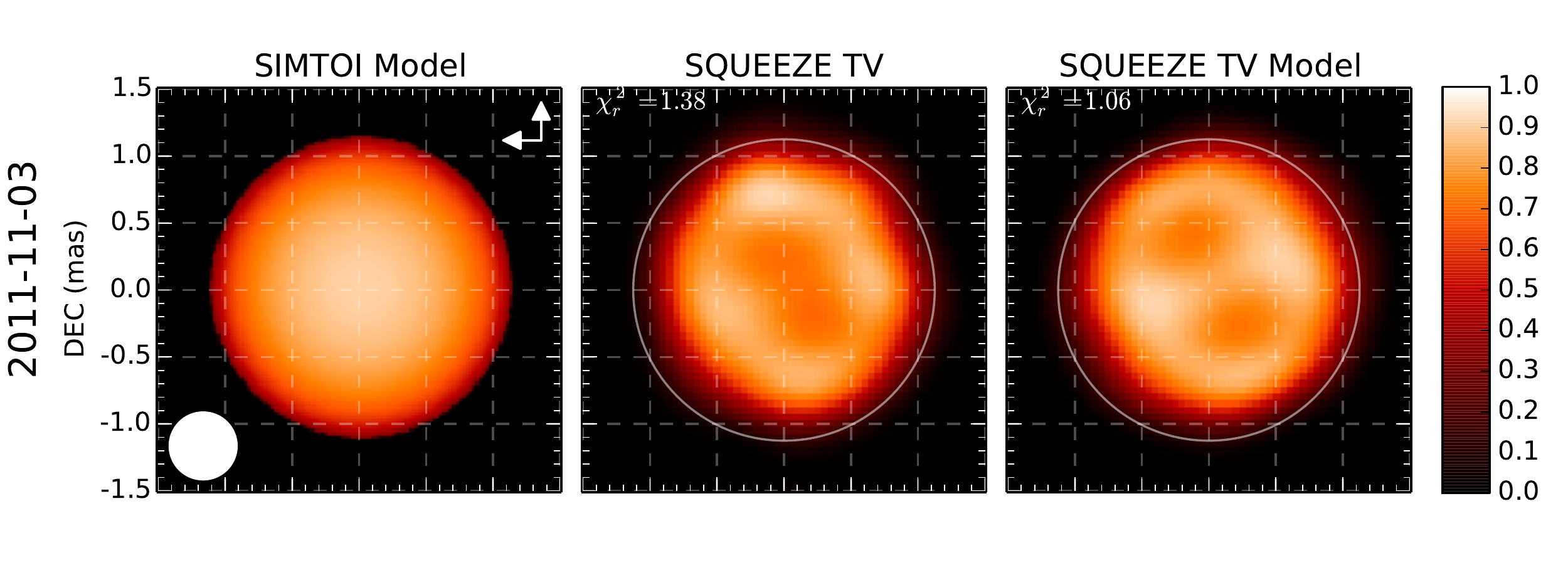}
\caption{(continued)}
\end{figure}

\end{landscape}

% Appendix figures

\appendix
\setcounter{figure}{0}
\renewcommand{\thefigure}{A\arabic{figure}}

%
% data/model plots
%
\clearpage
\begin{figure}
\includegraphics[height=\linewidth]{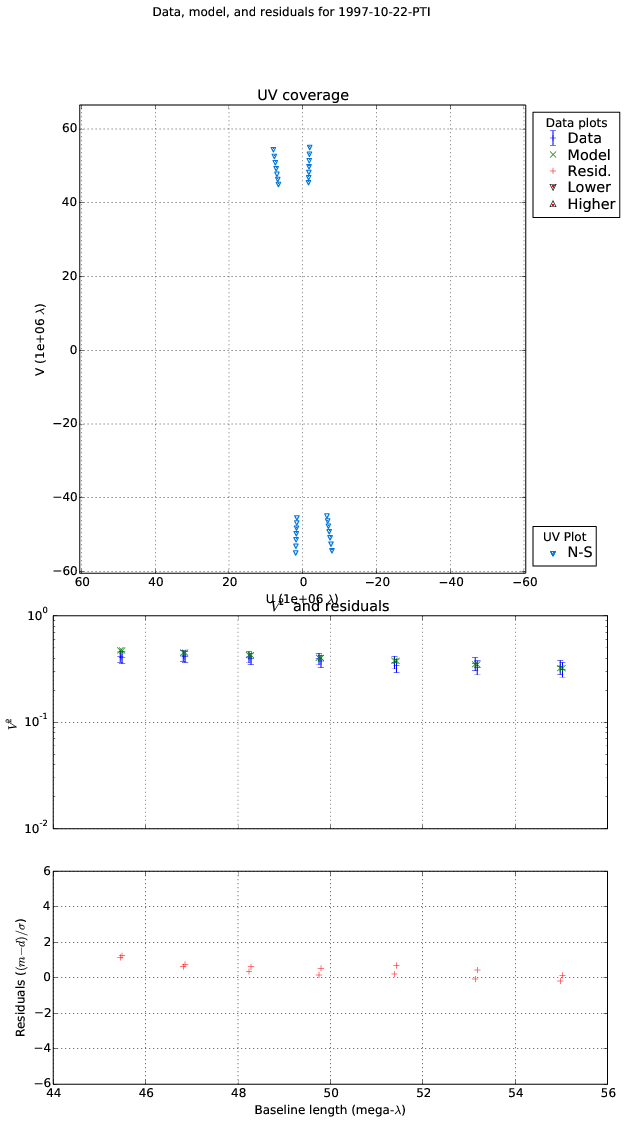}
\caption{UV coverage, data, model, and residual plots for the
1997-10-22-PTI data set. \label{data_plots_start}}
\end{figure}

\clearpage
\begin{figure}
\includegraphics[height=\linewidth]{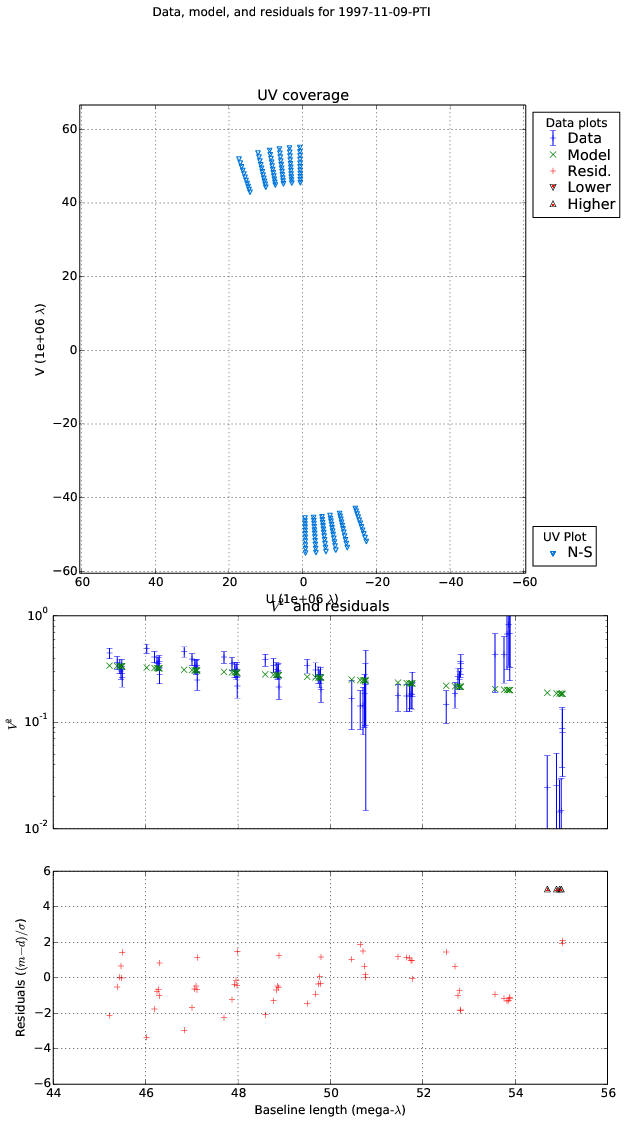}
\caption{UV coverage, data, model, and residual plots for the
1997-11-09-PTI data set.}
\end{figure}

\clearpage
\begin{figure}
\includegraphics[height=\linewidth]{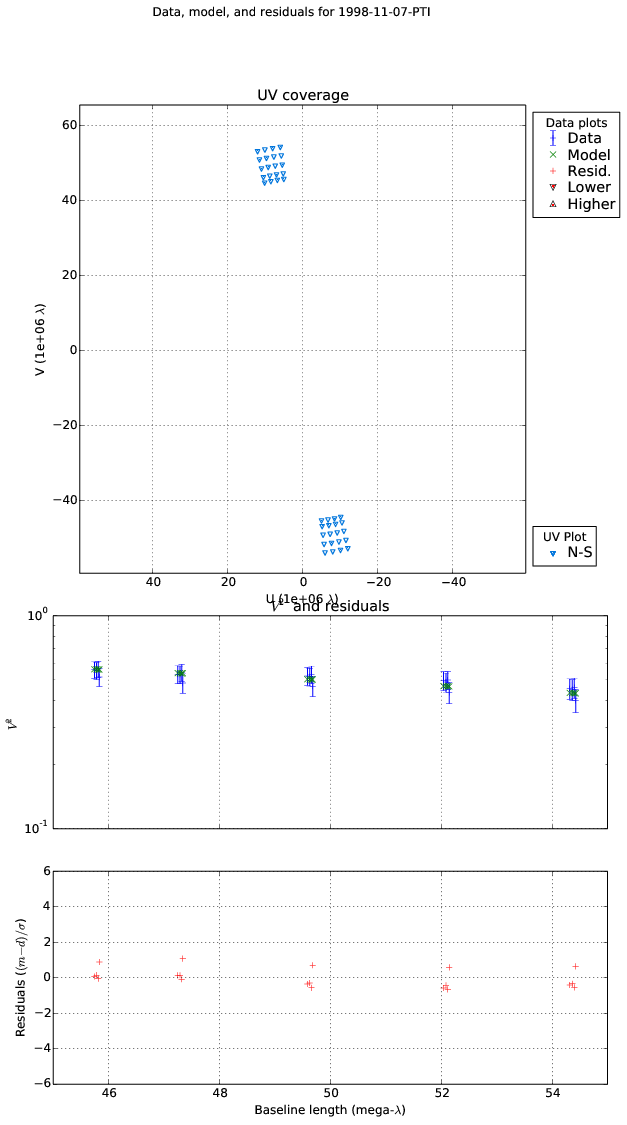}
\caption{UV coverage, data, model, and residual plots for the
1998-11-07-PTI data set.}
\end{figure}

\clearpage
\begin{figure}
\includegraphics[height=\linewidth]{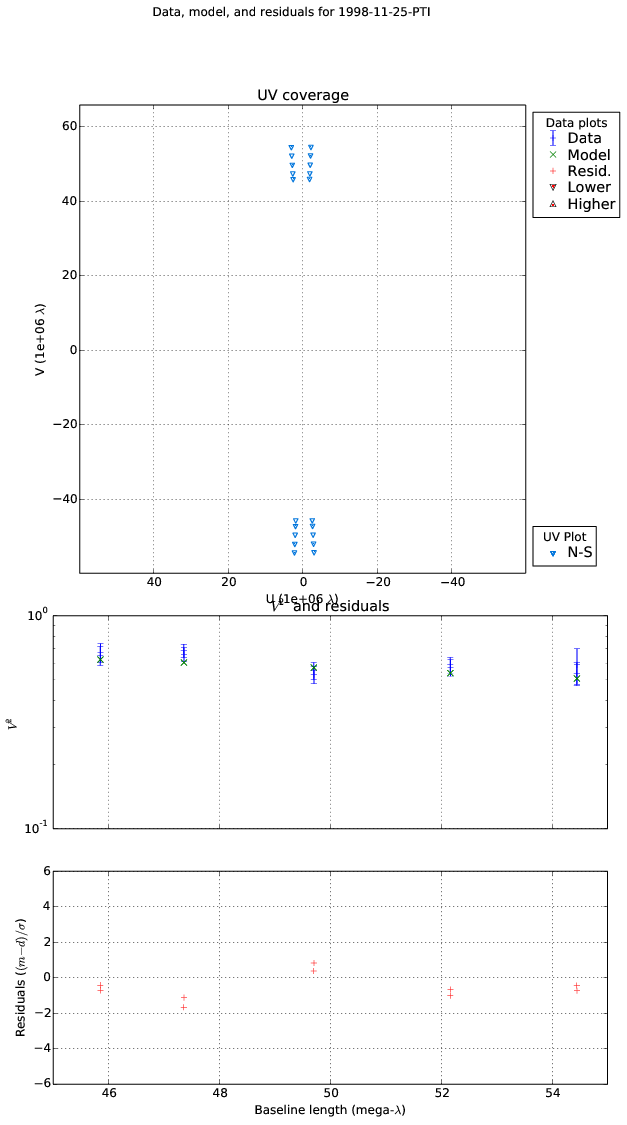}
\caption{UV coverage, data, model, and residual plots for the
1998-11-25-PTI data set.}
\end{figure}

\clearpage
\begin{figure}
\includegraphics[height=\linewidth]{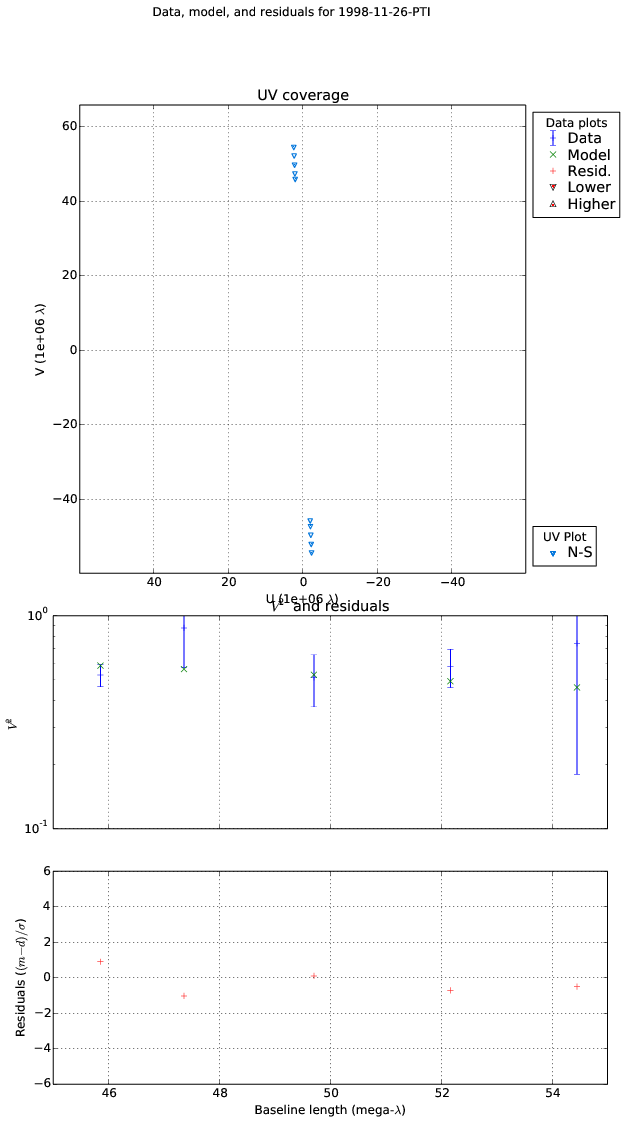}
\caption{UV coverage, data, model, and residual plots for the
1998-11-26-PTI data set.}
\end{figure}

\clearpage
\begin{figure}
\includegraphics[height=\linewidth]{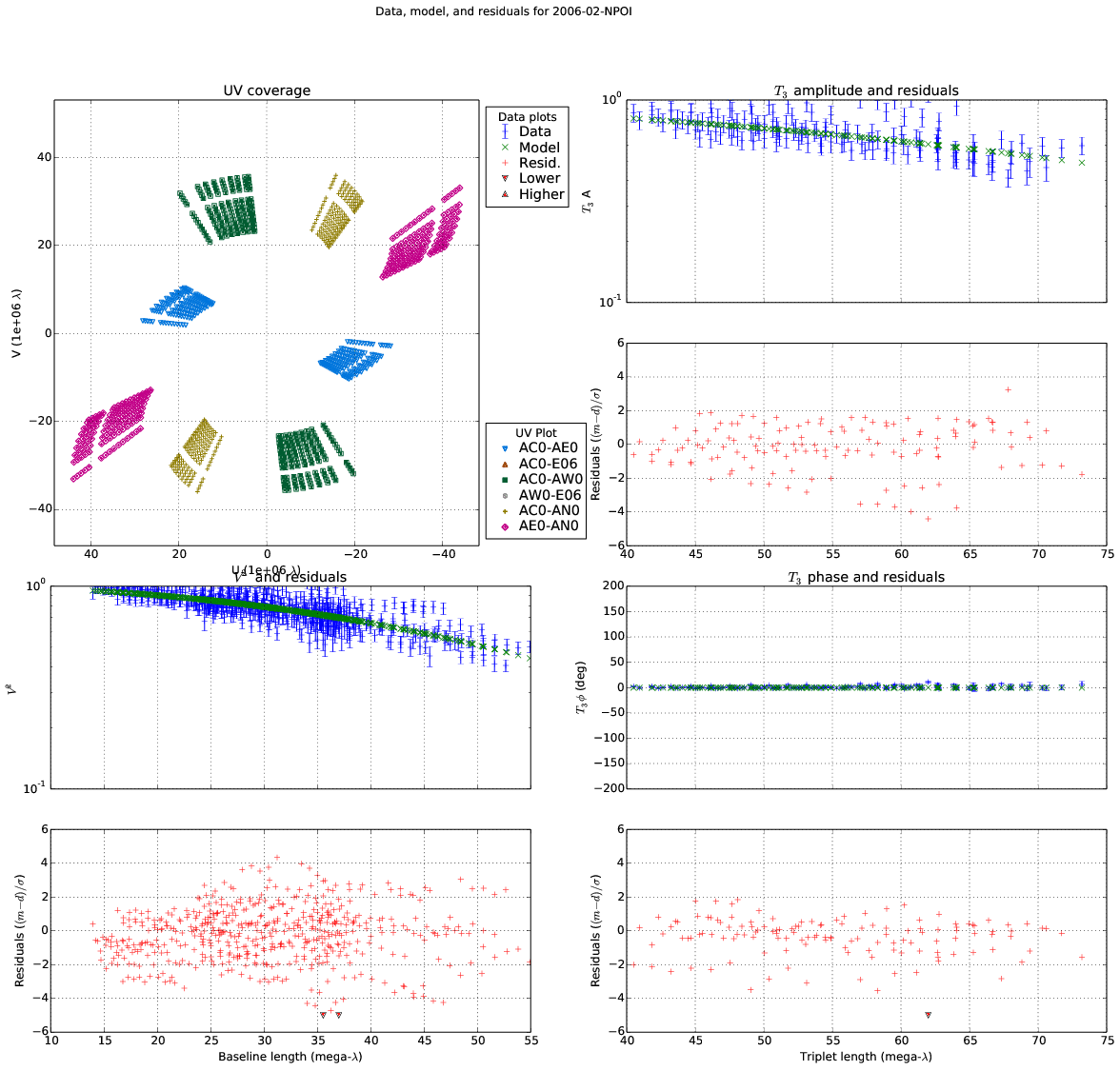}
\caption{UV coverage, data, model, and residual plots for the
2006-02-NPOI data set.}
\end{figure}

\clearpage
\begin{figure}
\includegraphics[height=\linewidth]{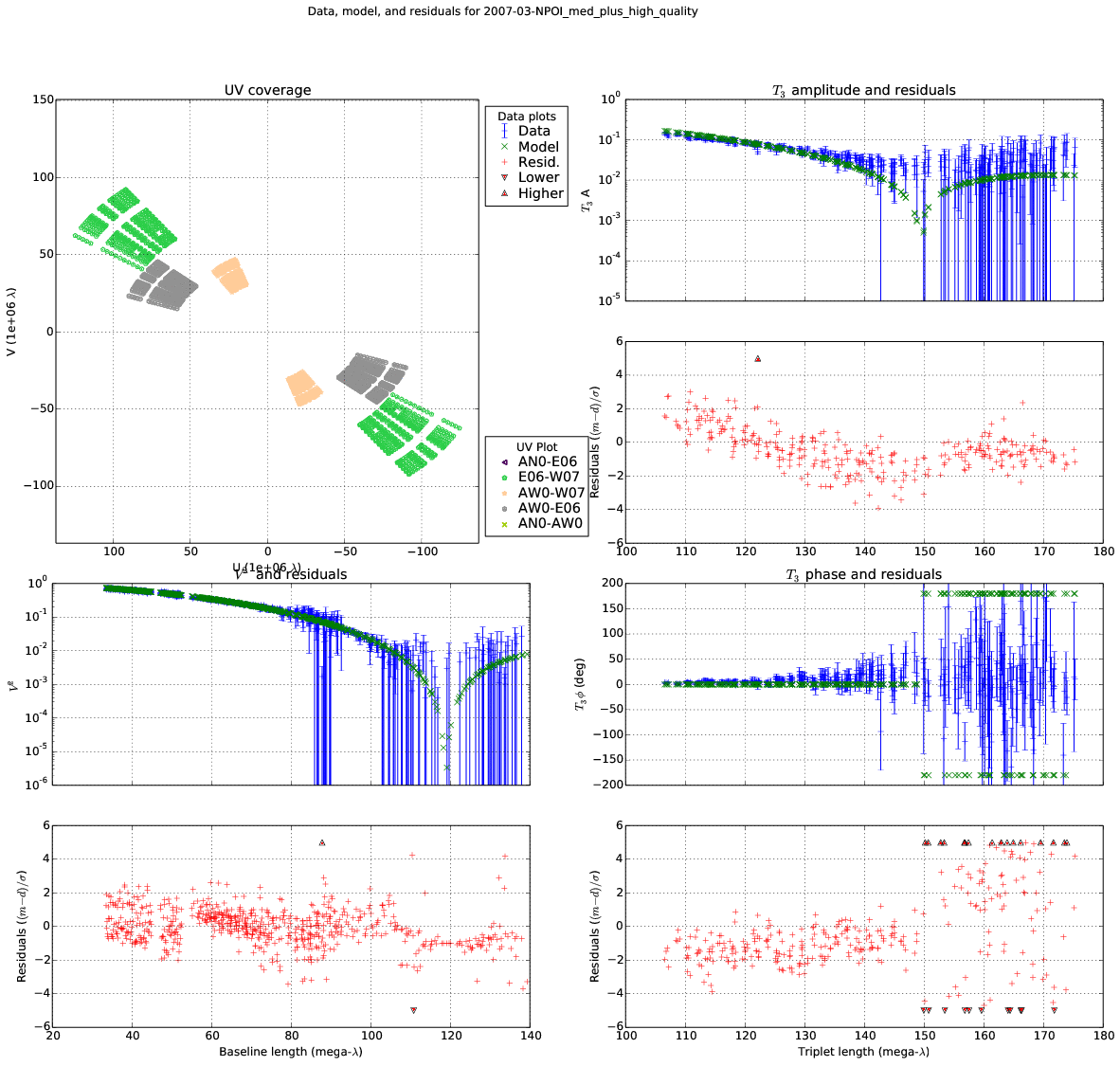}
\caption{UV coverage, data, model, and residual plots for the
2007-03-NPOI (high quality and medium baseline length) data set. The visibility
bias in NPOI data is clearly present near the first visibility null.}
\end{figure}

\clearpage
\begin{figure}
\includegraphics[height=\linewidth]{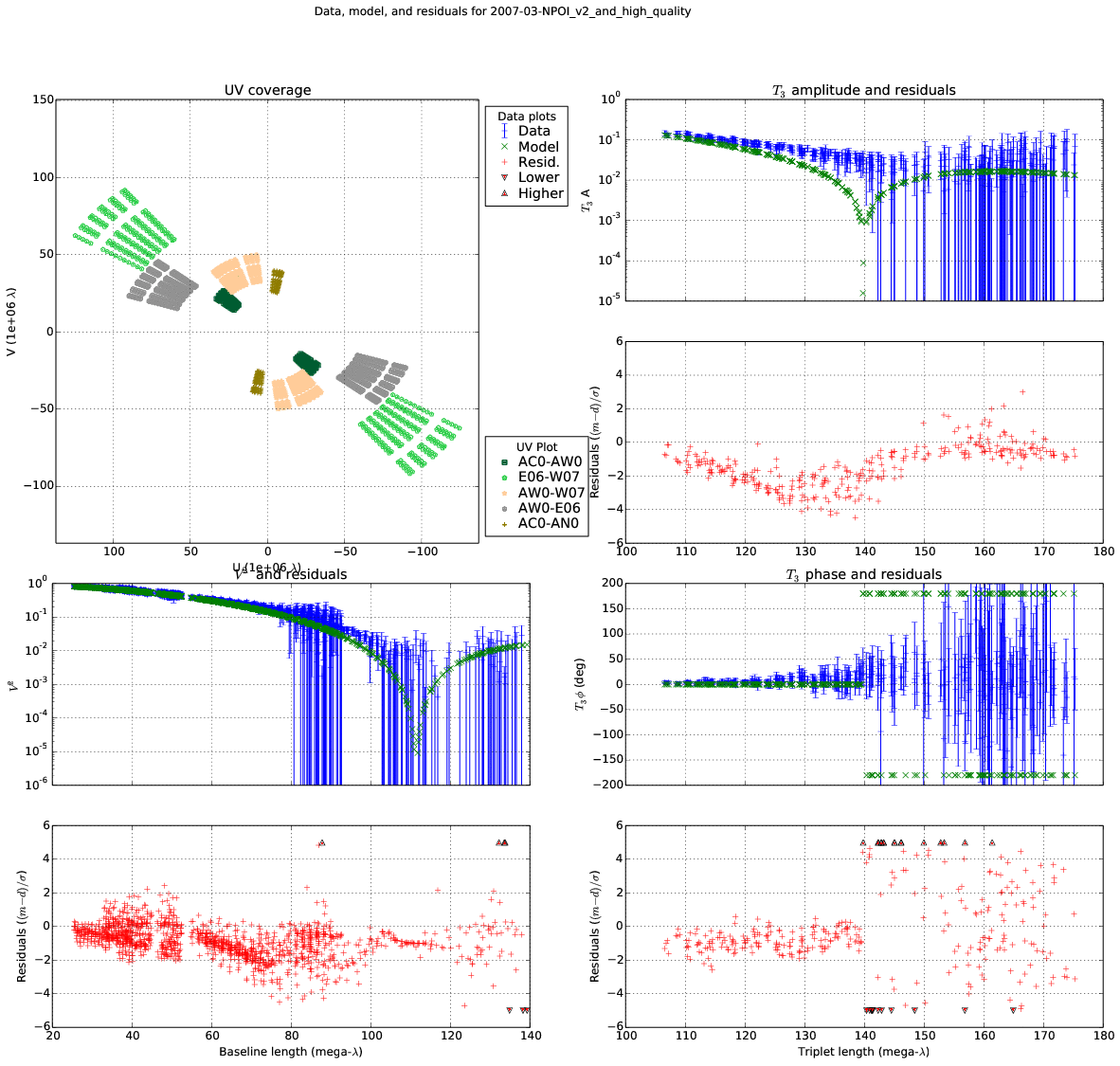}
\caption{UV coverage, data, model, and residual plots for the
2007-03-NPOI (high quality and all $V^2$) data set.}
\end{figure}

\clearpage
\begin{figure}
\includegraphics[height=\linewidth]{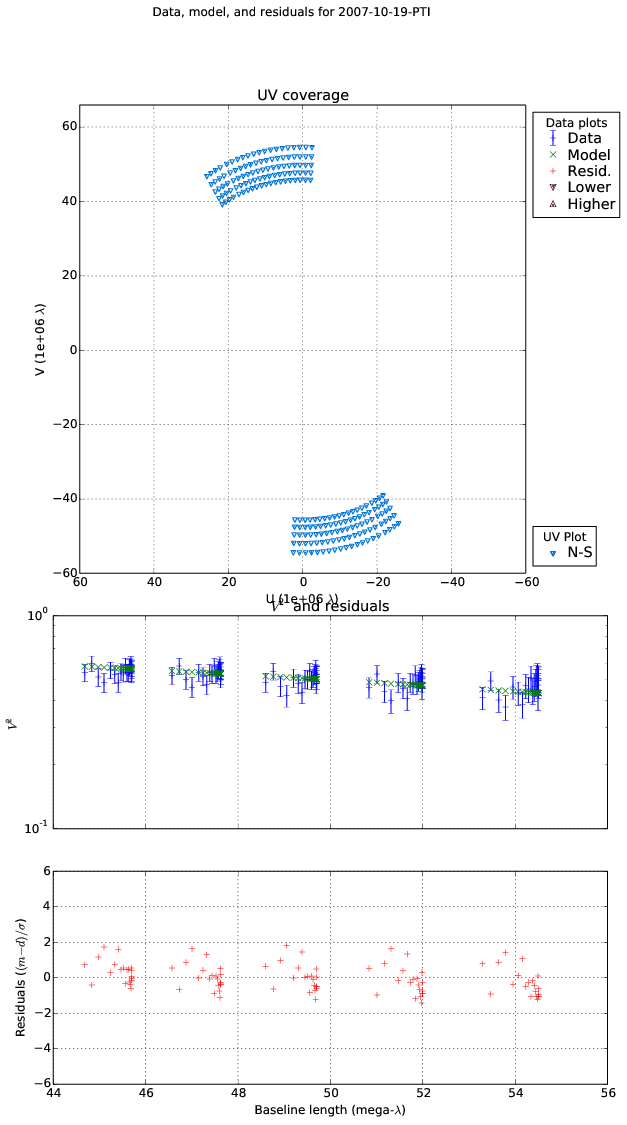}
\caption{UV coverage, data, model, and residual plots for the
2007-10-19-PTI data set.}
\end{figure}

\clearpage
\begin{figure}
\includegraphics[height=\linewidth]{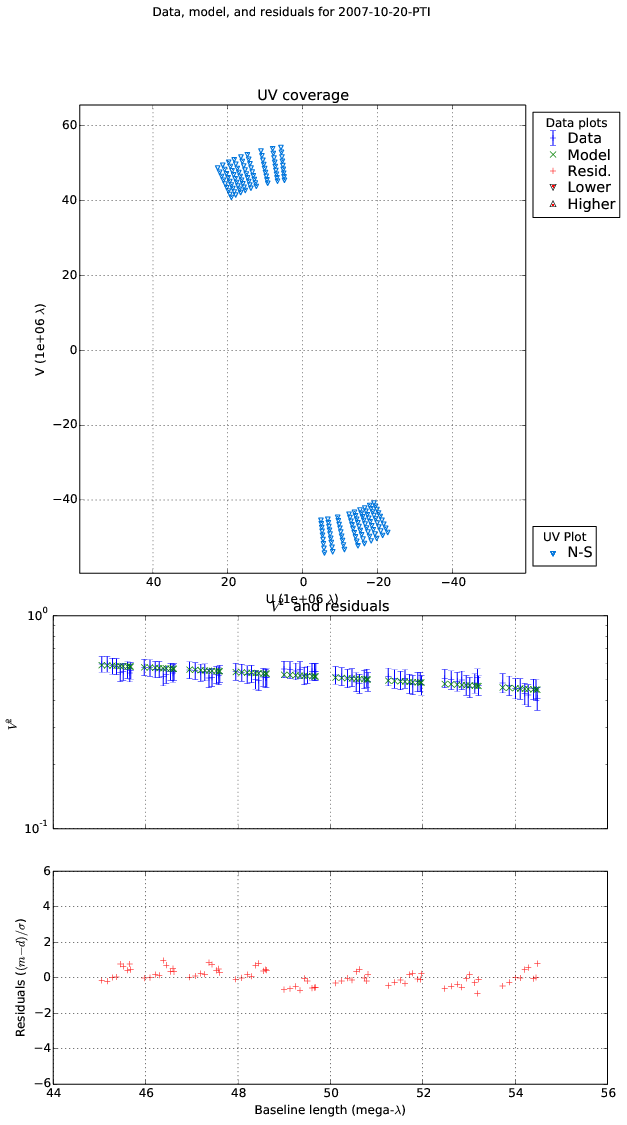}
\caption{UV coverage, data, model, and residual plots for the
2007-10-20-PTI data set.}
\end{figure}

\clearpage
\begin{figure}
\includegraphics[height=\linewidth]{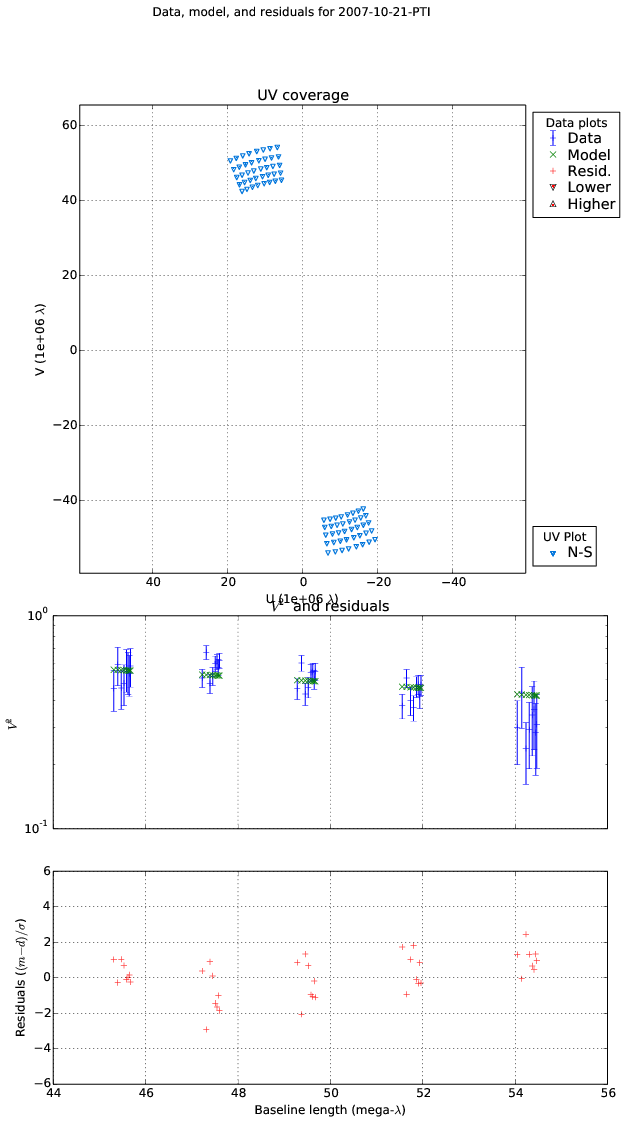}
\caption{UV coverage, data, model, and residual plots for the
2007-10-21-PTI data set.}
\end{figure}

\clearpage
\begin{figure}
\includegraphics[height=\linewidth]{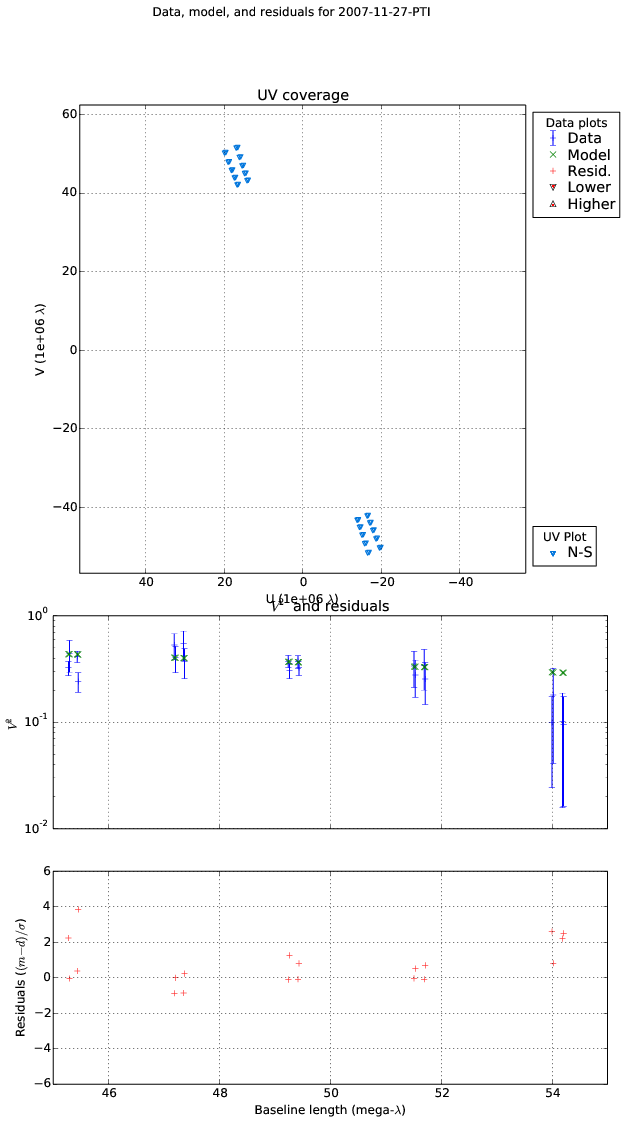}
\caption{UV coverage, data, model, and residual plots for the
2007-11-27-PTI data set.}
\end{figure}

\clearpage
\begin{figure}
\includegraphics[height=\linewidth]{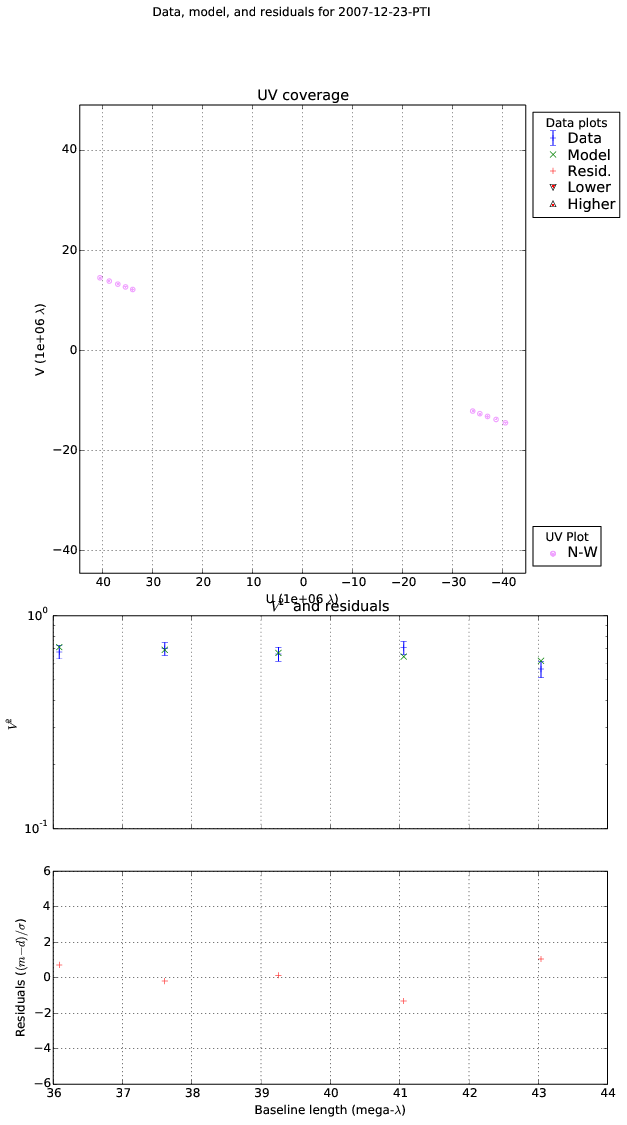}
\caption{UV coverage, data, model, and residual plots for the
2007-12-23-PTI data set.}
\end{figure}

\clearpage
\begin{figure}
\includegraphics[height=\linewidth]{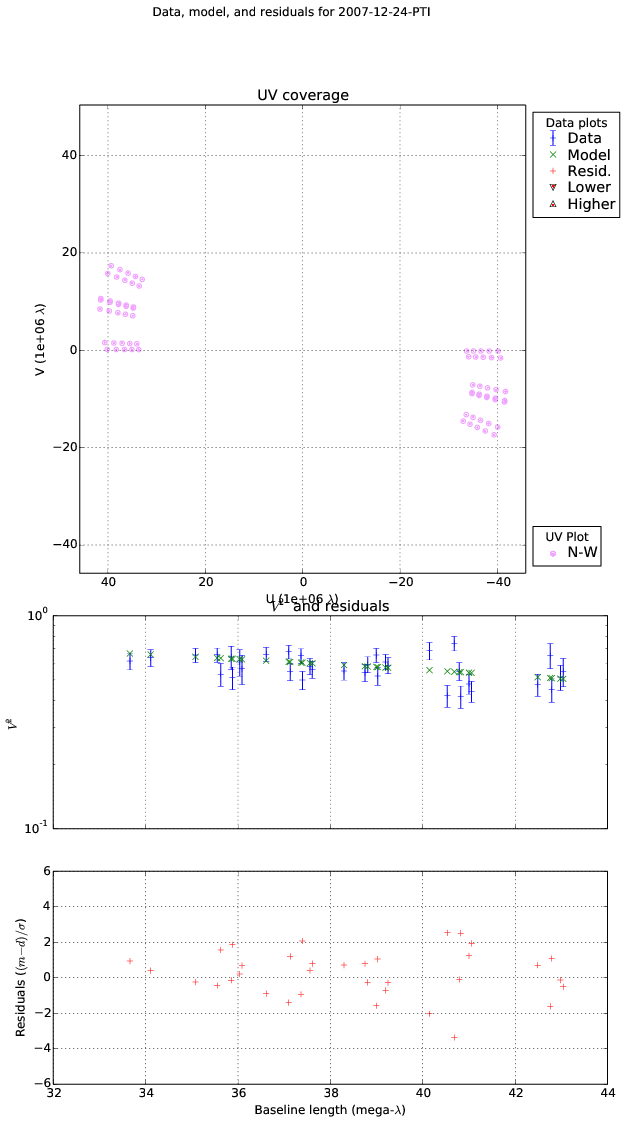}
\caption{UV coverage, data, model, and residual plots for the
2007-12-24-PTI data set.}
\end{figure}

\clearpage
\begin{figure}
\includegraphics[height=\linewidth]{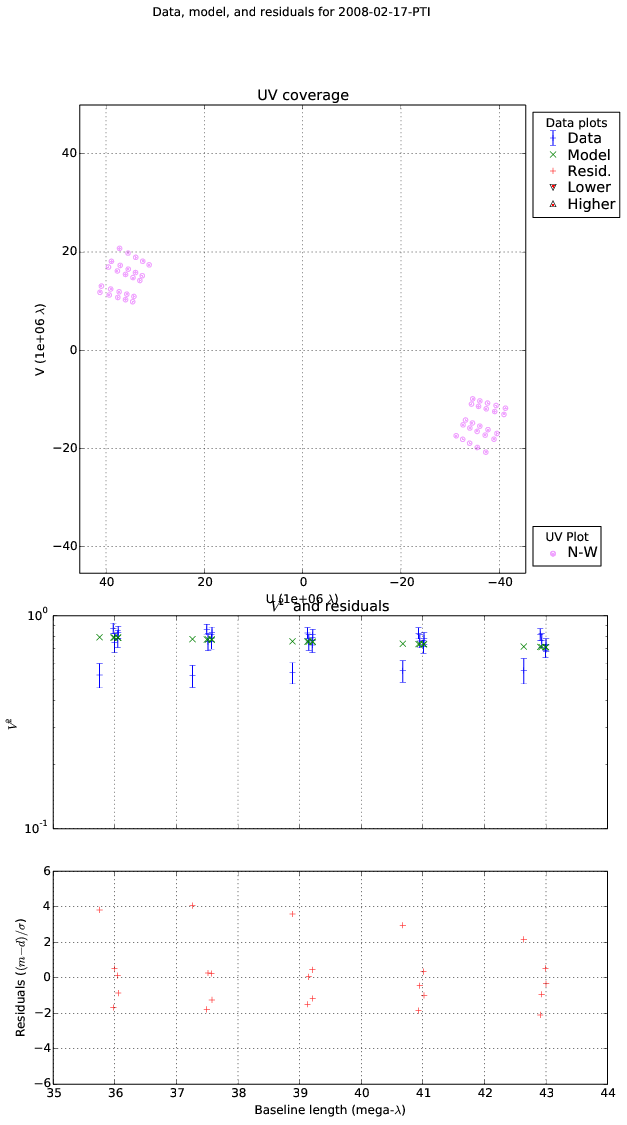}
\caption{UV coverage, data, model, and residual plots for the
2008-02-17-PTI data set.}
\end{figure}

\clearpage
\begin{figure}
\includegraphics[height=\linewidth]{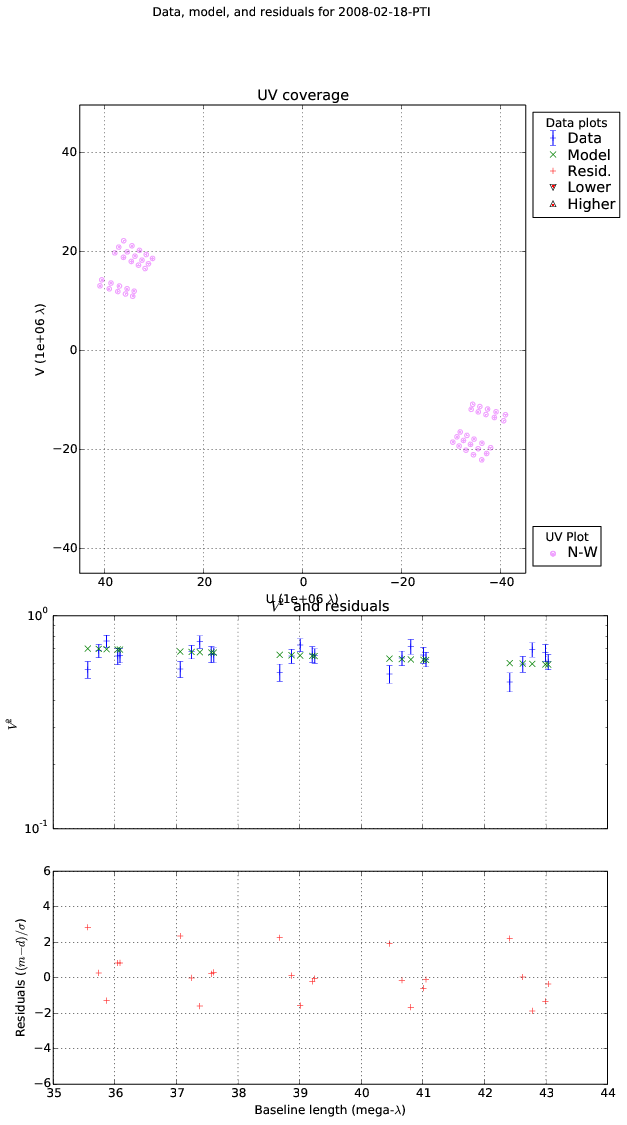}
\caption{UV coverage, data, model, and residual plots for the
2008-02-18-PTI data set.}
\end{figure}

\clearpage
\begin{figure}
\includegraphics[height=\linewidth]{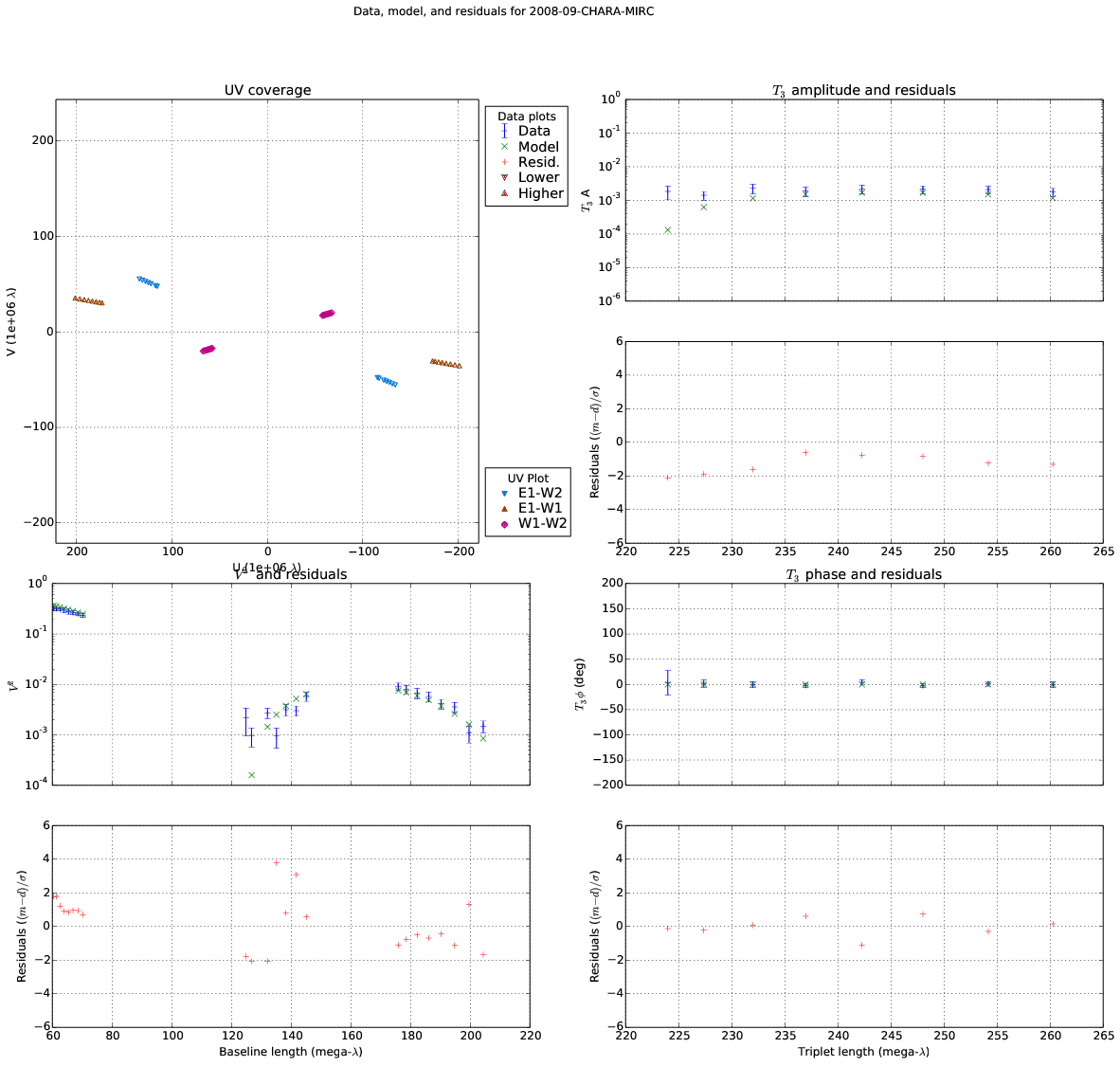}
\caption{UV coverage, data, model, and residual plots for the
2008-09-CHARA-MIRC data set.}
\end{figure}

\clearpage
\begin{figure}
\includegraphics[height=\linewidth]{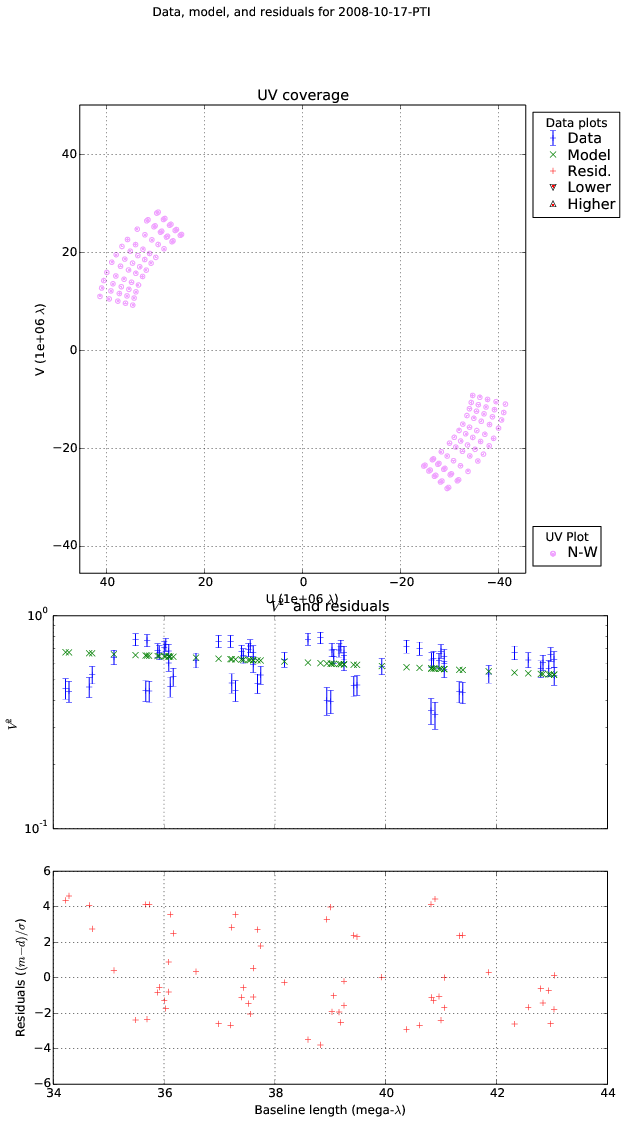}
\caption{UV coverage, data, model, and residual plots for the
2008-10-17-PTI data set.}
\end{figure}

\clearpage
\begin{figure}
\includegraphics[height=\linewidth]{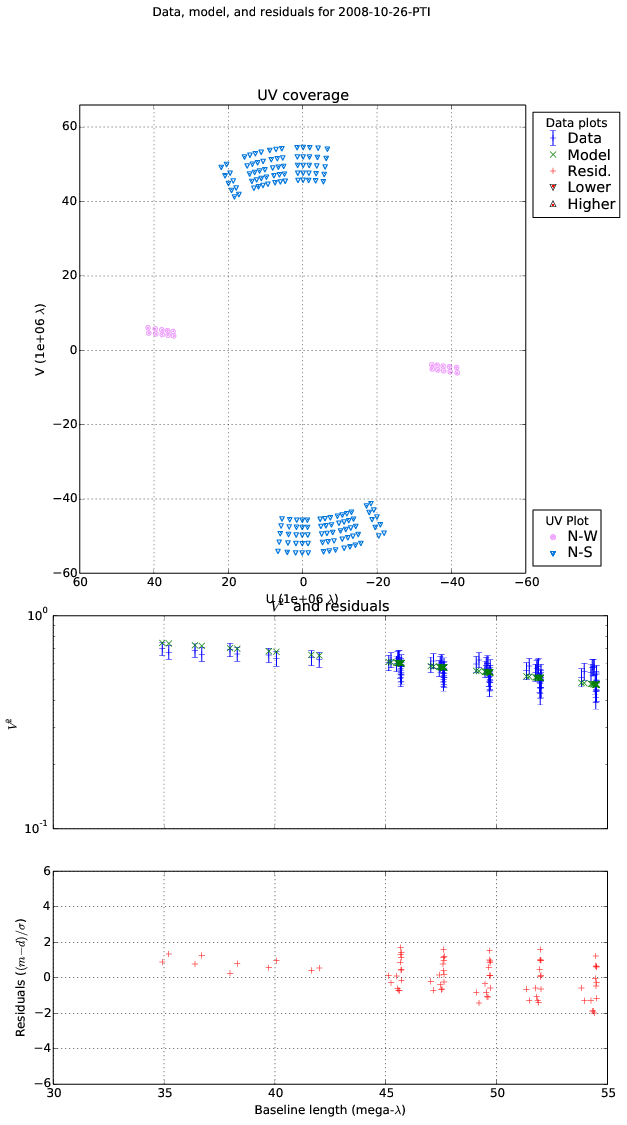}
\caption{UV coverage, data, model, and residual plots for the
2008-10-26-PTI data set.}
\end{figure}

\clearpage
\begin{figure}
\includegraphics[height=\linewidth]{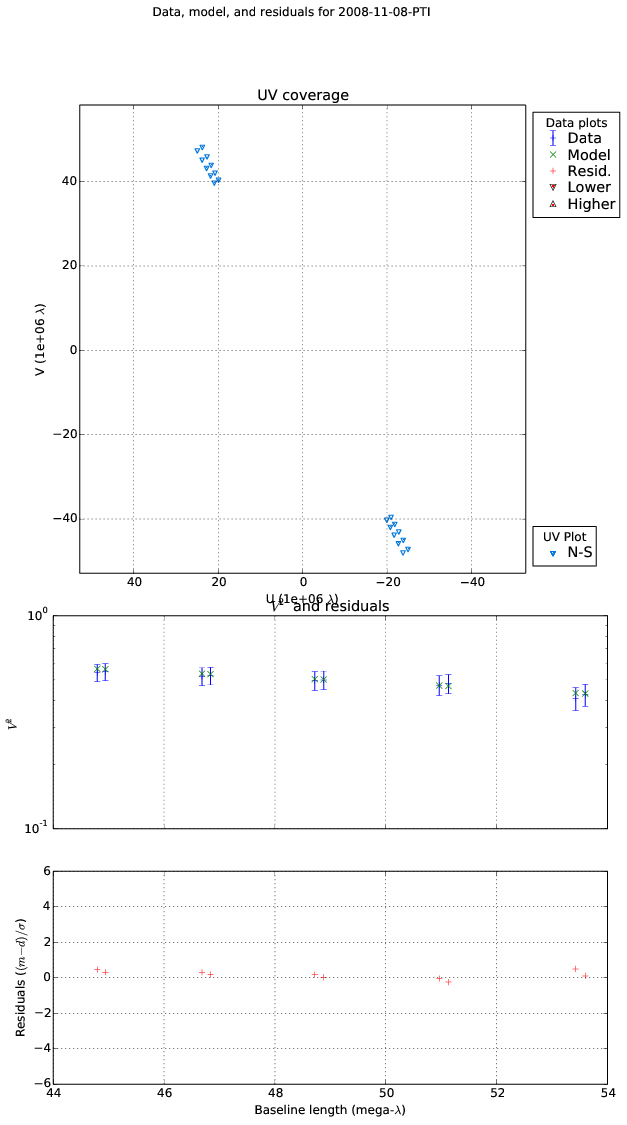}
\caption{UV coverage, data, model, and residual plots for the
2008-11-08-PTI data set.}
\end{figure}

\clearpage
\begin{figure}
\includegraphics[height=\linewidth]{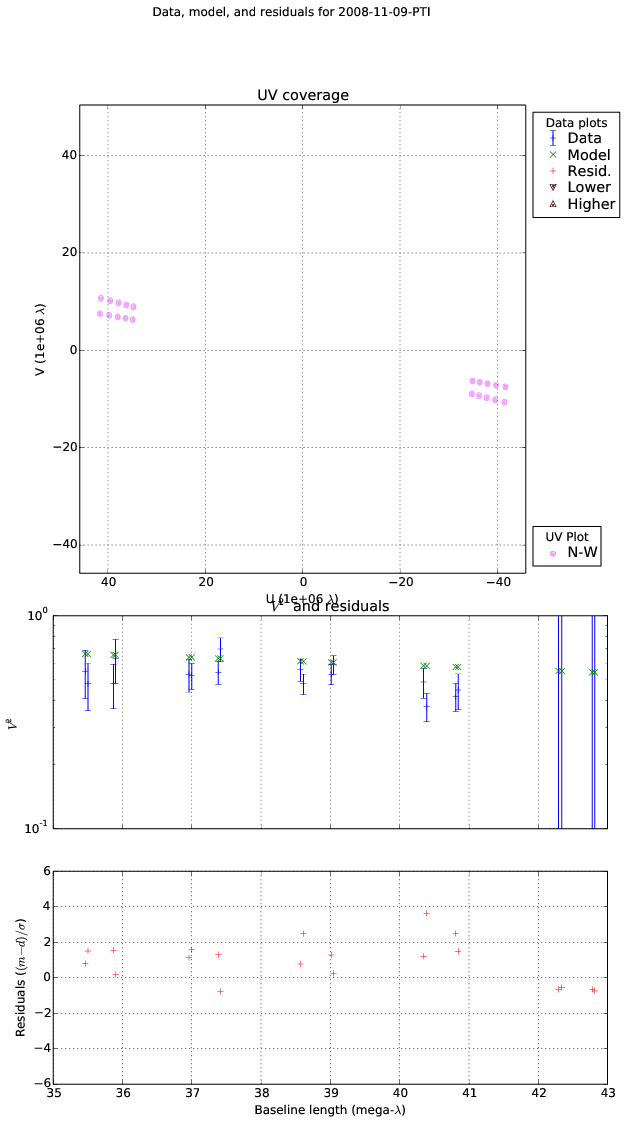}
\caption{UV coverage, data, model, and residual plots for the
2008-11-09-PTI data set.}
\end{figure}

\clearpage
\begin{figure}
\includegraphics[height=\linewidth]{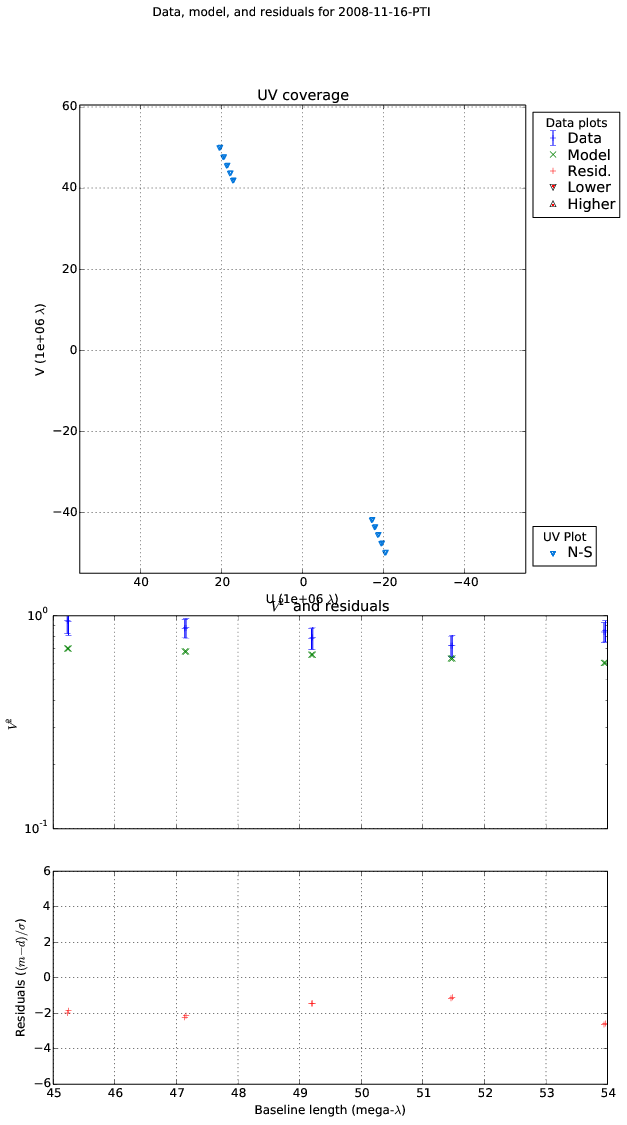}
\caption{UV coverage, data, model, and residual plots for the
2008-11-16-PTI data set.}
\end{figure}

\clearpage
\begin{figure}
\includegraphics[height=\linewidth]{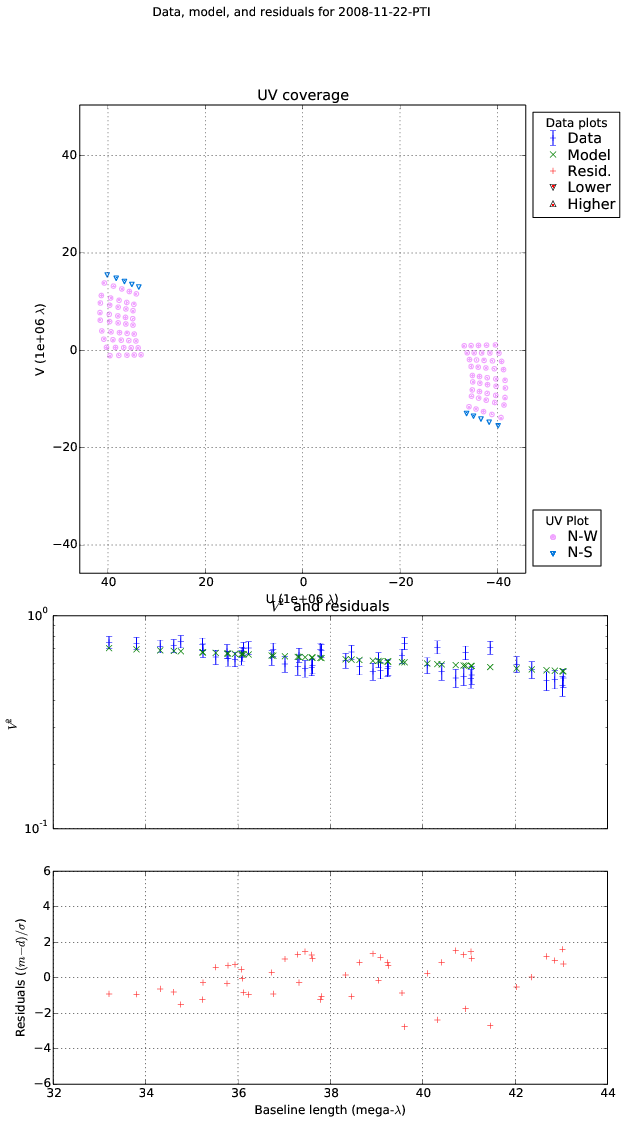}
\caption{UV coverage, data, model, and residual plots for the
2008-11-22-PTI data set.}
\end{figure}

\clearpage
\begin{figure}
\includegraphics[height=\linewidth]{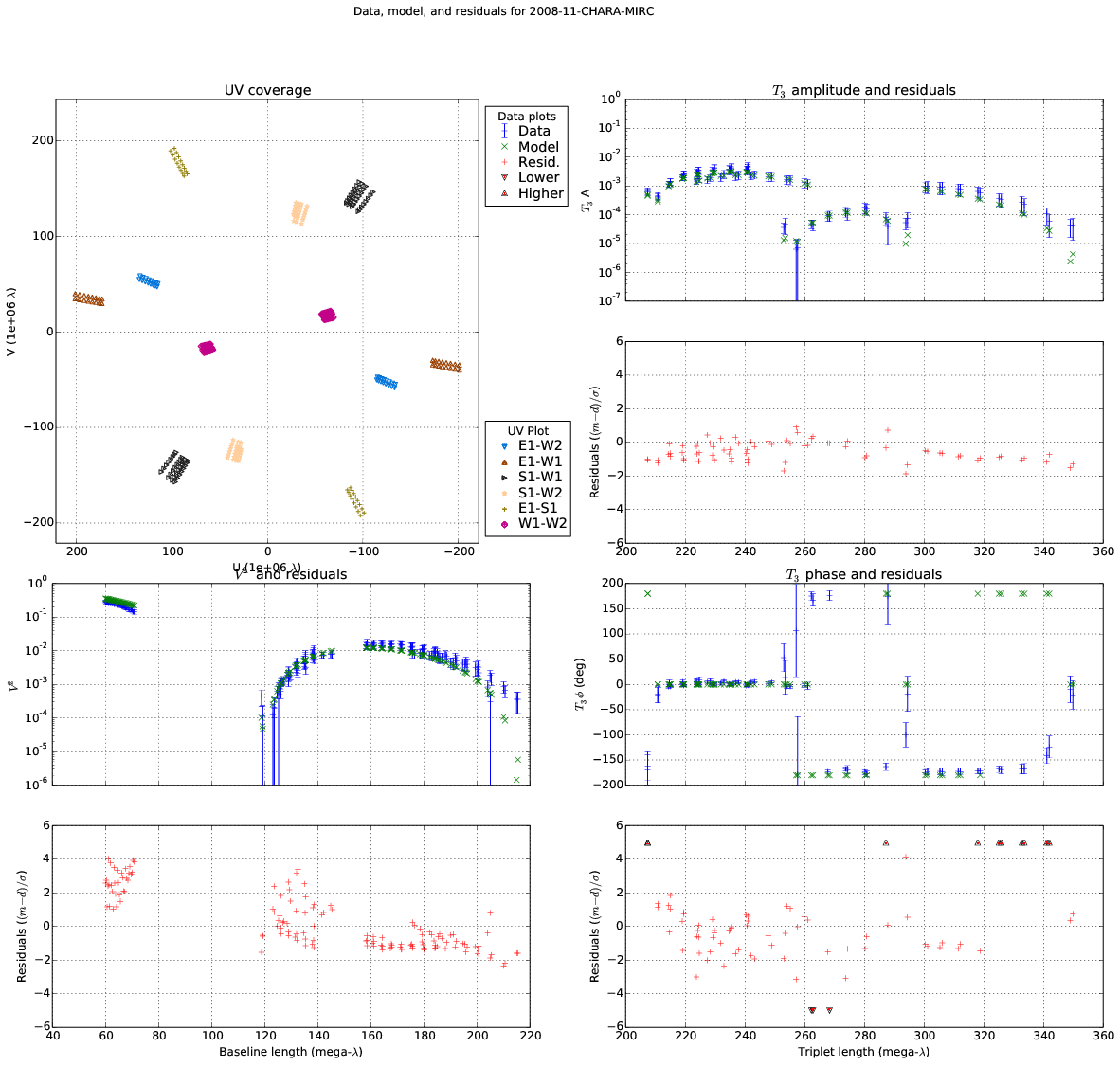}
\caption{UV coverage, data, model, and residual plots for the
2008-11-CHARA-MIRC data set.}
\end{figure}

\clearpage
\begin{figure}
\includegraphics[height=\linewidth]{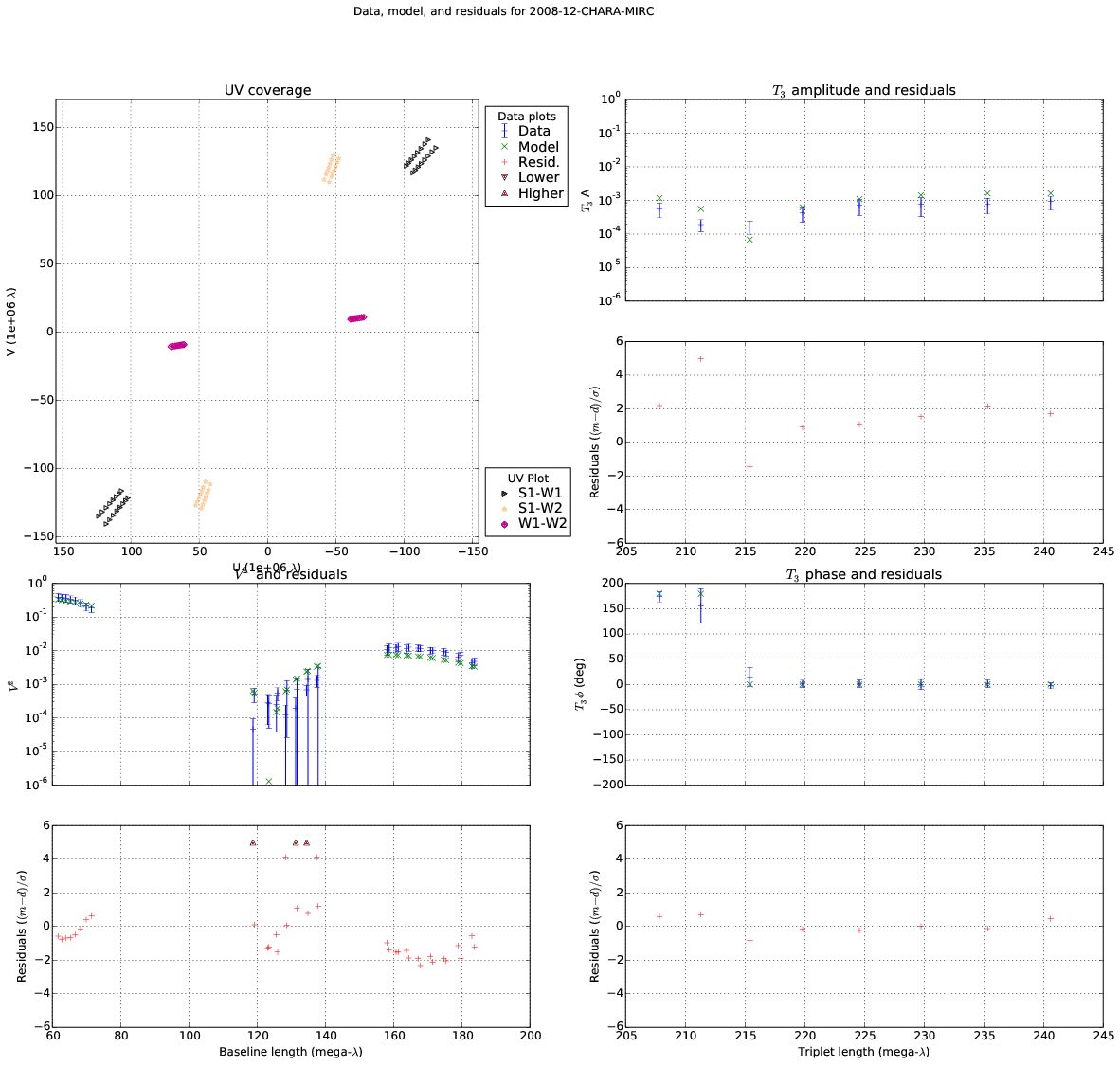}
\caption{UV coverage, data, model, and residual plots for the
2008-12-CHARA-MIRC data set.}
\end{figure}

\clearpage
\begin{figure}
\includegraphics[height=\linewidth]{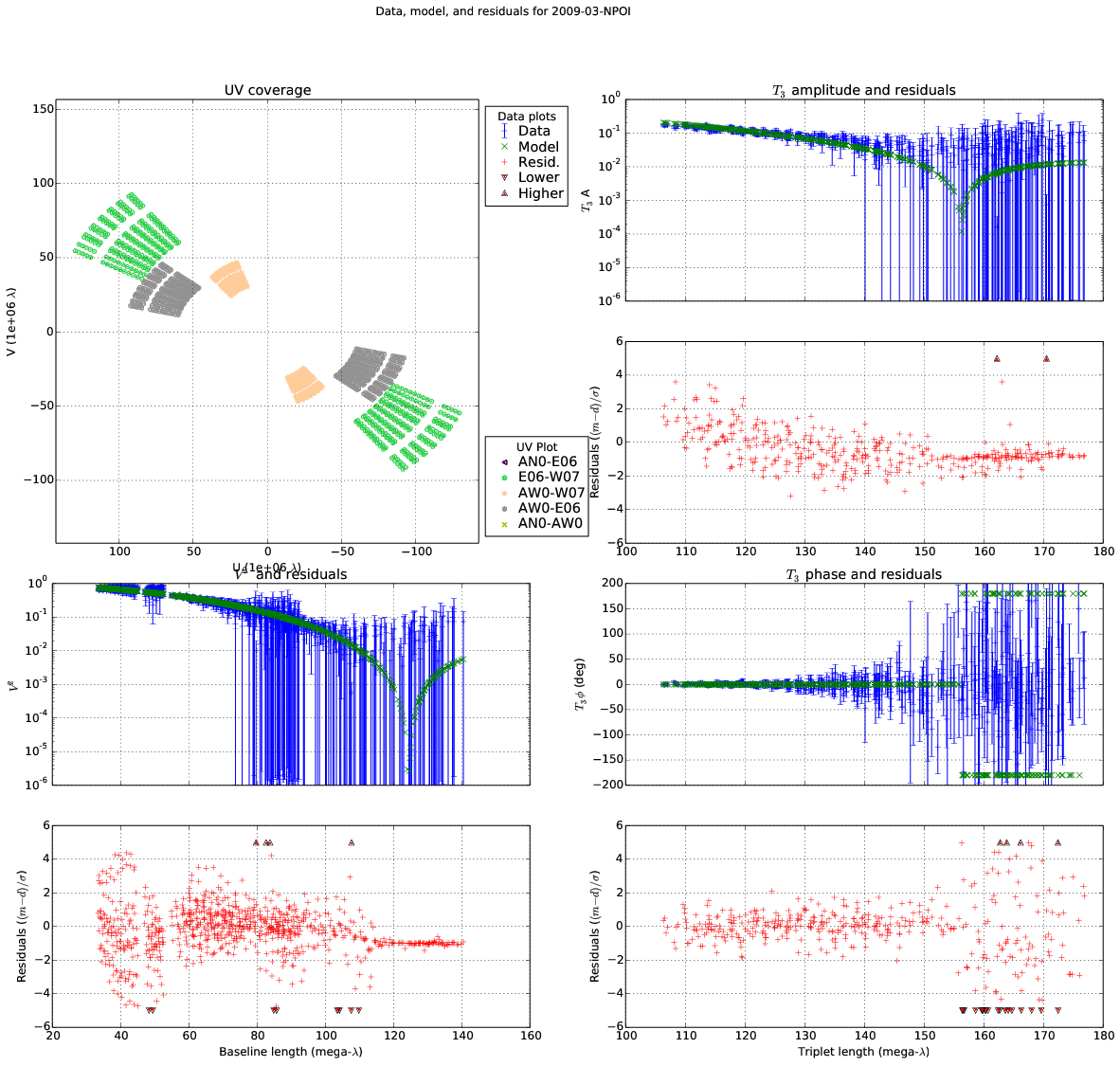}
\caption{UV coverage, data, model, and residual plots for the
2009-03-NPOI data set.}
\end{figure}

\clearpage
\begin{figure}
\includegraphics[height=\linewidth]{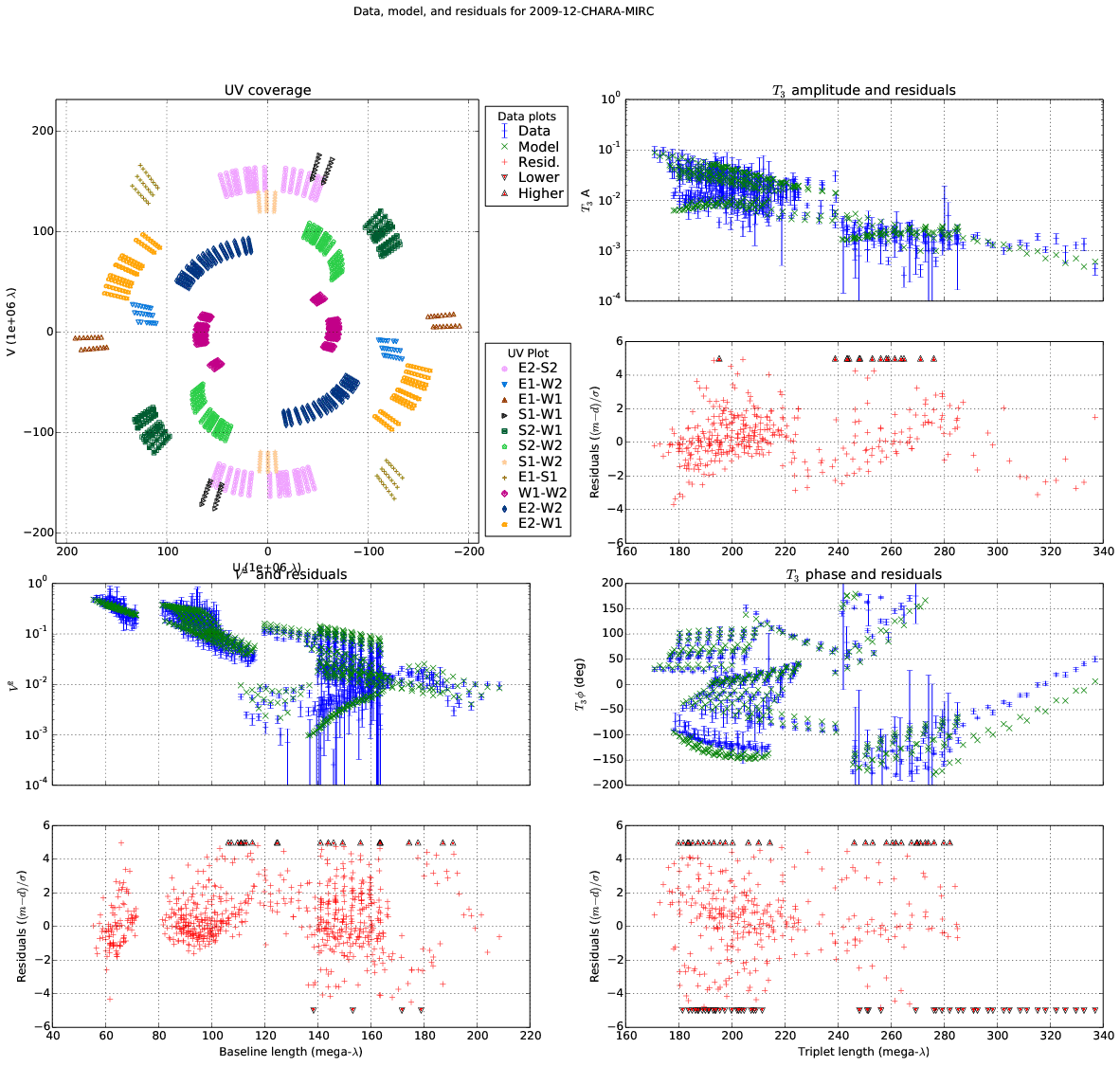}
\caption{UV coverage, data, model, and residual plots for the
2009-12-CHARA-MIRC data set. These data were previously published in \cite{Kloppenborg2010}.}
\end{figure}

\clearpage
\begin{figure}
\includegraphics[height=\linewidth]{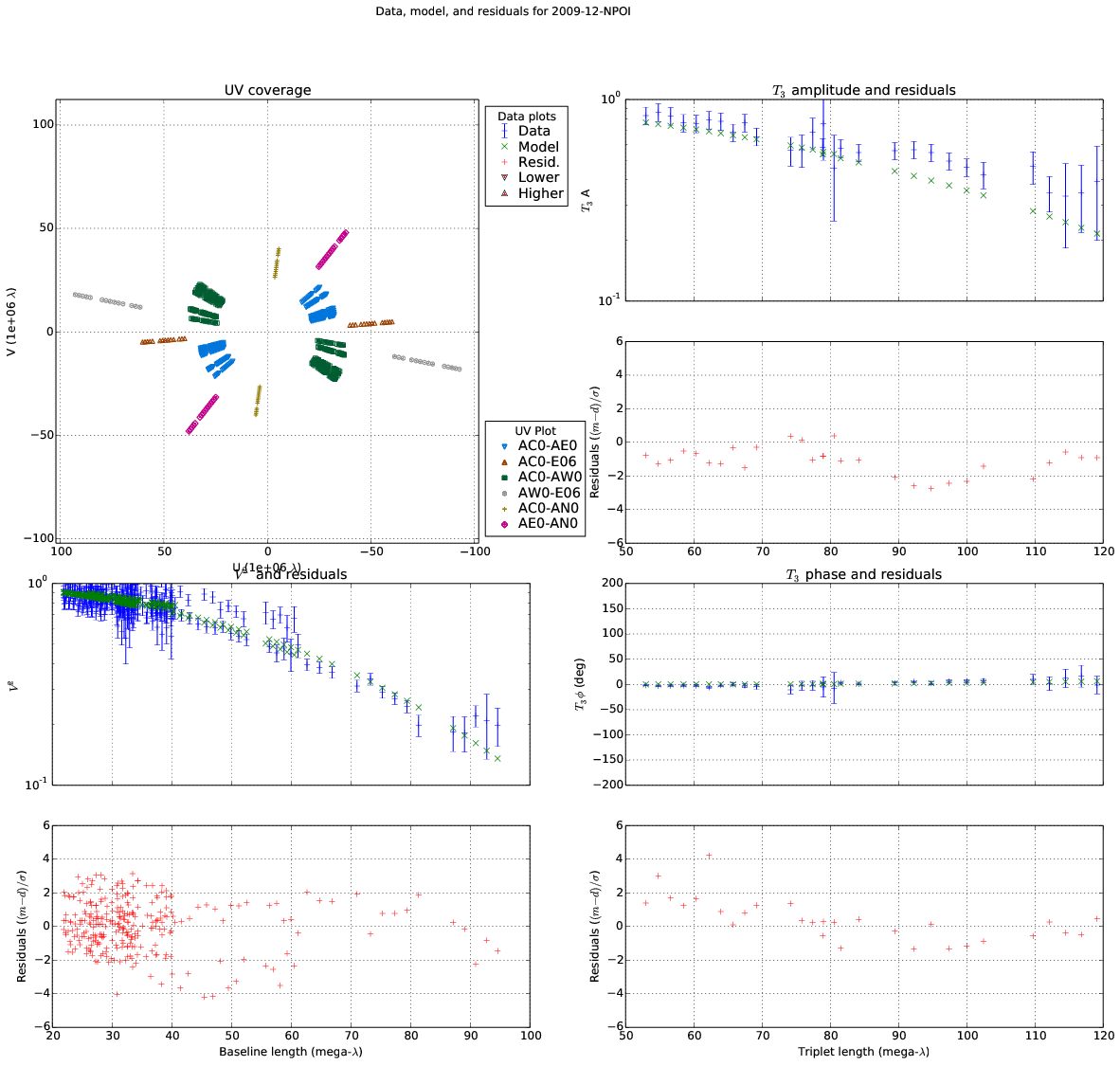}
\caption{UV coverage, data, model, and residual plots for the
2009-12-NPOI data set.}
\end{figure}

\clearpage
\begin{figure}
\includegraphics[height=\linewidth]{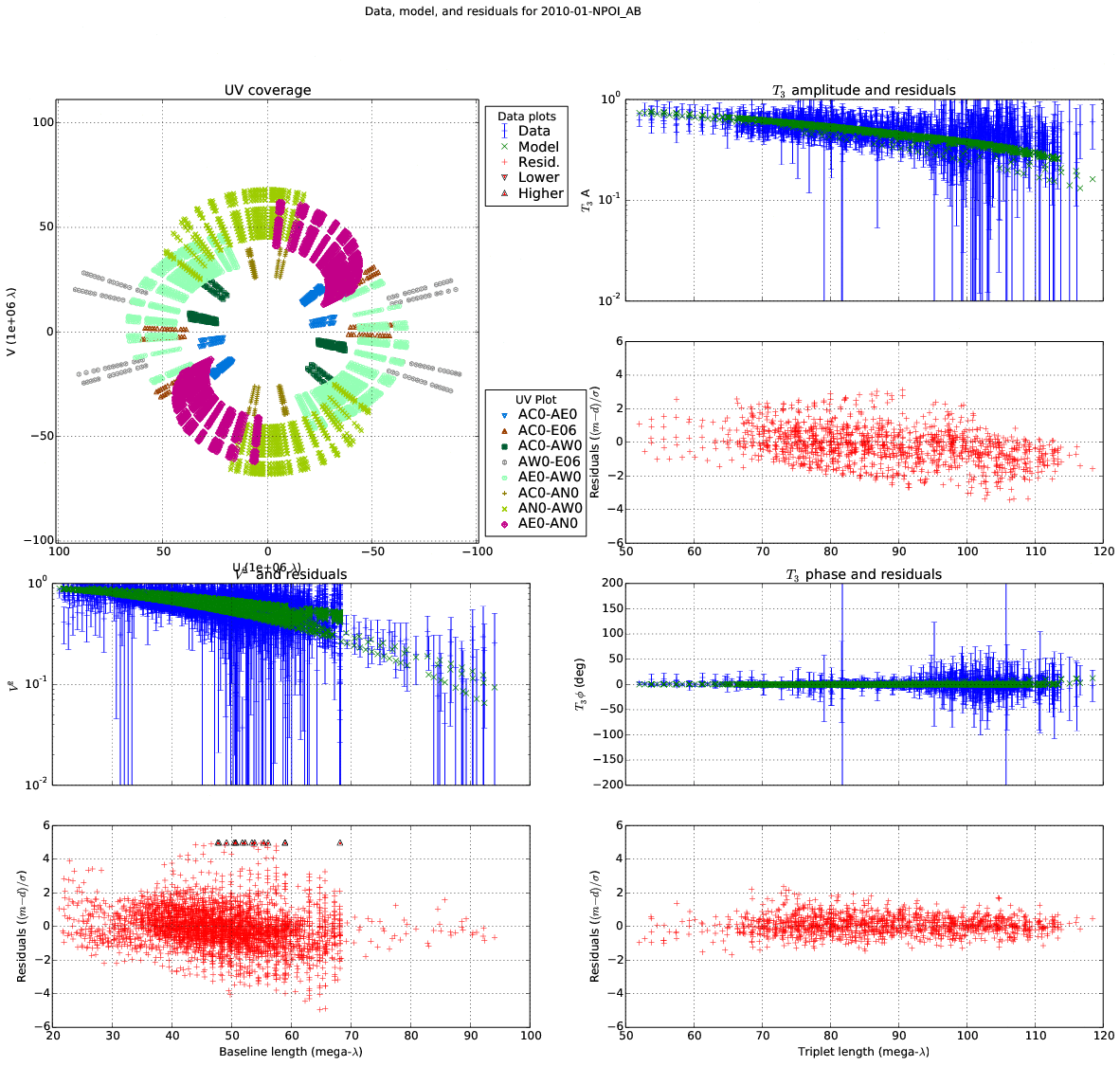}
\caption{UV coverage, data, model, and residual plots for the
2010-01-NPOI (AB) data set. Notice that although the baselines are short,
the visibilities are bifurcated at short baselines as a result of the disk
obstructing the southern hemisphere of the F-star.}
\end{figure}

\clearpage
\begin{figure}
\includegraphics[height=\linewidth]{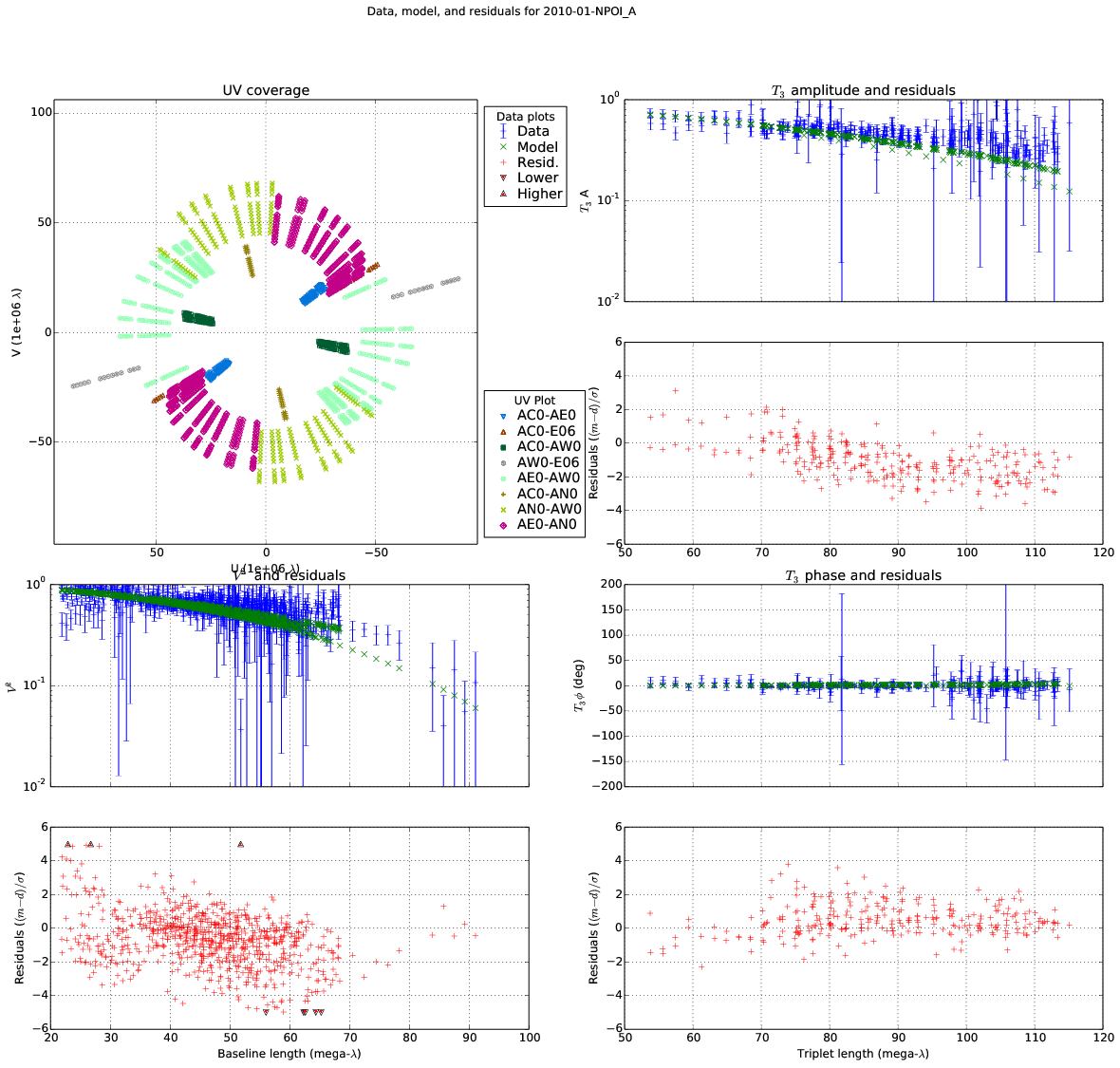}
\caption{UV coverage, data, model, and residual plots for the
2010-01-NPOI (A only) data set.}
\end{figure}

\clearpage
\begin{figure}
\includegraphics[height=\linewidth]{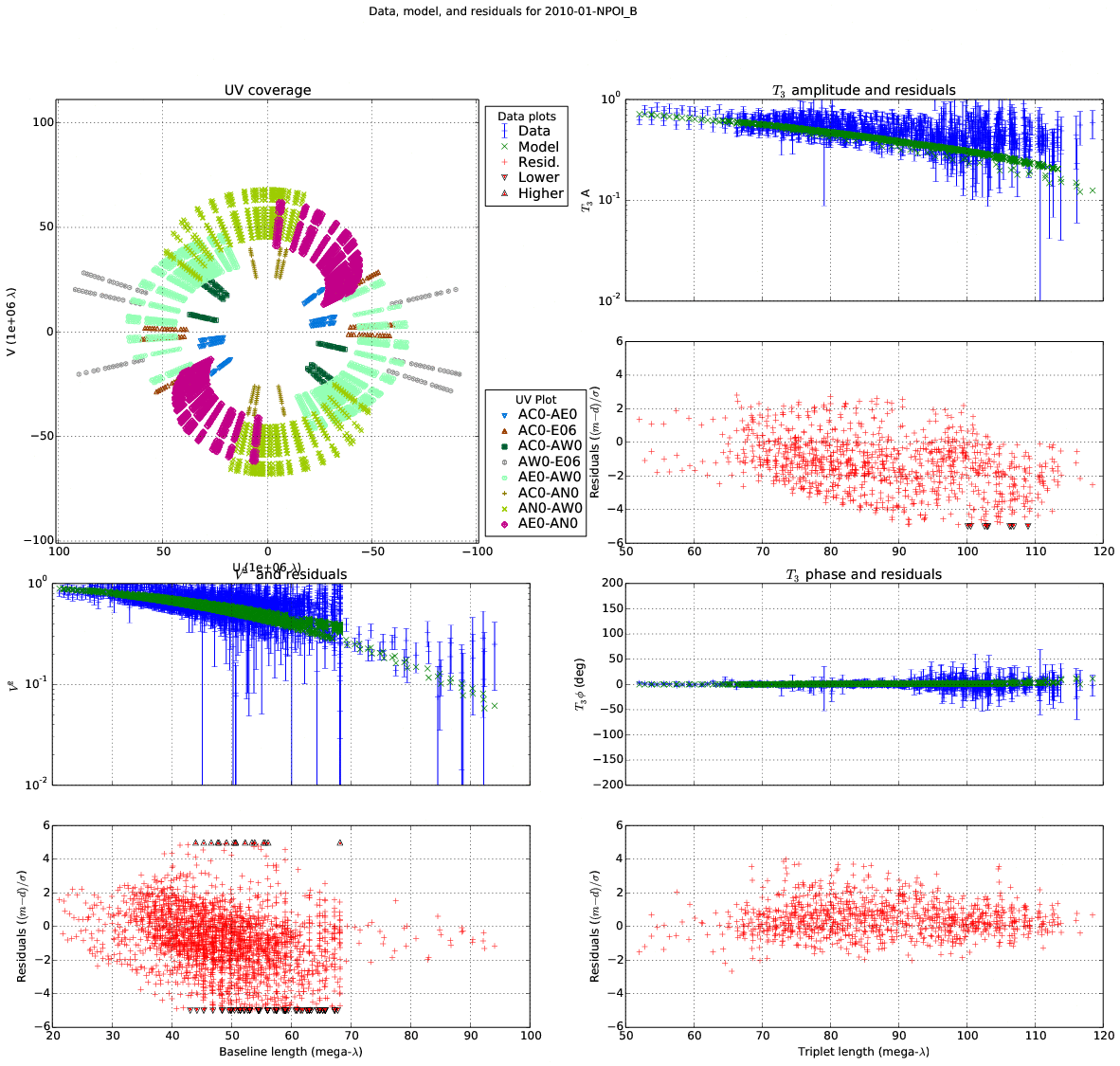}
\caption{UV coverage, data, model, and residual plots for the
2010-01-NPOI (B only) data set.}
\end{figure}

\clearpage
\begin{figure}
\includegraphics[height=\linewidth]{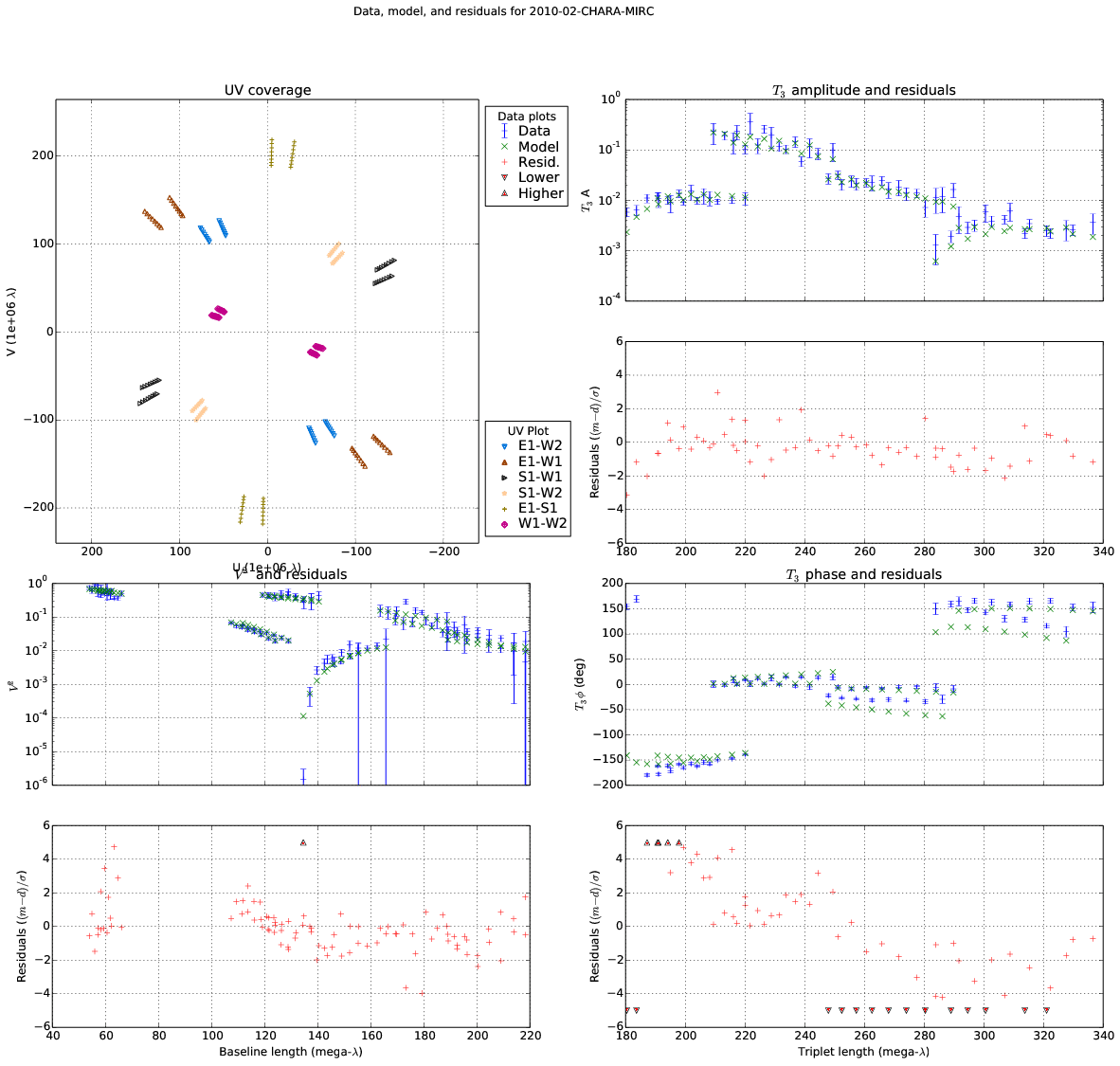}
\caption{UV coverage, data, model, and residual plots for the
2010-02-CHARA-MIRC data set.}
\end{figure}

\clearpage
\begin{figure}
\includegraphics[height=\linewidth]{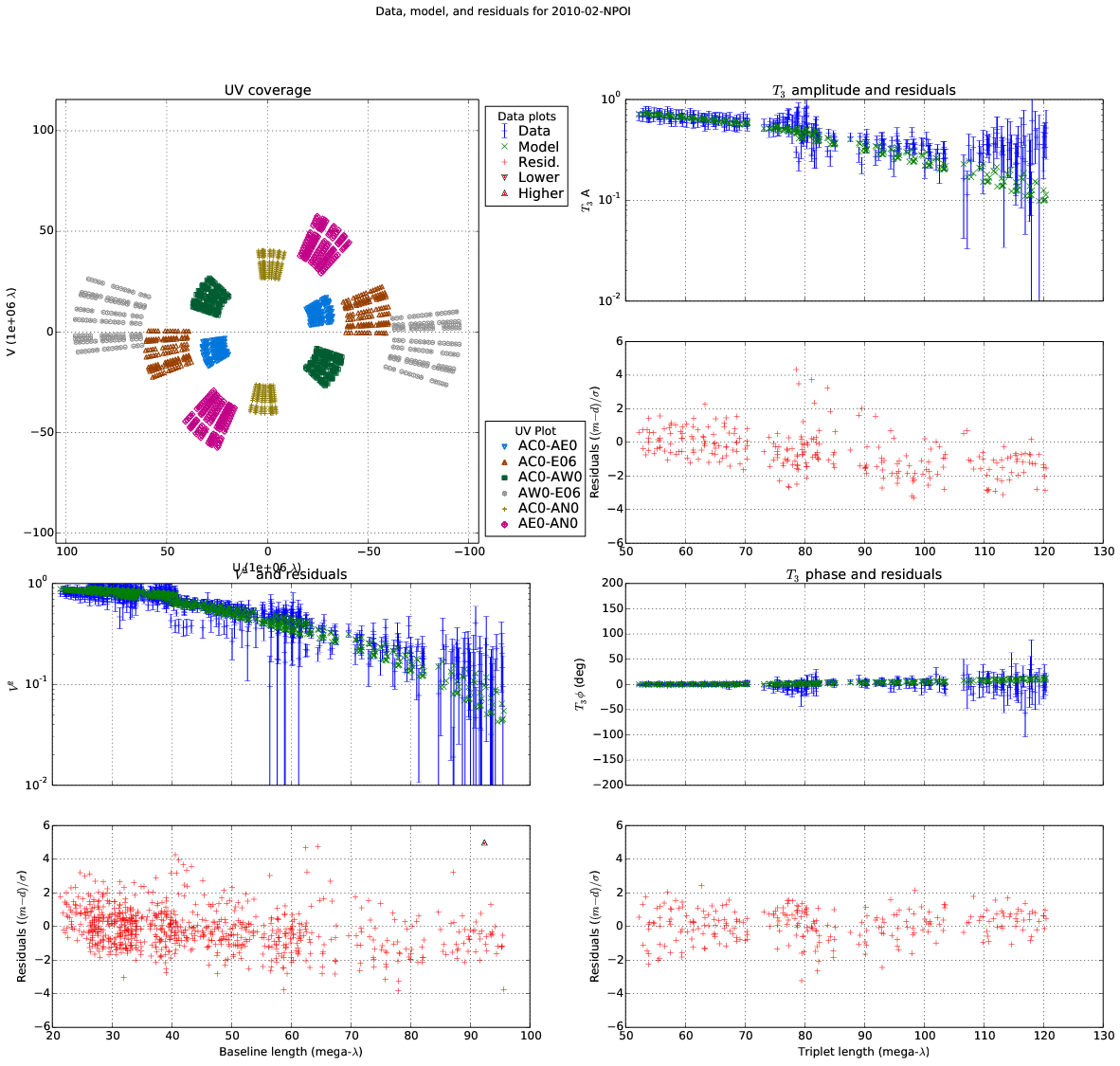}
\caption{UV coverage, data, model, and residual plots for the
2010-02-NPOI data set.}
\end{figure}

\clearpage
\begin{figure}
\includegraphics[height=\linewidth]{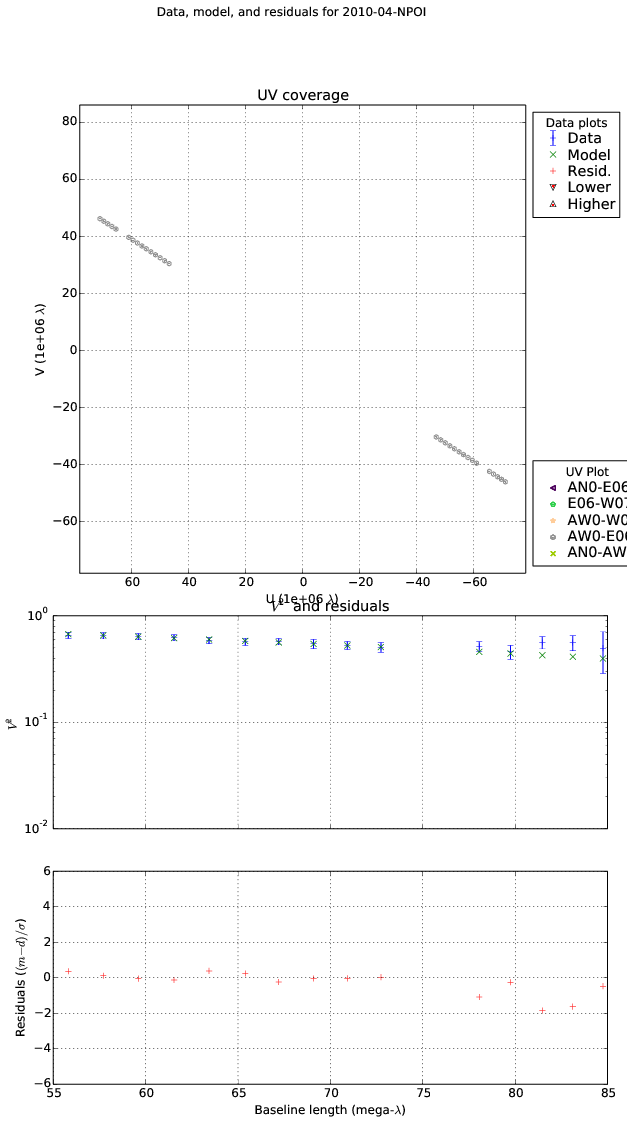}
\caption{UV coverage, data, model, and residual plots for the
2010-04-NPOI data set.}
\end{figure}

\clearpage
\begin{figure}
\includegraphics[height=\linewidth]{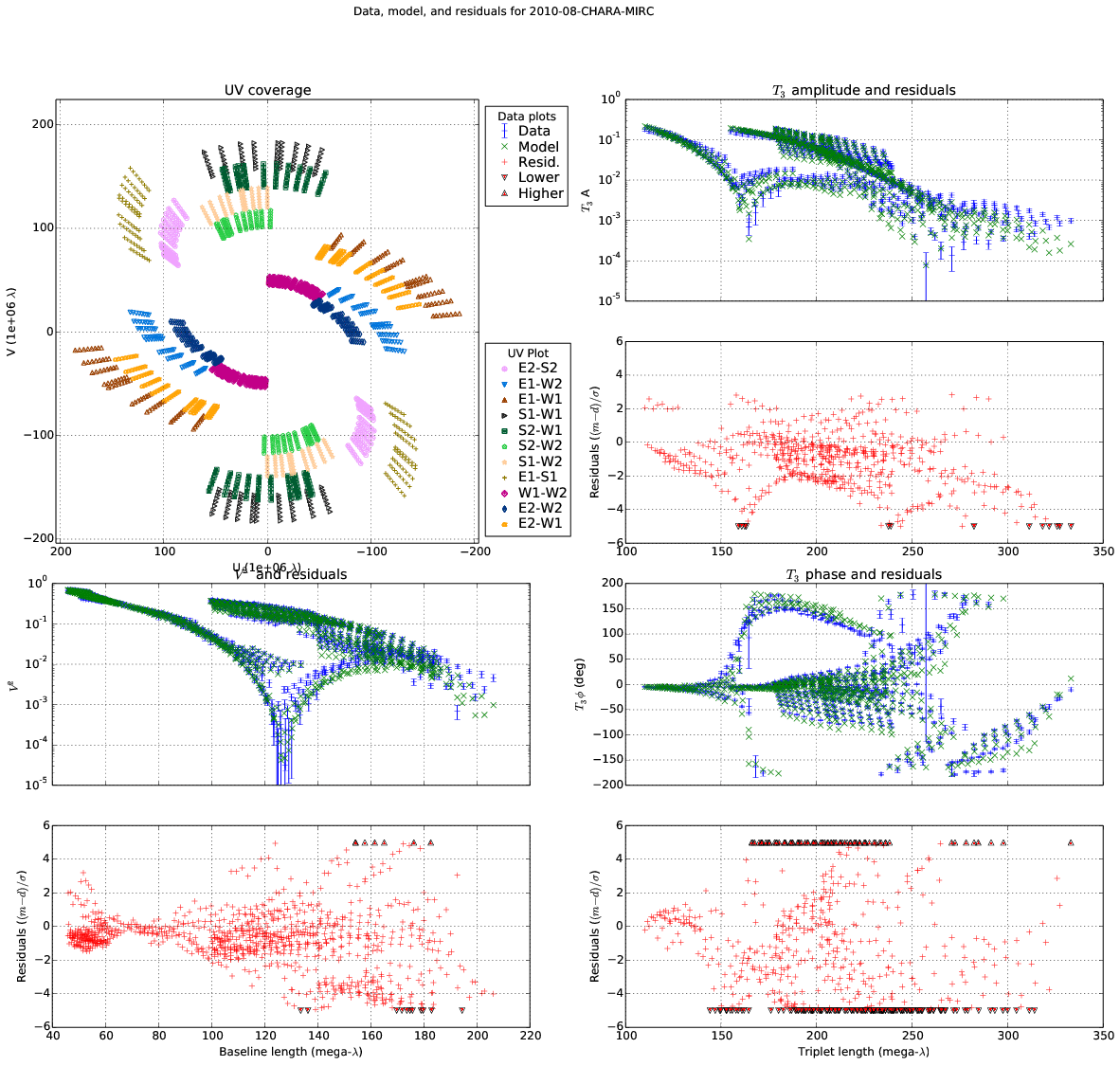}
\caption{UV coverage, data, model, and residual plots for the
2010-08-CHARA-MIRC data set.}
\end{figure}

\clearpage
\begin{figure}
\includegraphics[height=\linewidth]{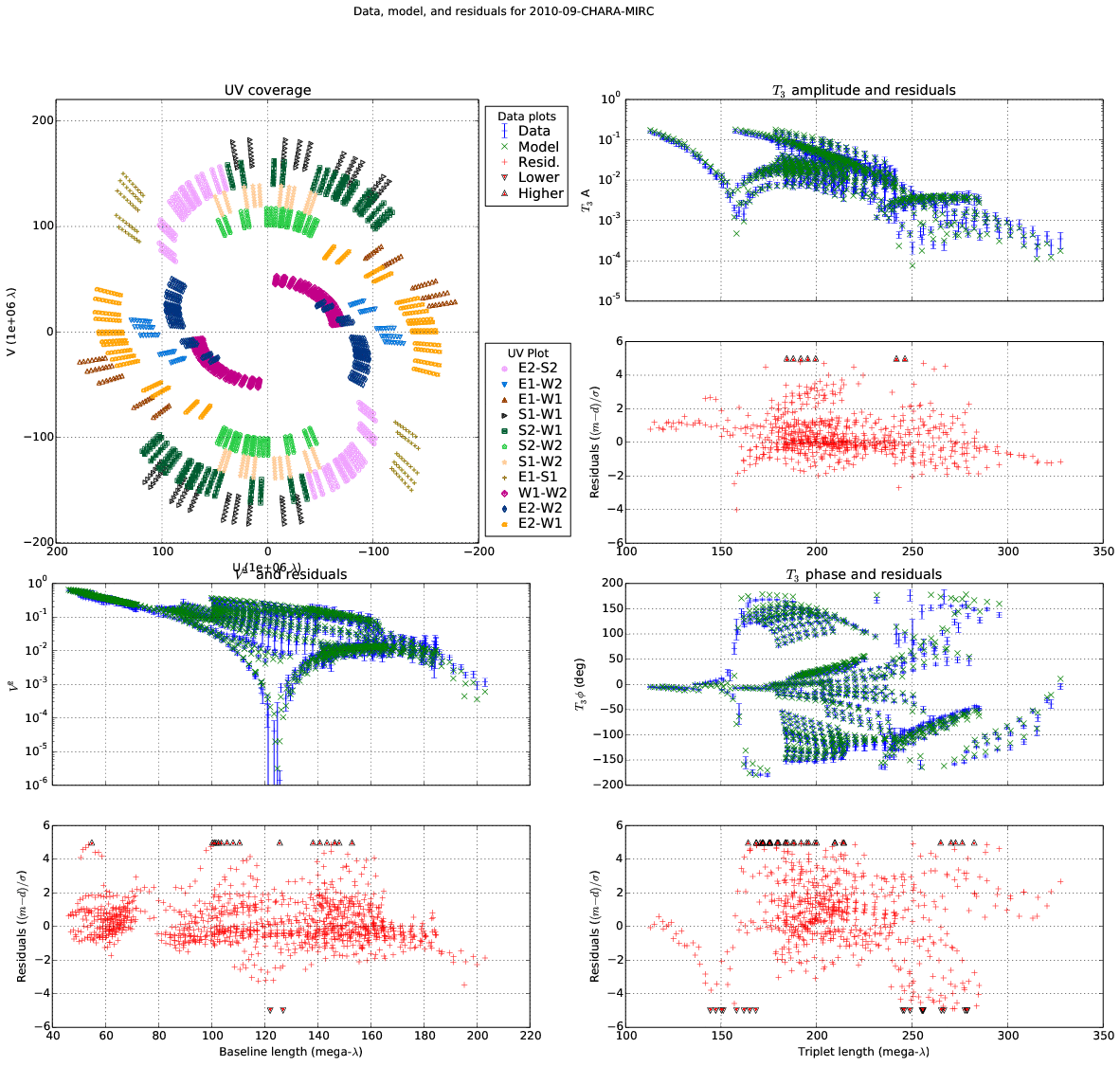}
\caption{UV coverage, data, model, and residual plots for the
2010-09-CHARA-MIRC data set.}
\end{figure}

\clearpage
\begin{figure}
\includegraphics[height=\linewidth]{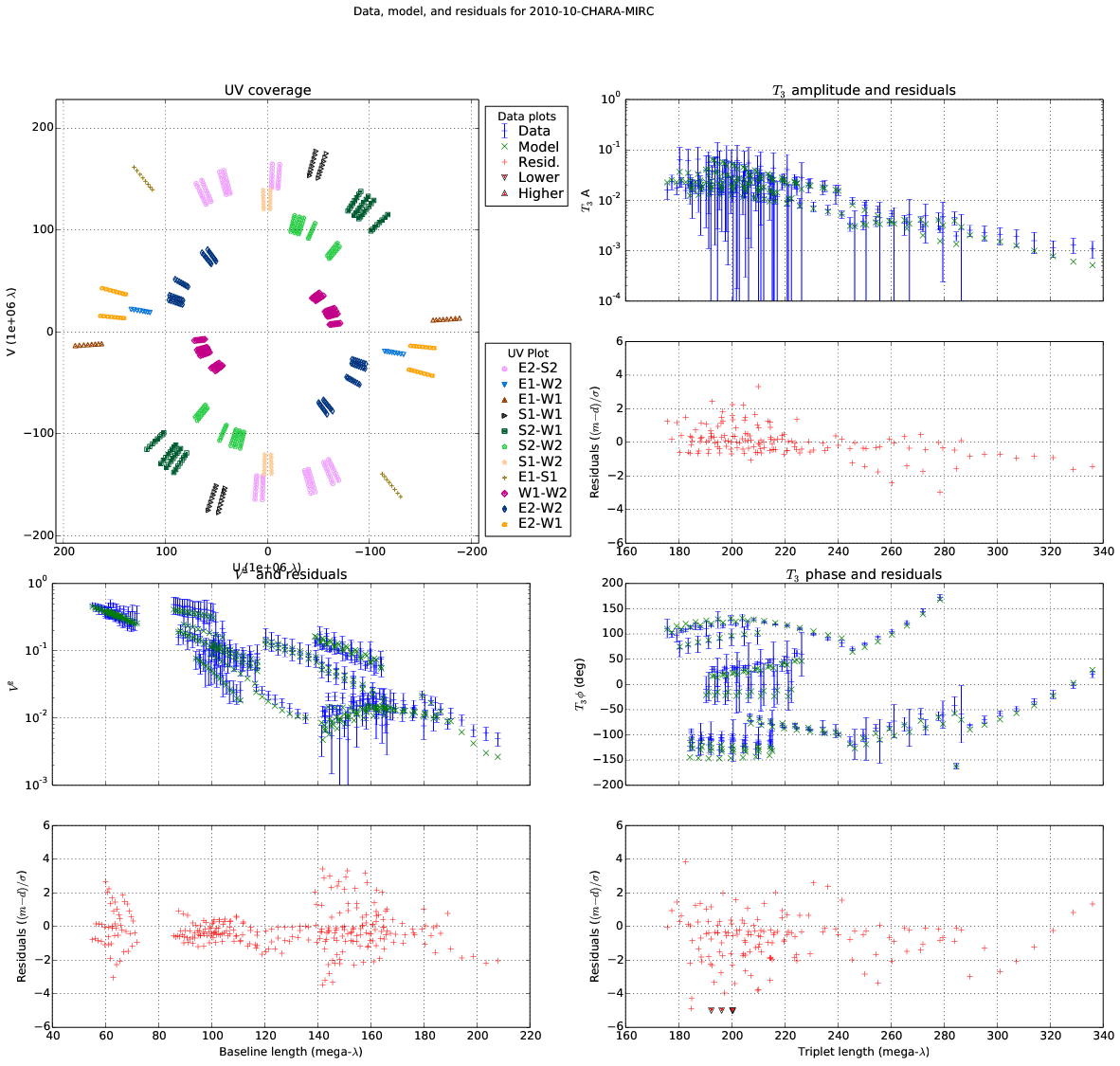}
\caption{UV coverage, data, model, and residual plots for the
2010-10-CHARA-MIRC data set.}
\end{figure}

\clearpage
\begin{figure}
\includegraphics[height=\linewidth]{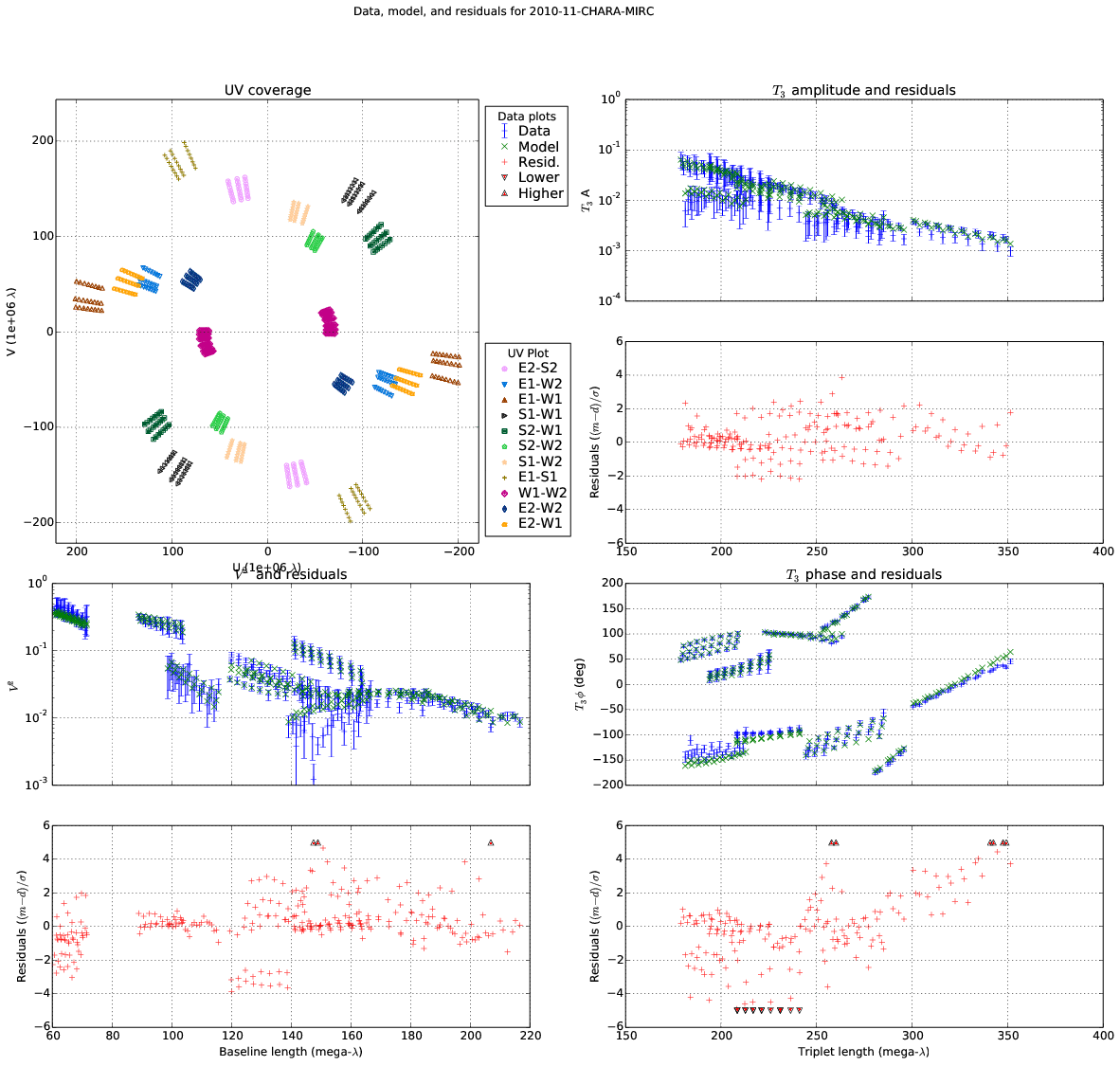}
\caption{UV coverage, data, model, and residual plots for the
2010-11-CHARA-MIRC data set.}
\end{figure}

\clearpage
\begin{figure}
\includegraphics[height=\linewidth]{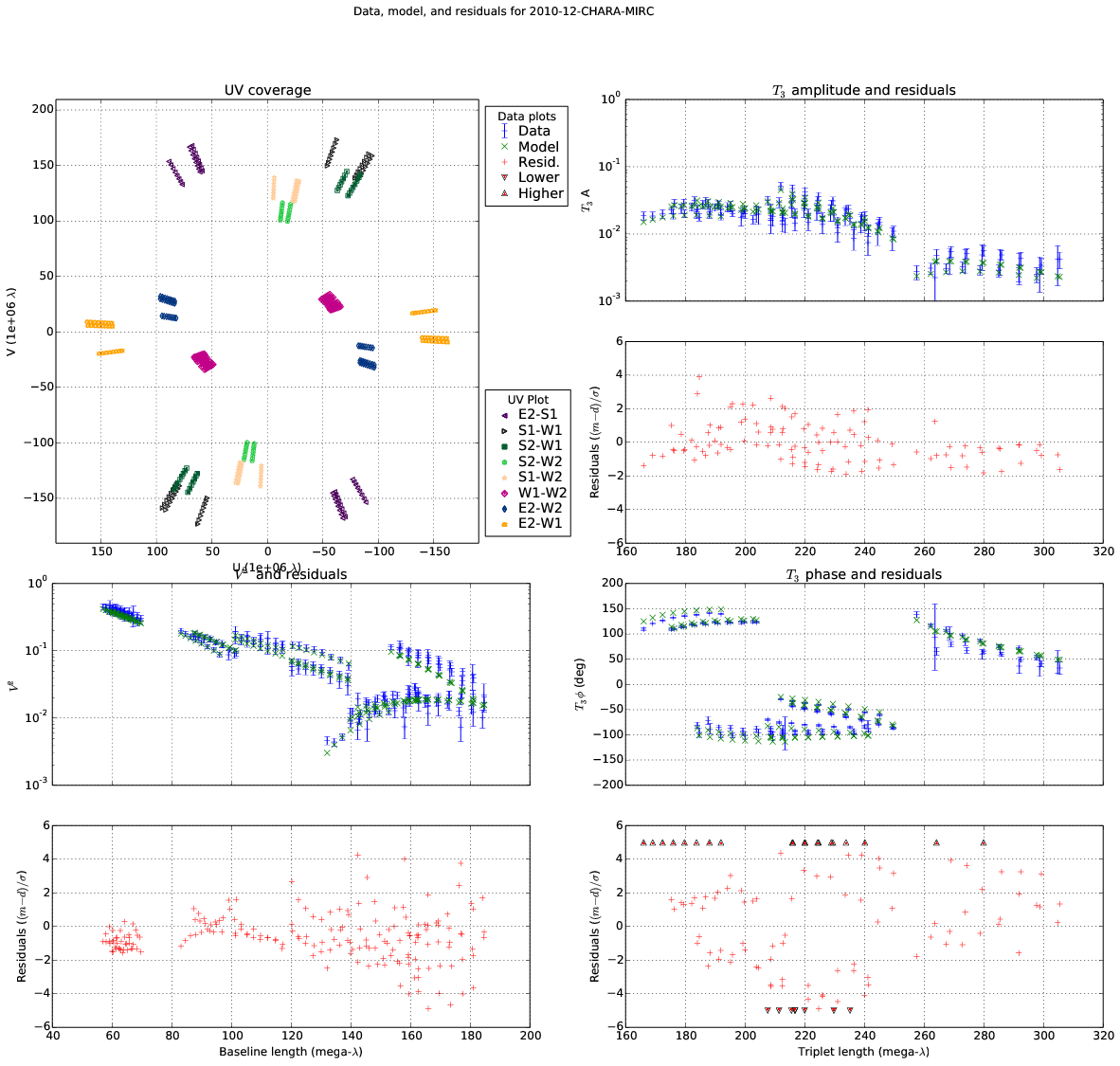}
\caption{UV coverage, data, model, and residual plots for the
2010-12-CHARA-MIRC data set.}
\end{figure}

\clearpage
\begin{figure}
\includegraphics[height=\linewidth]{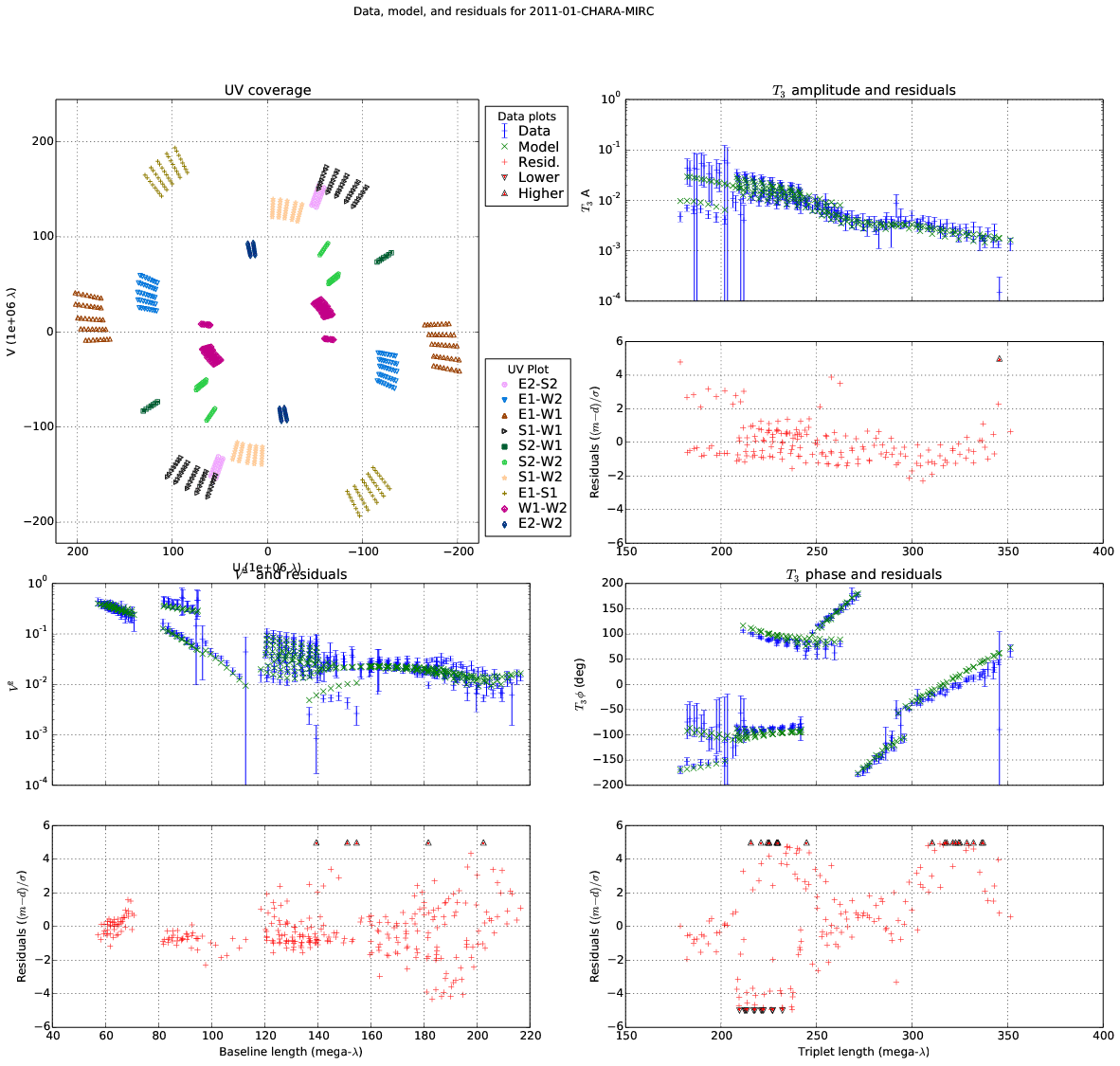}
\caption{UV coverage, data, model, and residual plots for the
2011-01-CHARA-MIRC data set.}
\end{figure}

\clearpage
\begin{figure}
\includegraphics[height=\linewidth]{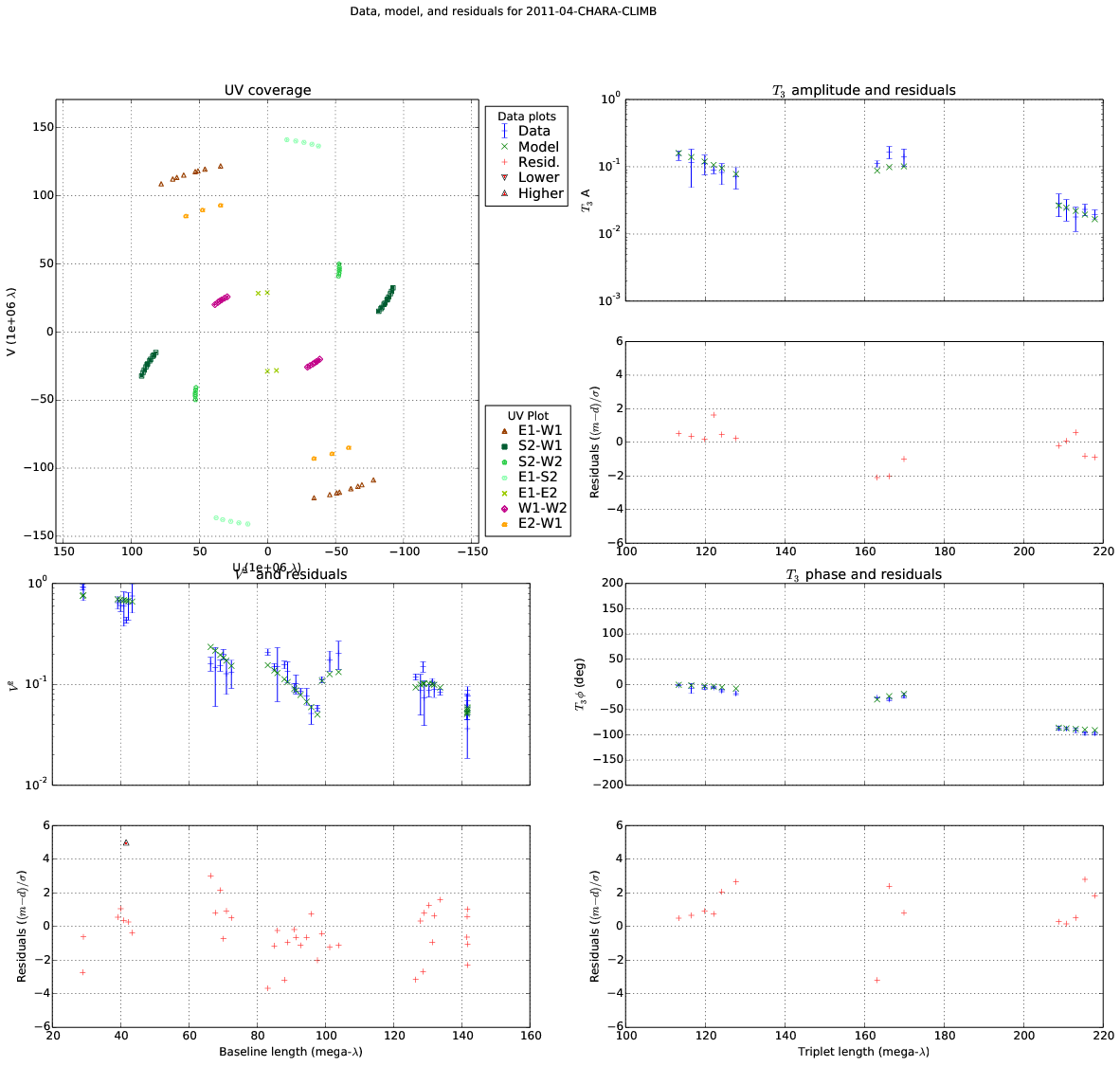}
\caption{UV coverage, data, model, and residual plots for the
2011-04-CHARA-CLIMB data set. These are the only data taken during the
egress phase.}
\end{figure}

\clearpage
\begin{figure}
\includegraphics[height=\linewidth]{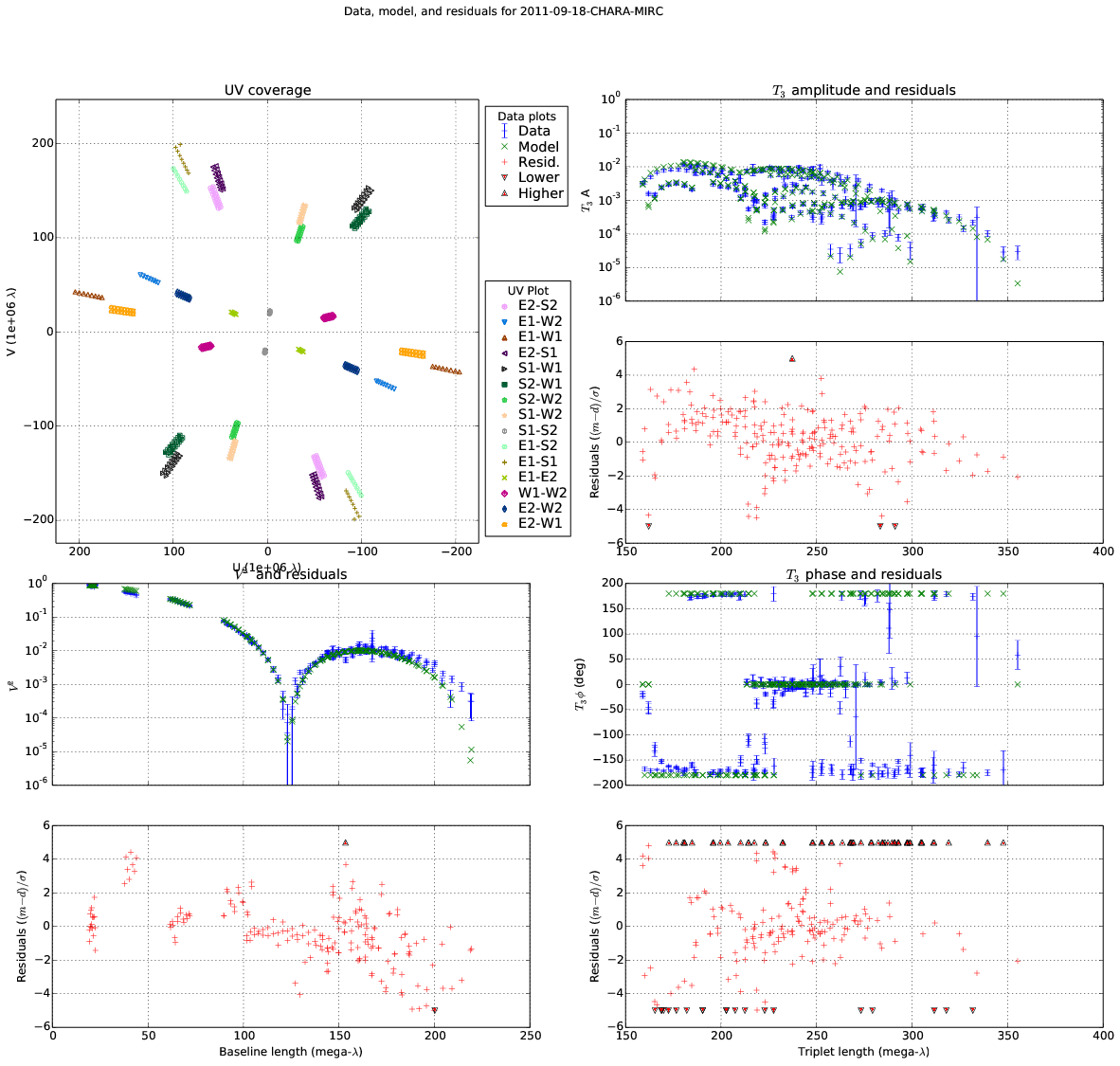}
\caption{UV coverage, data, model, and residual plots for the
2011-09-18-CHARA-MIRC data set. This is the first MIRC-6 observation of \epsAur{}}
\end{figure}

\clearpage
\begin{figure}
\includegraphics[height=\linewidth]{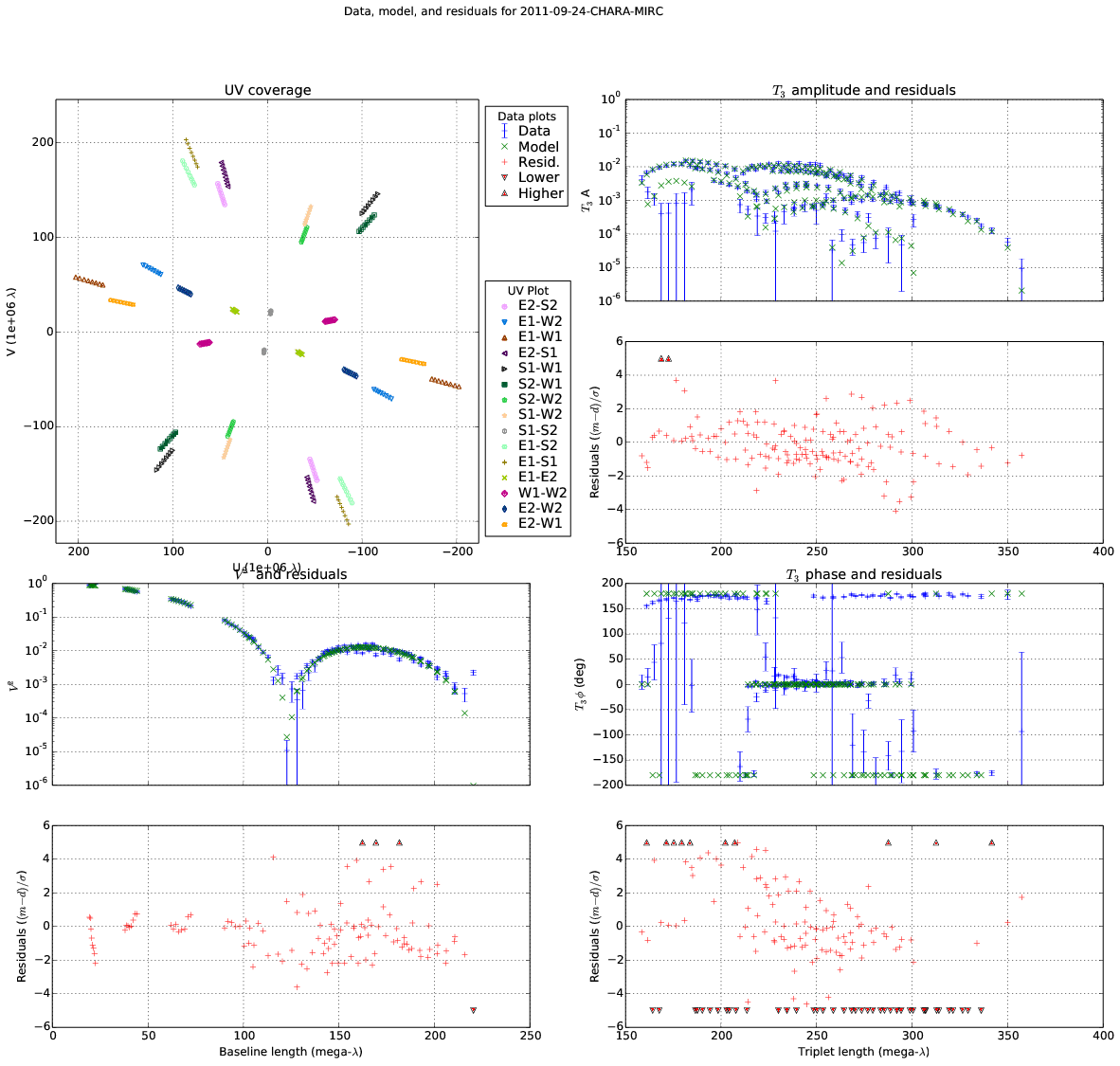}
\caption{UV coverage, data, model, and residual plots for the
2011-09-24-CHARA-MIRC data set.}
\end{figure}

\clearpage
\begin{figure}
\includegraphics[height=\linewidth]{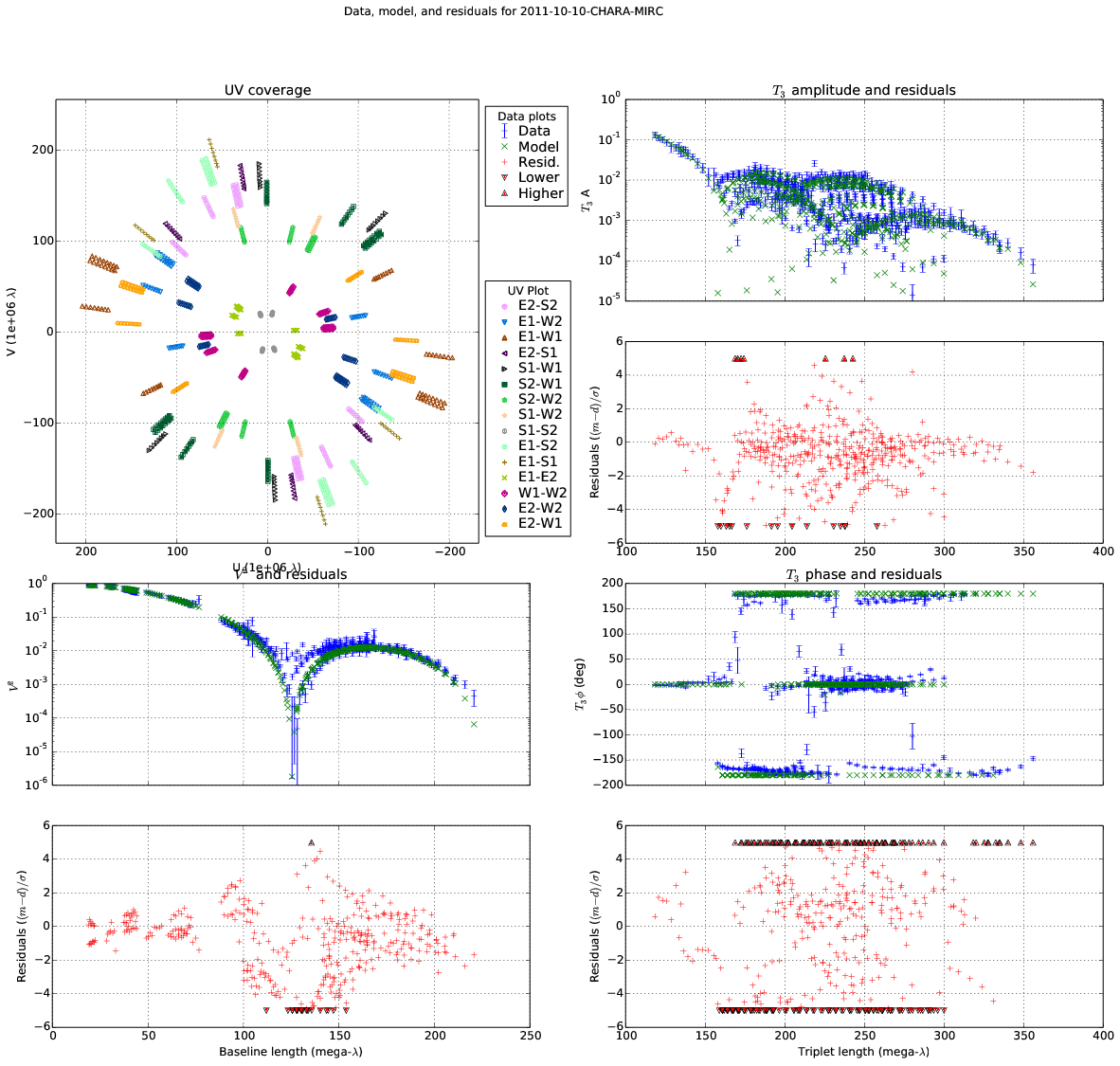}
\caption{UV coverage, data, model, and residual plots for the
2011-10-10-CHARA-MIRC data set.
\label{2011-10-10_data_plot}}
\end{figure}

\clearpage
\begin{figure}
\includegraphics[height=\linewidth]{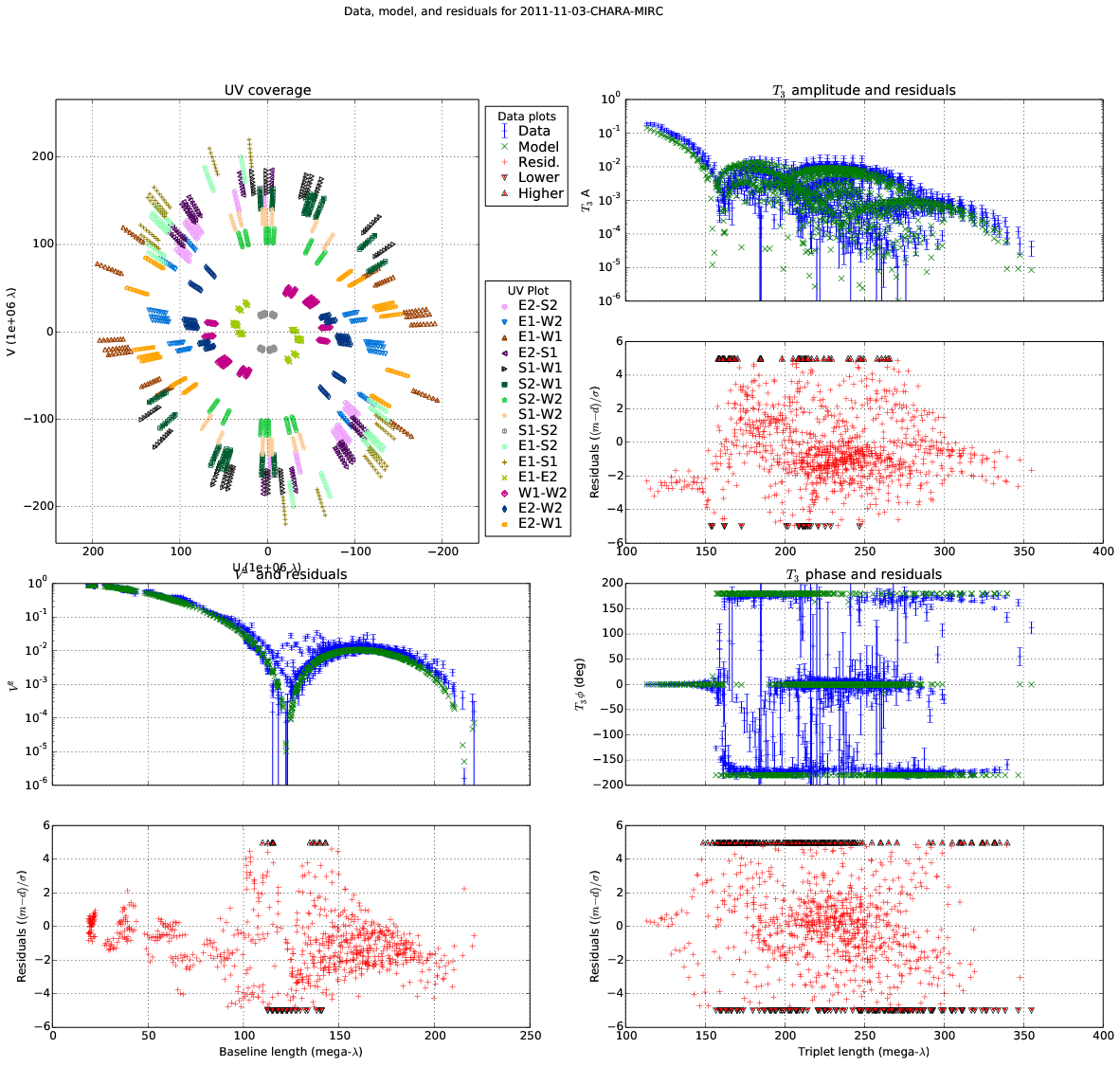}
\caption{UV coverage, data, model, and residual plots for the
2011-11-03-CHARA-MIRC data set. \label{2011-11-03_data_plot} \label{data_plots_end}}
\end{figure}

\begin{figure}
\includegraphics[width=\linewidth]{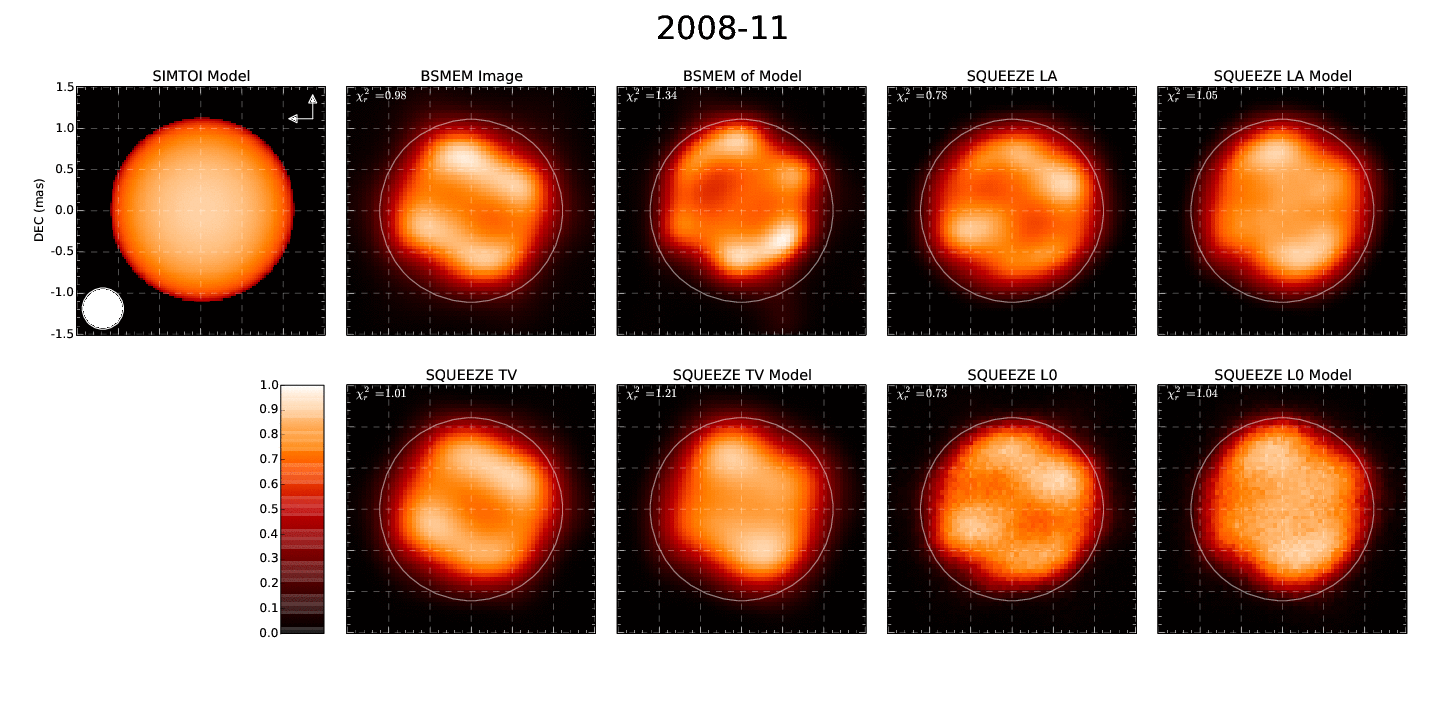}
\caption{Best-fit model and reconstructed images from the 2008-11 CHARA-MIRC epoch
\label{2008-11_image_grid}}
\end{figure}

\begin{figure}
\includegraphics[width=\linewidth]{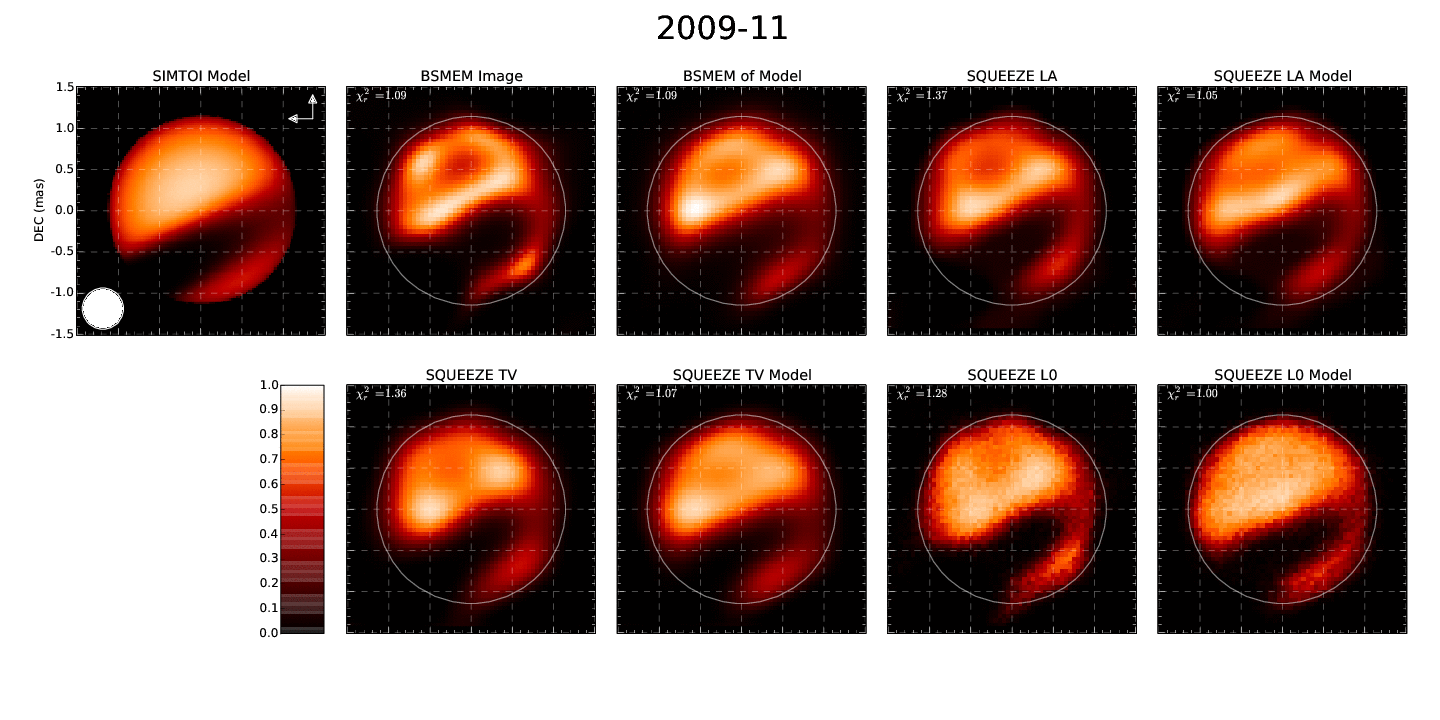}
\caption{Best-fit model and reconstructed images from the 2009-11 CHARA-MIRC epoch
\label{2009-11_image_grid}}
\end{figure}

\begin{figure}
\includegraphics[width=\linewidth]{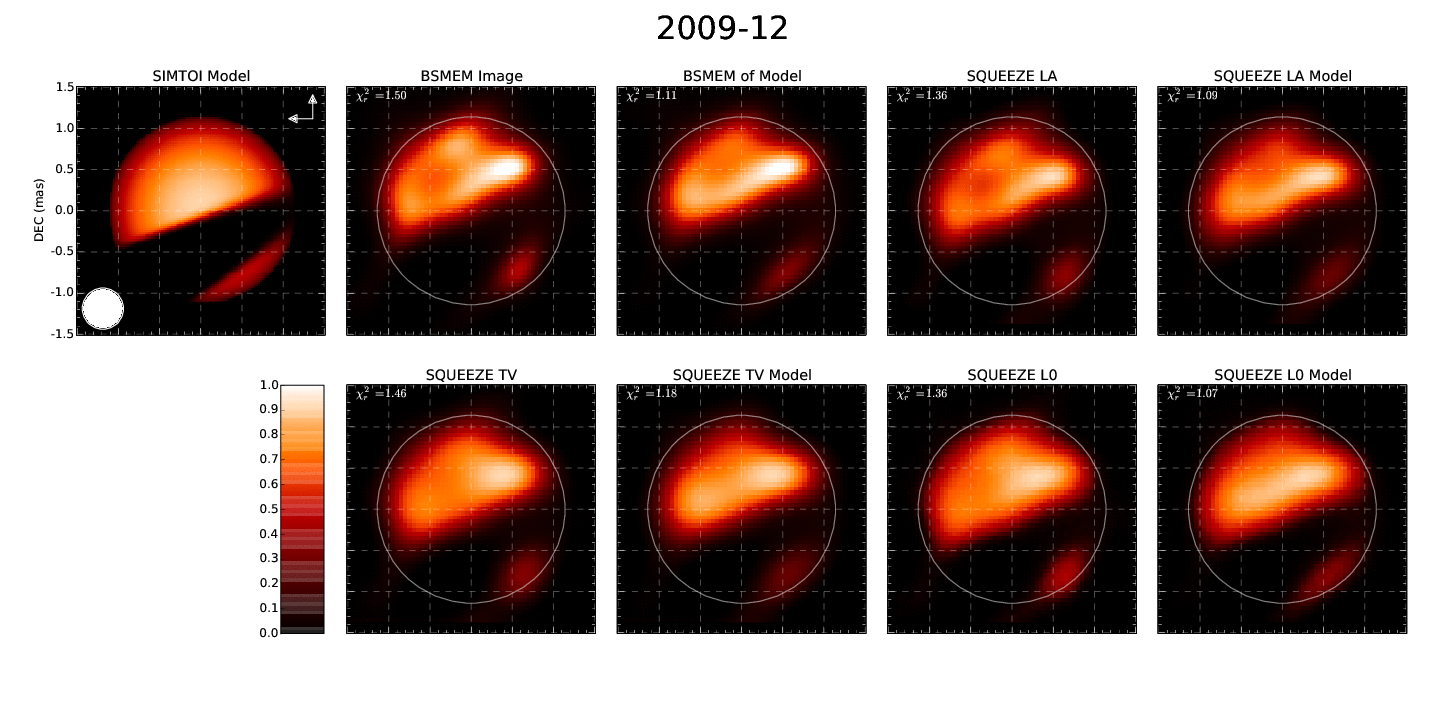}
\caption{Best-fit model and reconstructed images from the 2009-12 CHARA-MIRC epoch
\label{2009-12_image_grid}}
\end{figure}

\begin{figure}
\includegraphics[width=\linewidth]{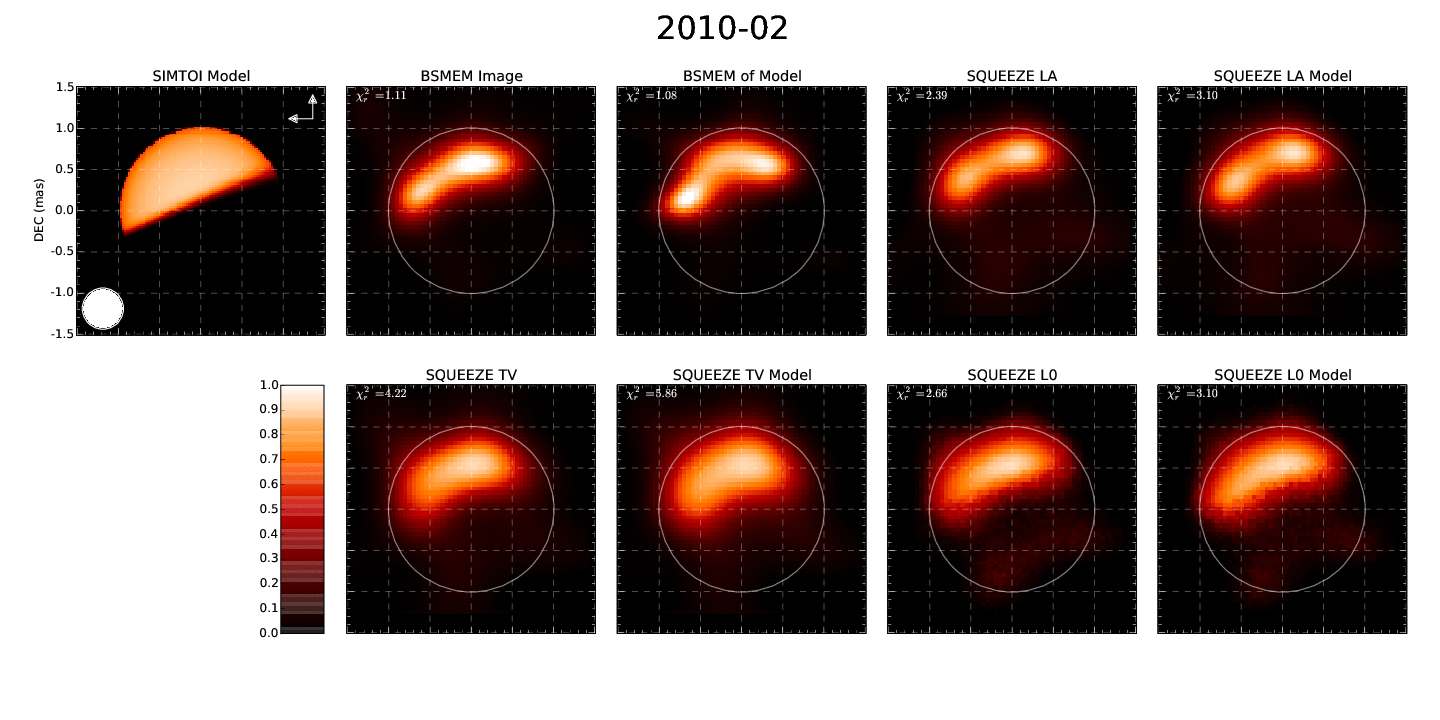}
\caption{Best-fit model and reconstructed images from the 2010-02 CHARA-MIRC epoch
\label{2010-02_image_grid}}
\end{figure}

% The following figure is split over three pages. Keep the figure number constant
\setcounter{figure}{49}
\begin{figure}
\includegraphics[width=\linewidth]{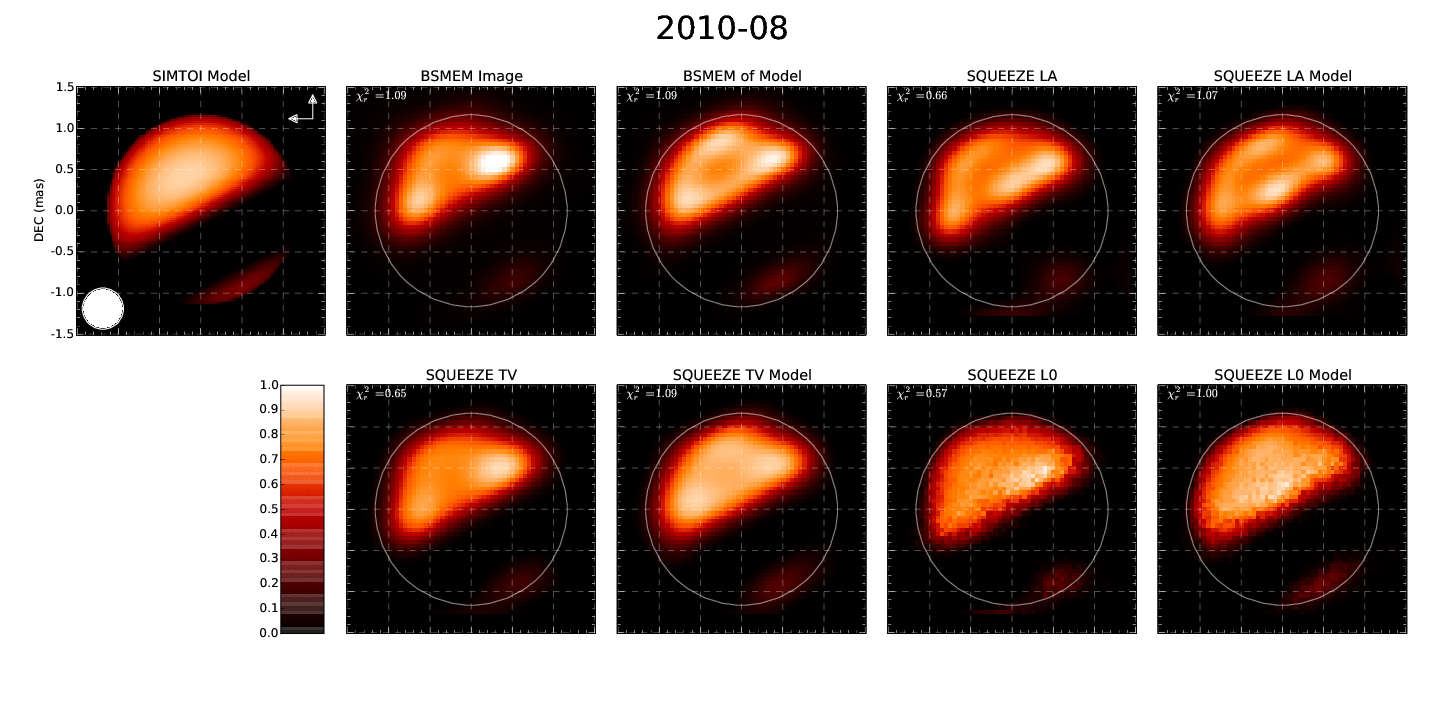}
\includegraphics[width=\linewidth]{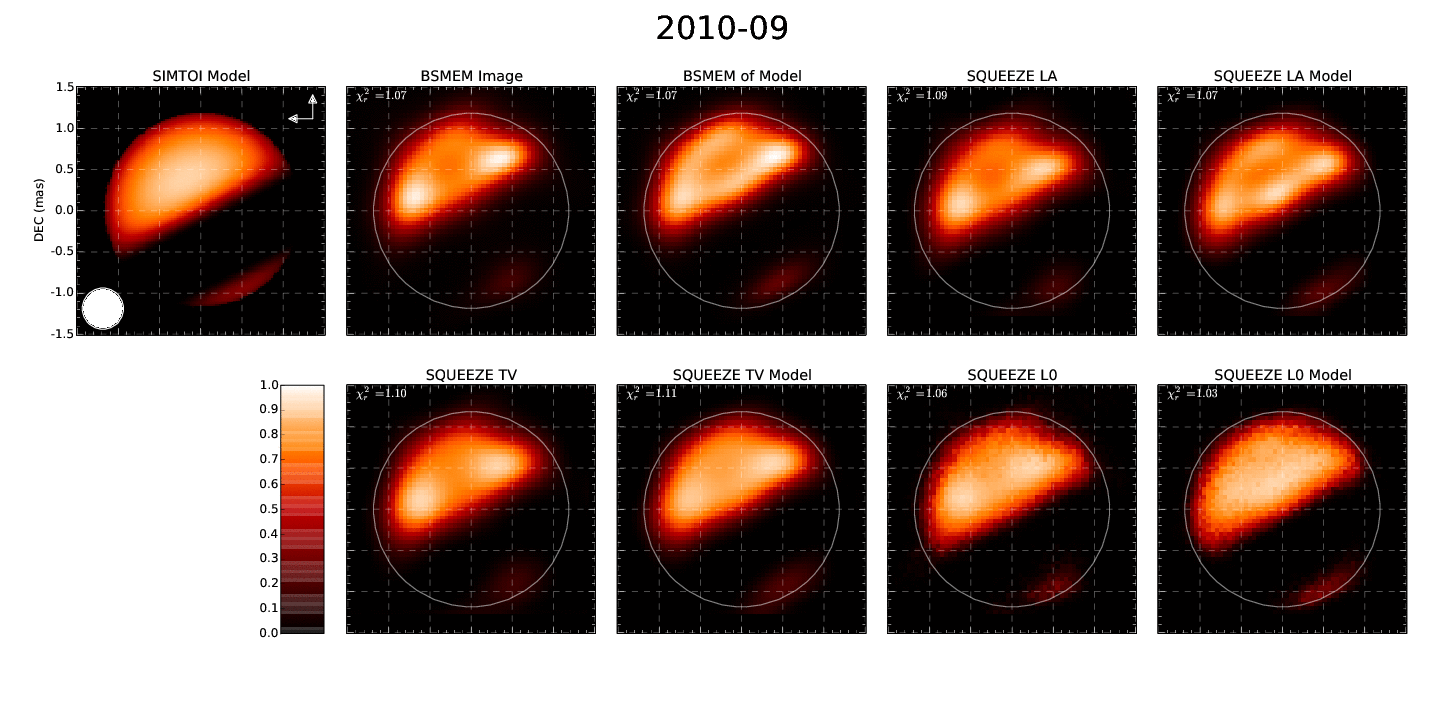}
\caption{Best-fit model and reconstructed images from the
2010-08, 2010-09, 2010-10, 2010-11, 2010-12, 2011-01 CHARA-MIRC epochs
\label{totality_image_grid}}
\end{figure}

\setcounter{figure}{49}
\begin{figure}
\includegraphics[width=\linewidth]{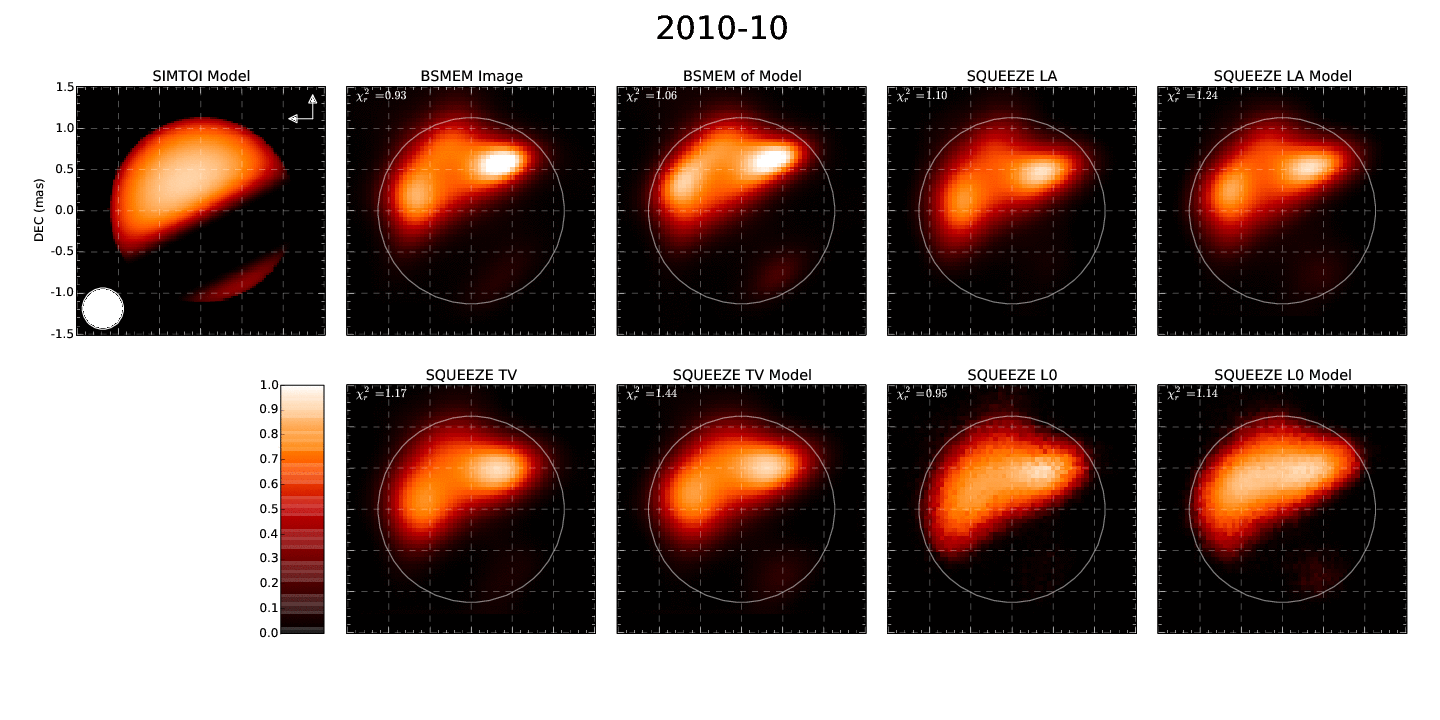}
\includegraphics[width=\linewidth]{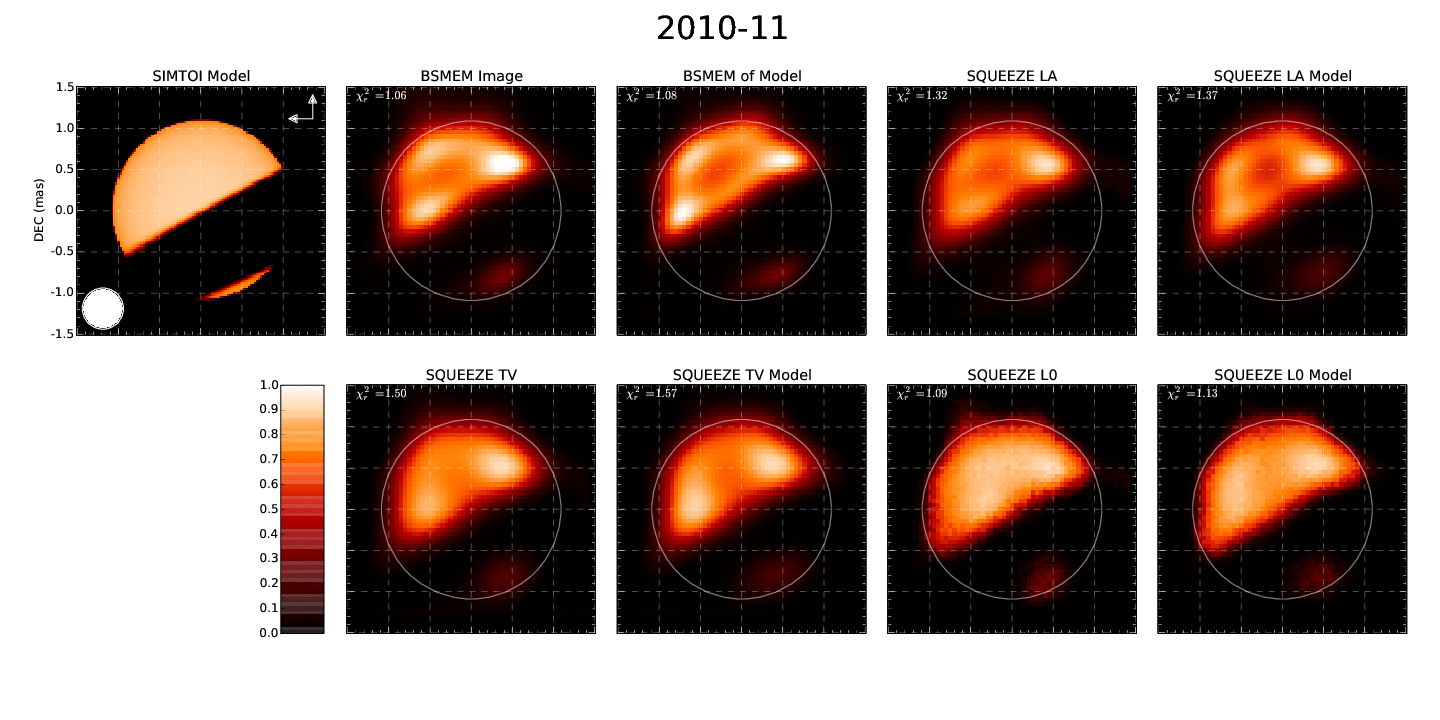}
\caption{(continued)}
\end{figure}

\setcounter{figure}{49}
\begin{figure}
\includegraphics[width=\linewidth]{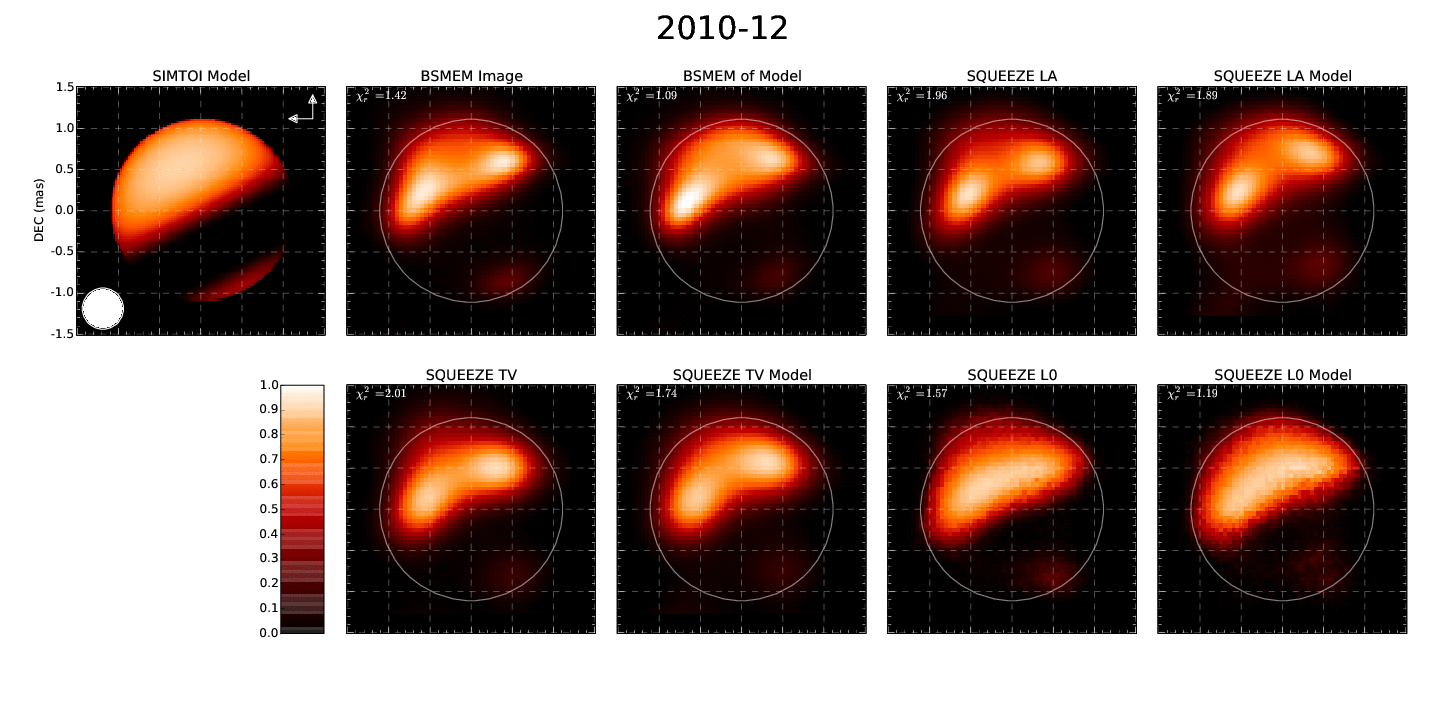}
\includegraphics[width=\linewidth]{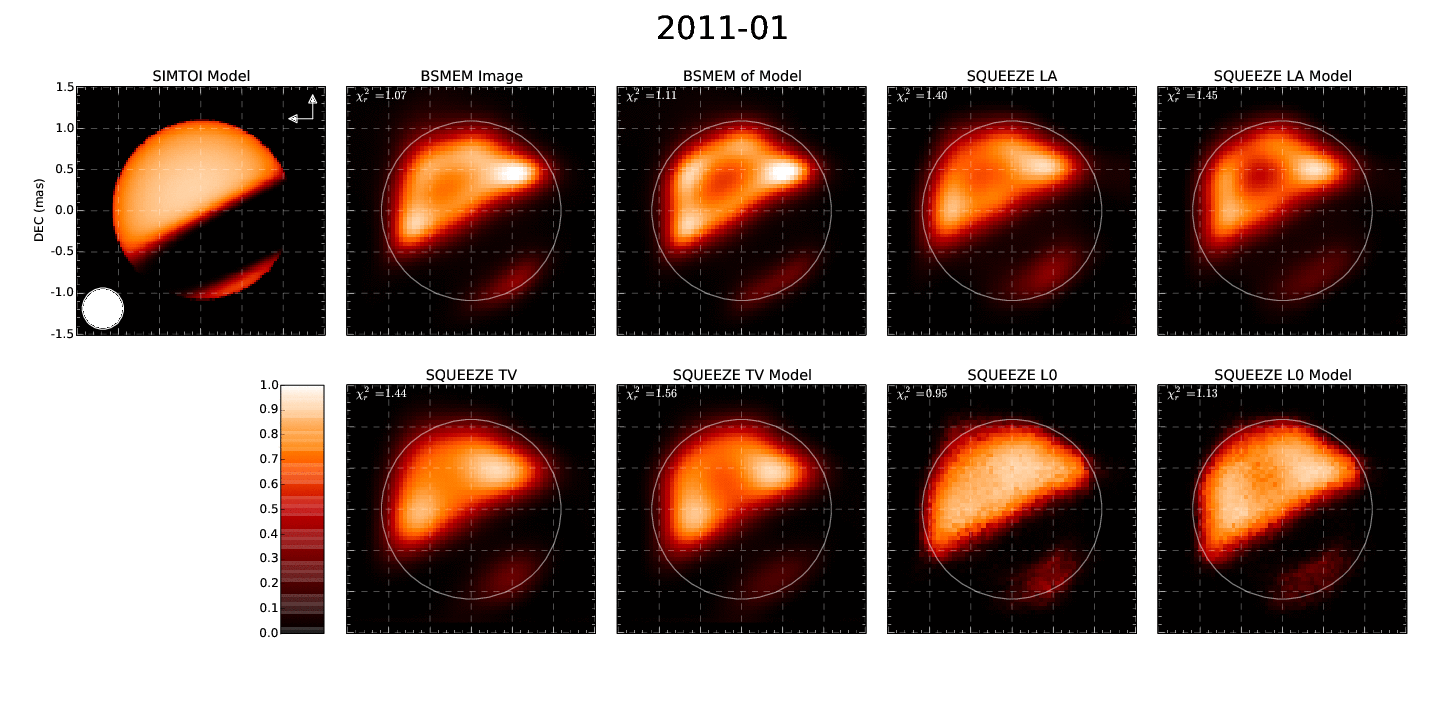}
\caption{(continued)}
\end{figure}

\begin{figure}
\includegraphics[width=\linewidth]{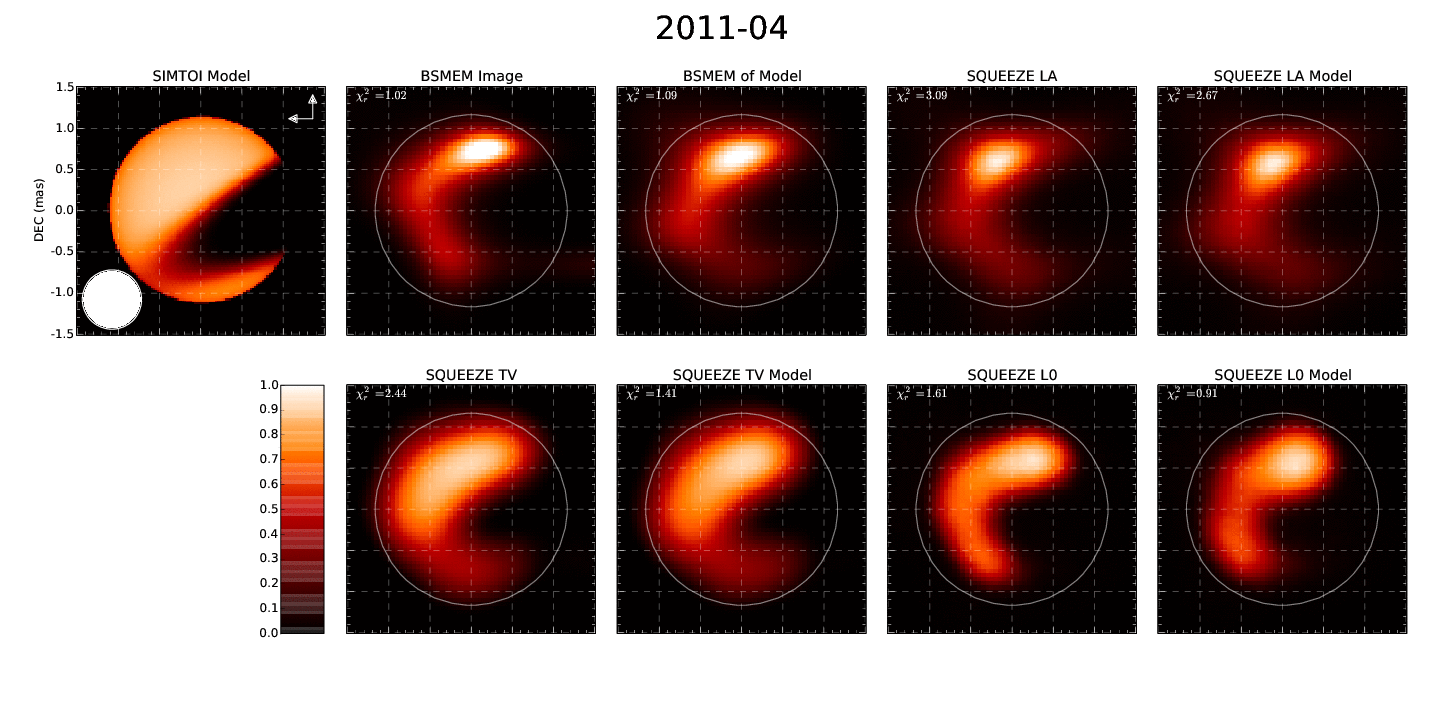}
\caption{Best-fit model and reconstructed images from the 2011-04 CHARA-CLIMB epoch
\label{2011-04_image_grid}}
\end{figure}

\setcounter{figure}{51}
\begin{figure}
\includegraphics[width=\linewidth]{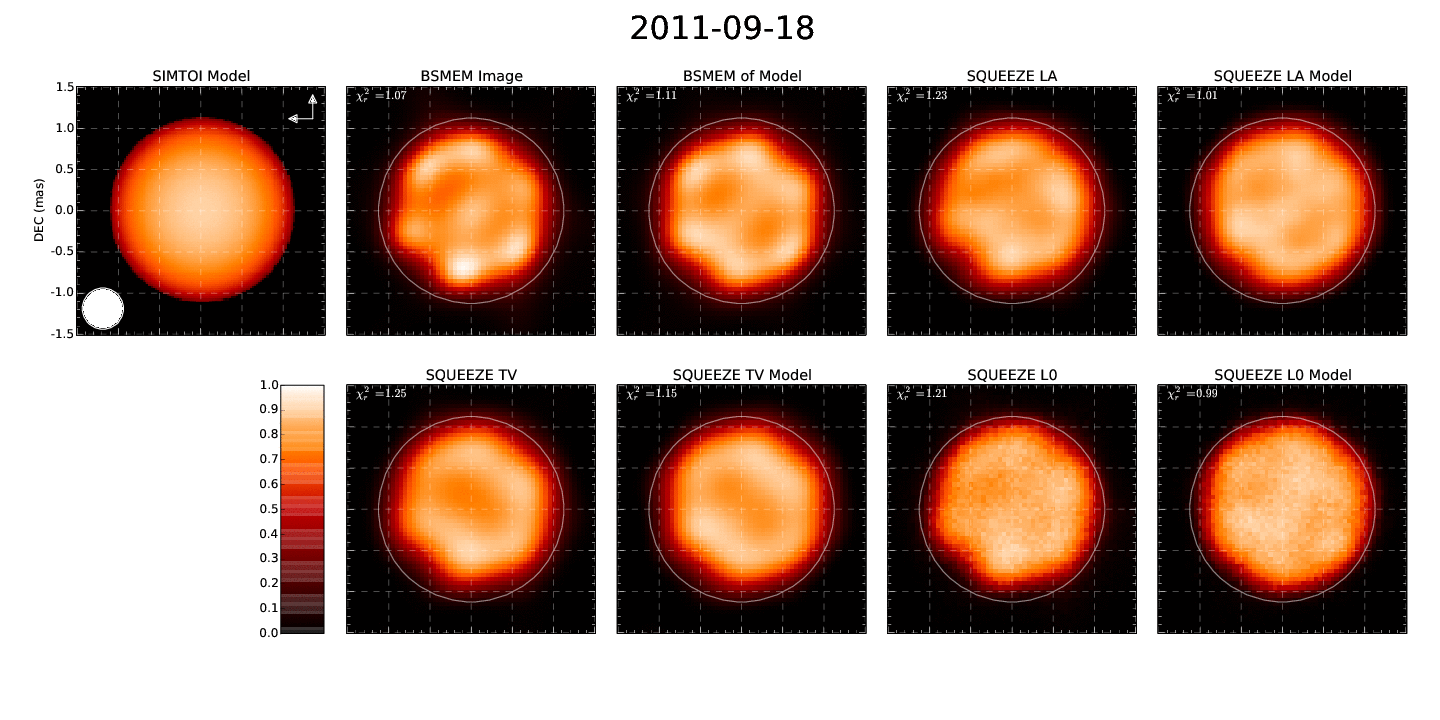}
\includegraphics[width=\linewidth]{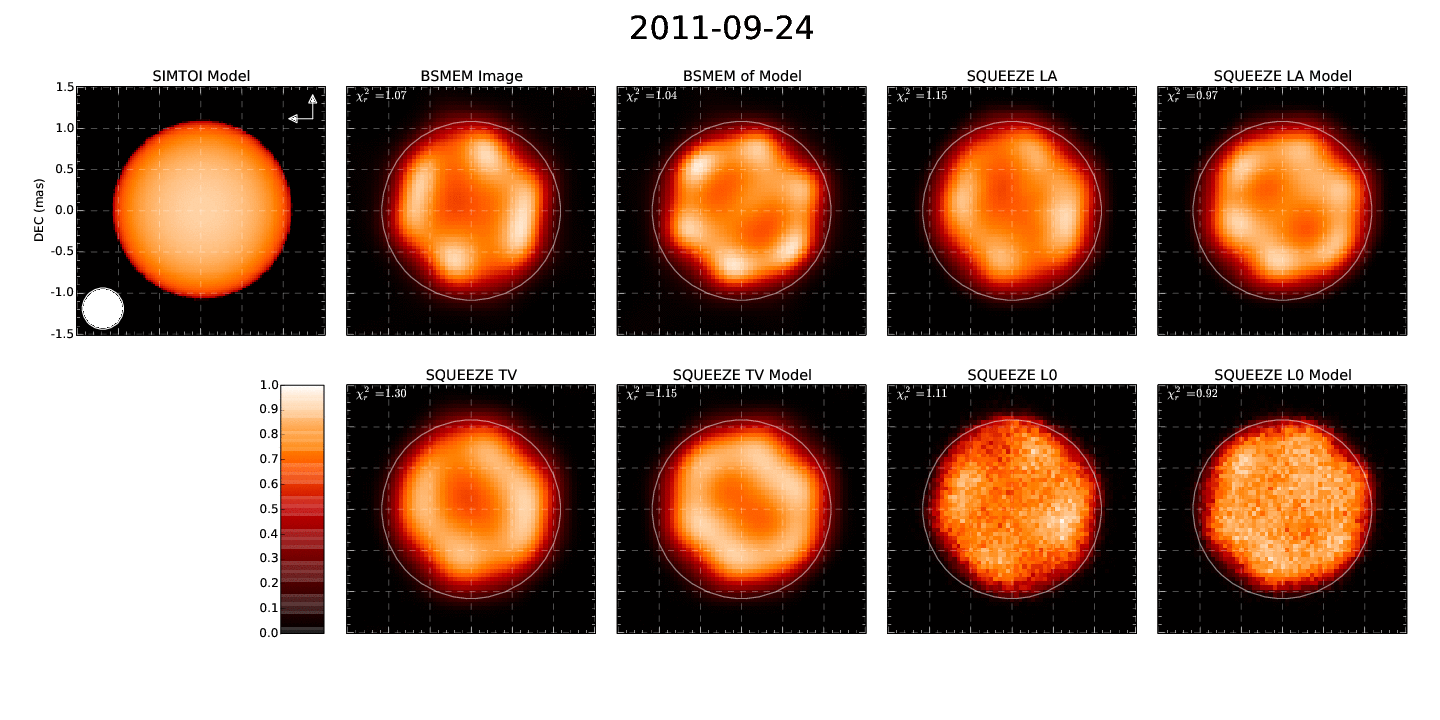}
\caption{Best-fit model and reconstructed images from the post-eclipse
2011-09-18, 2011-09-24, 2011-10-10, and 2011-11-03 CHARA-MIRC epochs
\label{post_eclipse_image_grid}}
\end{figure}

\setcounter{figure}{51}
\begin{figure}
\includegraphics[width=\linewidth]{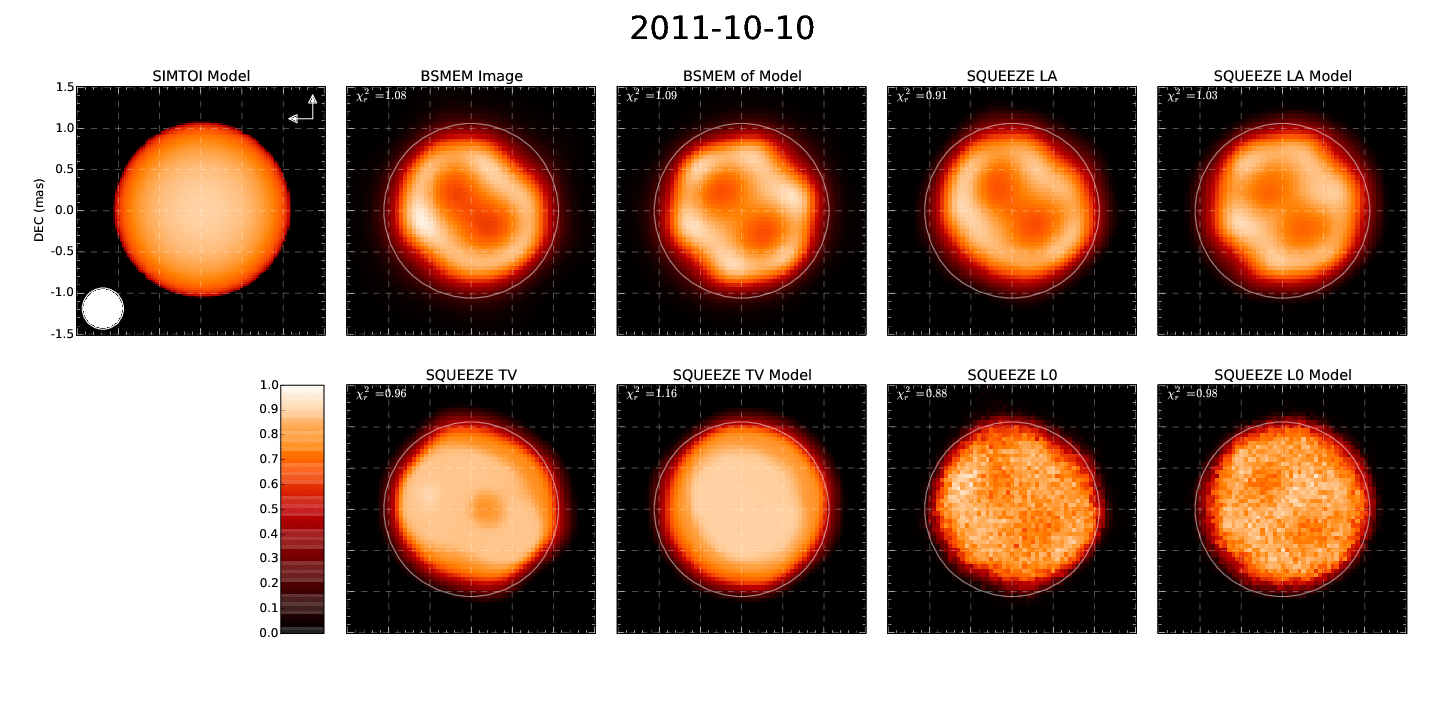}
\includegraphics[width=\linewidth]{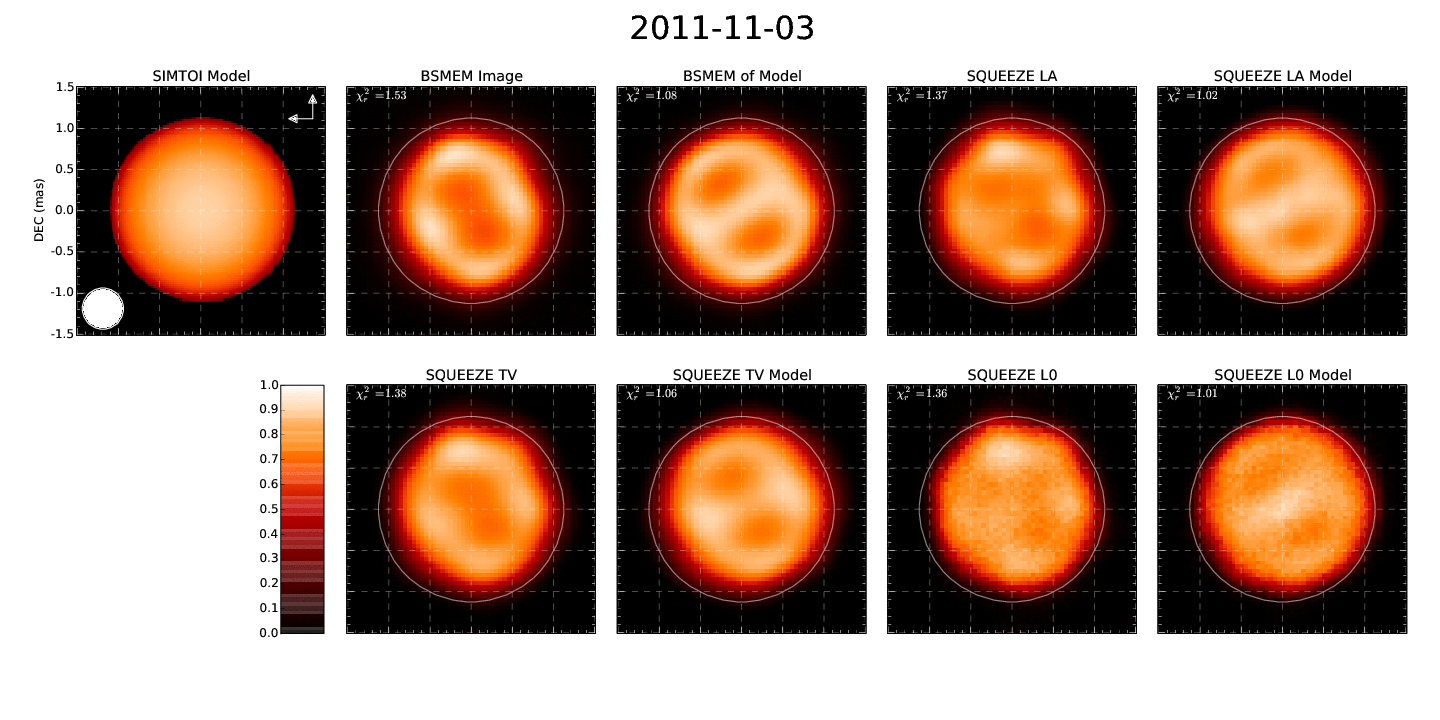}
\caption{(continued)}
\end{figure}

%% Tables may also be prepared as separate files. See the accompanying
%% sample file table.tex for an example of an external table file.
%% To include an external file in your main document, use the \input
%% command. Uncomment the line below to include table.tex in this
%% sample file. (Note that you will need to comment out the \documentclass,
%% \begin{document}, and \end{document} commands from table.tex if you want
%% to include it in this document.)

%
% tbl_observations
%
\begin{landscape}
\clearpage
\begin{deluxetable}{llp{2.8cm}llp{2cm}p{5cm}}
\tablecolumns{7}
\tablewidth{0pt}
\tabletypesize{\scriptsize}
\tablecaption{\label{tbl_observations}
List of observations from the MIRC, NPOI, CLIMB, and PTI beam combiners. 
Calibrator IDs may be cross-referenced with Table \ref{tbl_calibrators}}
\tablewidth{0pt}
\tablehead{
\colhead{Date} & 
\colhead{JD} & 
\colhead{Configuration} & 
\colhead{Array} & 
\colhead{Combiner / mode} & 
\colhead{Calibrators (HD)} & 
\colhead{Comment}
}
\startdata
1997-10-22 & 2,450,744.02 & N-S & PTI & K-band, 7 channels & 32630, 33167 &  \\ 
1997-11-09 & 2,450,761.95 & N-S & PTI & K-band, 11 channels & 32630 &  \\ 
1998-11-07 & 2,451,124.95 & N-S & PTI & K-band, 5 channels & 32630 &  \\ 
1998-11-25 & 2,451,142.94 & N-S & PTI & K-band, 5 channels & 30823 &  \\ 
1998-11-26 & 2,451,143.93 & N-S & PTI & K-band, 5 channels & 42807 &  \\ 
2005-12-11 & 2,453,715.78 & N-W & PTI & K-band, 9 channels & 29645 &  \\ 
2006-01-31 & \multicolumn{1}{l}{} &  & PTI & K-band & no calibrators &  \\ 
2006-02-25 & 2,453,791.74 & E06-AC0-AW0, AC0-AE0-AN0 & NPOI & 16 channels & 32630 &  \\ 
2006-02-26 & 2,453,792.75 & E06-AC0-AW0, AC0-AE0-AN0 & NPOI & 16 channels & 32630 &  \\ 
2006-03-03 & 2,453,432.68 & E06-AC0-AW0, AC0-AE0-AN0 & NPOI & 16 channels & 32630 & Noisy data on longest baseline \\ 
2007-03-02 & 2,454,161.63 & W07-AC0-AN0, AW0-AC0-W07 & NPOI & 16 channels & 32630 &  \\ 
2007-03-05 & 2,454,164.67 & E06-AN0-AW0, E06-AW0-W07 & NPOI & 16 channels & 32630 &  \\ 
2007-03-10 & 2,454,169.65 & E06-AN0-AW0, E06-AW0-W07 & NPOI & 16 channels & 32630 &  \\ 
2007-03-12 & 2,454,898.63 &  & NPOI & 16 channels & 32630 & Calibrator scans incoherent, unusable \\ 
2007-03-13 & 2,454,172.63 & E06-AN0-AW0, E06-AW0-W07 & NPOI & 16 channels & 32630 &  \\ 
2007-03-14 & 2,454,173.63 & E06-AN0-AW0, E06-AW0-W07 & NPOI & 16 channels & 32630 &  \\ 
2007-03-15 & 2,454,174.63 & E06-AN0-AW0, E06-AW0-W07 & NPOI & 16 channels & 32630 &  \\ 
\tablebreak
2007-03-16 & 2,454,175.64 & E06-AN0-AW0, E06-AW0-W07 & NPOI & 16 channels & 32630 &  \\ 
2007-10-19 & 2,454,392.99 & N-S & PTI & K-band, 5 channels & 29645 &  \\ 
2007-10-20 & 2,454,393.97 & N-S & PTI & K-band, 9 channels & 30138, 32630 & Incoherent v2 \\ 
2007-10-21 & 2,454,394.98 & N-S & PTI & K-band, 5 channels & 30138 &  \\ 
2007-11-27 & 2,454,431.85 & N-S & PTI & K-band, 5 channels & 27946, 30138, 32537 &  \\ 
2007-12-23 & 2,454,457.78 & N-W & PTI & K-band, 5 channels & 30138, 32630, 33167 & Incoherent v2 \\ 
2007-12-24 & 2,454,458.75 & N-W & PTI & K-band, 5 channels & 32630 &  \\ 
2008-02-16 & 2,454,512.63 & N-W & PTI & K-band, 5 channels & 30138, 32630 &  \\ 
2008-02-17 & 2,454,513.64 & N-W & PTI & K-band, 5 channels & 30138, 32630 &  \\ 
2008-02-18 & 2,454,514.65 & N-W & PTI & K-band, 5 channels & 30138, 32630 &  \\ 
2008-09-19 & 2,454,729.02 & S1-E1-W1-W2 & CHARA & MIRC, 4T, LR-H, choppers & 3360, 22928, 219080 &  \\ 
2008-10-17 & 2,454,757.00 & N-W & PTI & K-band, 5 channels & 32630 &  \\ 
2008-10-26 & 2,454,765.98 & N-S, N-W & PTI & K-band, 5 channels & 29645, 30138, 30823, 32630 & Incoherent v2 \\ 
2008-11-07 & 2,454,777.89 & S1-E1-W1-W2 & CHARA & MIRC, 4T, LR-H, choppers & 5448, 101501, 24398, 50019 &  \\ 
2008-11-08 & 2,454,778.88 & S1-E1-W1-W2 & CHARA & MIRC, 4T, LR-H, choppers & 5448, 101501, 24398, 50019 &  \\ 
2008-11-08 & 2,454,778.87 & N-S & PTI & K-band, 5 channels & 29645, 30138, 30823, 32630 &  \\ 
2008-11-09 & 2,454,779.87 & N-W & PTI & K-band, 5 channels & 29645, 30138, 30823, 32630 &  \\ 
2008-11-15 & 2,454,786.87 & N-S & PTI & K-band, 5 channels & 42807, 73262 & Closest calibrator, HD 32406, is a bad calibrator. Had to use these cals which are $>3$ hours away. \\ 
\tablebreak
2008-11-22 & 2,454,792.82 & N-S, N-W & PTI & K-band, 5 channels & 30823, 32630 &  \\ 
2008-12-10 & 2,454,810.82 & S1-E1-W1-W2 & CHARA & MIRC, 4T, LR-H, choppers &  101501, 24398 &  \\ 
2008-12-20 & \multicolumn{1}{l}{} &  & PTI &  & 32630 & No calibrators in spectral records. Will not calibrate. \\ 
2008-12-21 & \multicolumn{1}{l}{} &  & PTI &  & 32630 & Only one record, will not calibrate. \\ 
2009-03-08 & 2,454,898.63 & E06-AN0-AW0, E06-AW0-W07 & NPOI & 16 channels & 32630 & Poor weather \\ 
2009-03-12 & 2,454,902.63 & E06-AN0-AW0, E06-AW0-W07 & NPOI & 16 channels & 32630 & Poor weather, not used. \\ 
2009-11-02 & 2,455,137.80 & S1-E1-W1-W2 & CHARA & MIRC, 4T, LR-H & 32630 &  \\ 
2009-11-02 & 2,455,137.95 & S2-E2-W1-W2 & CHARA & MIRC, 4T, LR-H & 32630 &  \\ 
2009-11-03 & 2,455,138.79 & S1-E1-W1-W2 & CHARA & MIRC, 4T, LR-H &  3360, 24760, 32630 & MIRC + PAVO \\ 
2009-11-03 & 2,455,138.98 & S2-E2-W1-W2 & CHARA & MIRC, 4T, LR-H &  24760, 32630 & MIRC + PAVO \\ 
2009-11-04 & 2,455,139.75 & S1-E1-W1-W2 & CHARA & MIRC, 4T, LR-H &  3360, 24760, 32630 &  \\ 
2009-11-04 & 2,455,139.93 & S2-E2-W1-W2 & CHARA & MIRC, 4T, LR-H &  24760, 32630 &  \\ 
2009-12-02 & 2,455,167.77 & S1-E1-W1-W2 & CHARA & MIRC, 4T, LR-H & 32630 &  \\ 
2009-12-02 & 2,455,167.89 & S2-E2-W1-W2 & CHARA & MIRC, 4T, LR-H &  24760, 32630 &  \\ 
2009-12-03 & 2,455,168.74 & S1-E1-W1-W2 & CHARA & MIRC, 4T, LR-H & 24760 &  \\ 
2009-12-03 & 2,455,168.93 & S2-E2-W1-W2 & CHARA & MIRC, 4T, LR-H &  32630, 41636 &  \\ 
2009-12-04 & 2,455,169.88 & S2-E2-W1-W2 & CHARA & MIRC, 4T, LR-H &  24760, 32630, 41636 &  \\ 
2009-12-18 & 2,455,183.77 & AW0-AC0, AE0-AC0 & NPOI & 16 channels & 32630 &  \\ 
2009-12-19 & 2,455,184.73 & AW0-AC0, AE0-AC0 & NPOI & 16 channels & 32630 &  \\ 
2009-12-20 & 2,455,185.80 & AW0-AC0, AE0-AC0 & NPOI & 16 channels & 32630 &  \\ 
2009-12-21 & 2,455,186.81 & AC0-E06-AW0, AE0-AC0-AN0 & NPOI & 16 channels & 32630 &  \\ 
\tablebreak
2009-12-25 & 2,455,190.71 & AW0-AC0, AE0-AC0 & NPOI & 16 channels & 32630 & Very noisy, little data, not used. \\ 
2009-12-27 & 2,455,192.69 & AW0-AC0, AE0-AC0 & NPOI & 16 channels & 32630 &  \\ 
2010-01-03 & 2,455,199.67 & AC0-E06-AW0, AE0-AC0-AN0 & NPOI & 16 channels & 32630 &  \\ 
2010-01-04 & 2,455,289.60 &  & NPOI & 16 channels & 32630 & All incoherent scans, unusable \\ 
2010-01-05 & 2,455,201.68 & AC0-E06-AW0, AE0-AC0-AN0 & NPOI & 16 channels & 32630 &  \\ 
2010-01-06 & 2,455,202.79 & AE0-AN0-AW0, AW0-AE0-AN0 & NPOI & 16 channels & 32630 &  \\ 
2010-01-07 & 2,455,203.78 & AE0-AN0-AW0, AW0-AE0-AN0 & NPOI & 16 channels & 32630 &  \\ 
2010-01-08 & 2,455,204.66 & AE0-AN0-AW0, AW0-AE0-AN0 & NPOI & 16 channels & 32630 &  \\ 
2010-01-09 & 2,455,205.86 & AE0-AN0-AW0, AW0-AE0-AN0 & NPOI & 16 channels & 32630 &  \\ 
2010-01-10 & 2,455,206.83 & AE0-AN0-AW0, AW0-AE0-AN0 & NPOI & 16 channels & 32630 &  \\ 
2010-01-12 & 2,455,208.77 & AE0-AN0-AW0, AW0-AE0-AN0 & NPOI & 16 channels & 32630 & Erradic behavior on long baselines, possibly real signal. \\ 
2010-01-13 & 2,455,209.70 & AE0-AN0-AW0, AW0-AE0-AN0 & NPOI & 16 channels & 32630 &  \\ 
2010-01-16 & 2,455,212.69 & AC0-E06-AW0, AE0-AC0-AN0 & NPOI & 16 channels & 32630 &  \\ 
2010-02-14 & 2,455,241.62 & AC0-E06-AW0, AE0-AC0-AN0 & NPOI & 16 channels & 32630 &  \\ 
2010-02-15 & 2,455,242.62 & AC0-E06-AW0, AE0-AC0-AN0 & NPOI & 16 channels & 32630 &  \\ 
\tablebreak
2010-02-16 & 2,455,243.62 & AC0-E06-AW0, AE0-AC0-AN0 & NPOI & 16 channels & 32630 &  \\ 
2010-02-17 & 2,455,244.62 & AC0-E06-AW0, AE0-AC0-AN0 & NPOI & 16 channels & 32630 &  \\ 
2010-02-18 & 2,455,245.74 & S2-E2-W1-W2 & CHARA & MIRC, 4T, LR-H &  32630, 41636 &  \\ 
2010-02-19 & 2,455,246.70 & S1-E1-W1-W2 & CHARA & MIRC, 4T, LR-H & No calibrators, used closure phase only & See Text \\ 
2010-04-03 & 2,455,289.60 & AN0-E06-AW0, AW0-E06-W07 & NPOI & 16 channels & 32630 &  \\ 
2010-08-20 & 2,455,428.96 & S1-E1-W1-W2 & CHARA & MIRC, 4T, LR-H &  3360, 21770, 24760, 32630 &  \\ 
2010-08-21 & 2,455,429.96 & S2-E2-W1-W2 & CHARA & MIRC, 4T, LR-H &  3360, 21770, 24760, 32630 &  \\ 
2010-08-22 & 2,455,430.99 & S1-E1-W1-W2 & CHARA & MIRC, 4T, LR-H &  21770, 32630 &  \\ 
2010-08-23 & 2,455,431.96 & S2-E2-W1-W2 & CHARA & MIRC, 4T, LR-H &  3360, 21770, 24760, 32630, 219080 &  \\ 
2010-09-23 & 2,455,462.92 & S1-E2-W1-W2 & CHARA & MIRC, 4T, LR-H &  24760, 32630 &  \\ 
2010-09-24 & 2,455,463.92 & S2-E2-W1-W2 & CHARA & MIRC, 4T, LR-H &  3360, 6961, 21770, 219080, 24760, 32630 &  \\ 
2010-09-26 & 2,455,465.98 & W1-W2-S2-E2 & CHARA & MIRC, 4T, LR-H & 32630 & MIRC + PAVO \\ 
2010-09-27 & 2,455,466.99 & W1-W2-S2-E2 & CHARA & MIRC, 4T, LR-H & 32630 & MIRC + PAVO \\ 
2010-09-28 & 2,455,467.97 & W1-W2-E2-S2 & CHARA & MIRC, 4T, LR-H & 32630 & MIRC + PAVO \\ 
2010-10-26 & 2,455,495.85 & S1-E1-W1-W2 & CHARA & MIRC, 4T, LR-H &  24760, 32630 &  \\ 
2010-10-26 & 2,455,495.98 & S2-E2-W1-W2 & CHARA & MIRC, 4T, LR-H & 32630 &  \\ 
\tablebreak
2010-10-27 & 2,455,496.86 & S1-E1-W1-W2 & CHARA & MIRC, 4T, LR-H & 32630 &  \\ 
2010-10-27 & 2,455,496.91 & S2-E2-W1-W2 & CHARA & MIRC, 4T, LR-H & 32630 &  \\ 
2010-11-04 & 2,455,504.95 & S2-E2-W1-W2 & CHARA & MIRC, 4T, LR-H & 32630 &  \\ 
2010-11-05 & 2,455,505.89 & S1-E1-W1-W2 & CHARA & MIRC, 4T, LR-H &  21770, 32630, 41636 &  \\ 
2010-12-12 & 2,455,542.77 & W1-W2-E2-S1 & CHARA & MIRC, 4T, LR-H &  32630, 50019 & MIRC + PAVO \\ 
2010-12-13 & 2,455,543.75 & W1-W2-E2-S1 & CHARA & MIRC, 4T, LR-H &  24760, 32630 & MIRC + PAVO \\ 
2010-12-14 & 2,455,544.75 & W1-W2-S2-E1 & CHARA & MIRC, 4T, LR-H & 32630 & MIRC + PAVO \\ 
2011-01-18 & 2,455,579.67 & S1-E1-W1-W2 & CHARA & MIRC, 4T, LR-H & 32630 &  \\ 
2011-01-18 & 2,455,579.89 & S2-E2-W1-W2 & CHARA & MIRC, 4T, LR-H &  32630, 50019 &  \\ 
2011-01-19 & 2,455,580.66 & S1-E1-W1-W2 & CHARA & MIRC, 4T, LR-H & 32630 &  \\ 
2011-01-19 & 2,455,580.78 & S2-E2-W1-W2 & CHARA & MIRC, 4T, LR-H &  32630, 41636 &  \\ 
2011-03-18 & \multicolumn{1}{l}{} & E1-E2-W1 & CHARA & CLIMB, H & 32630 & Non-standard readout mode and bad conditions. Not usable. \\ 
2011-04-01 & 2,455,652.68 & E1-E2-W1 & CHARA & CLIMB, K & 32630 &  \\ 
2011-04-03 & 2,455,654.67 & S2-W2-W1 & CHARA & CLIMB, K & 32630 &  \\ 
2011-04-05 & 2,455,656.66 & S2-E1-W1 & CHARA & CLIMB, K & 32630 &  \\ 
2011-09-18 & 2,455,823.03 & W1-S2-S1-E1-E2-W2 & CHARA & MIRC, 6T, LR-H & 27396 &  \\ 
2011-09-24 & 2,455,829.03 & W1-S2-S1-E1-E2-W2 & CHARA & MIRC, 6T, LR-H & 32630 & See Text \\ 
2011-10-10 & 2,455,844.94 & W1-S2-S1-E1-E2-W2 & CHARA & MIRC, 6T, LR-H & 32630 &  \\ 
2011-11-03 & 2,455,868.85 & W1-S2-S1-E1-E2-W2 & CHARA & MIRC, 6T, LR-H &  21770, 24760, 32630 &  \\ 
\enddata
\end{deluxetable}

\end{landscape}

%
% tbl_calibrators
%
\begin{landscape}
\clearpage
\setlength{\tabcolsep}{0.05in}
\begin{deluxetable}{llllllllllllllp{2cm}}
\tablecolumns{15}
\tablewidth{0pt}
\tabletypesize{\tiny}
%\rotate
\tablecaption{\label{tbl_calibrators}
Calibrators and adopted uniform disk diameters (\UDD)}
\tablehead{
% Row 1, top headers
\multicolumn{2}{c}{} &
\multicolumn{2}{c}{Position (J2000)} &
\multicolumn{3}{c}{} &
\multicolumn{3}{c}{Published diameters} &
\multicolumn{2}{c}{Adopted} &
\multicolumn{3}{c}{} \\
\cmidrule(lr){3-4} \cmidrule(lr){8-10} \cmidrule(lr){11-12}
% Row 2, column headers
\colhead{HD} &
\colhead{Name} &
\colhead{RA} &
\colhead{DEC} &
\colhead{$\mu_{\alpha}$} &
\colhead{$\mu_{\delta}$} &
\colhead{$\pi$} &
\colhead{$\theta_\text{UDD-H}$} &
\colhead{$\theta_\text{UDD-K}$} &
\colhead{$\sigma_{\theta_\text{UDD}}$} &
\colhead{\UDD{}} &
\colhead{$\sigma_{\theta_\text{UDD}}$} &
\colhead{Array} &
\colhead{Ref.} & 
\colhead{Notes} \\
% Row 3, units
\colhead{} &
\colhead{} &
\colhead{(HH MM SS.SS)} &
\colhead{(DD MM SS.SS)} &
\colhead{(\mas{} yr$^{-1}$)} &
\colhead{(\mas{} yr$^{-1}$)} &
\colhead{(\mas{})} &
\colhead{(\mas{})} &
\colhead{(\mas{})} &
\colhead{(\mas{})} &
\colhead{(\mas{})} &
\colhead{(\mas{})} &
\colhead{} &
\colhead{} &
\colhead{} \\
}
\startdata
3360 & zet Cas & 00 36 58.28419 & +53 53 48.8673 & 17.38 & -9.86 & 5.5 & 0.287 & 0.288 & 0.02 & 0.287 & 0.020 & C & 2 &  \\ 
5448 & 37 And & 00 56 45.21211 & +38 29 57.6380 & 153.48 & 36.49 & 25.14 & 0.593 & 0.594 & 0.042 & 0.593 & 0.042 & C & 2 &  \\ 
6961 & tet Cas & 01 11 06.16225 & +55 08 59.6472 & 226.77 & -18.75 & 24.42 & $\ldots$ & $\ldots$ & $\ldots$ & 0.471 & 0.033 & C & 4 &  \\ 
21770 & 36 Per & 03 32 26.26028 & +46 03 24.6965 & -52.11 & -75.26 & 27.53 & $\ldots$ & $\ldots$ & $\ldots$ & 0.582 & 0.041 & C & 4 &  \\ 
22928 & del Per & 03 42 55.50426 & +47 47 15.1746 & 25.58 & -43.06 & 6.32 & $\ldots$ & $\ldots$ & $\ldots$ & 0.549 & 0.038 & C & 4 &  \\ 
24398 & zet Per & 03 54 07.92248 & +31 53 01.0812 & 5.77 & -9.92 & 4.34 & $\ldots$ & $\ldots$ & $\ldots$ & 0.700 & 0.030 & C & 1 &  \\ 
27396 & 53 Per & 04 21 33.16557 & +46 29 55.9554 & 20.06 & -35.45 & 6.43 & $\ldots$ & $\ldots$ & $\ldots$ & 0.285 & 0.020 & C & 2 &  \\ 
27946 & 67 Tau & 04 25 25.01518 & +22 11 59.9876 & 111.97 & -47.71 & 22.03 & 0.442 & 0.443 & 0.031 & 0.443 & 0.031 & P & 2 &  \\ 
28052 & 71 Tau & 04 26 20.74092 & +15 37 05.7652 & 114.31 & -32.19 & 20.37 & $\ldots$ & $\ldots$ & $\ldots$ & 0.542 & 0.038 & P & 4 &  \\ 
29645 & HR 1489 & 04 41 50.25660 & +38 16 48.6622 & 241.65 & -97.15 & 31.38 & 0.521 & 0.522 & 0.037 & 0.523 & 0.037 & P & 2 &  \\ 
30138 & HR 1514 & 04 46 44.47871 & +40 18 45.3270 & 9.07 & -36.83 & 7.53 & 0.856 & 0.86 & 0.061 & 0.826 & 0.061 & P & 4 &  \\ 
30823 & HR 1550 & 04 52 47.75706 & +42 35 11.8569 & -10.63 & 0.46 & 7.2 & 0.317 & 0.317 & 0.022 & 0.317 & 0.022 & P & 2 & Get cal estimates $\sim 0.1$ \mas{} larger \\ 
32537 & 9 Aur & 05 06 40.62967 & +51 35 51.8025 & -30.49 & -172.89 & 38.04 & $\ldots$ & $\ldots$ & $\ldots$ & 0.589 & 0.041 & P & 4 &  \\ 
32630 & eta Aur & 05 06 30.89337 & +41 14 04.1127 & 31.45 & -67.87 & 13.4 & $\ldots$ & $\ldots$ & $\ldots$ & 0.453 & 0.012 & C,P,N & 3 &  \\ 
33167 & HR 1668 & 05 10 42.92081 & +46 57 43.4550 & 58.34 & -149.99 & 20.54 & $\ldots$ & $\ldots$ & $\ldots$ & 0.498 & 0.035 & P & 4 &  \\ 
37147 & 122 Tau & 05 37 03.73543 & +17 02 25.1776 & 42.24 & -33.69 & 20.58 & $\ldots$ & $\ldots$ & $\ldots$ & 0.375 & 0.026 & P & 4 &  \\ 
41636 & HR 2153 & 06 08 23.13611 & +41 03 20.6194 & 1.36 & -48.55 & 7.82 & $\ldots$ & $\ldots$ & $\ldots$ & 0.765 & 0.054 & C & 4 &  \\ 
42807 & HR 2208 & 06 13 12.50242 & +10 37 37.7095 & 77.38 & -298 & 55.71 & 0.473 & 0.475 & 0.034 & 0.475 & 0.034 & P & 2 &  \\ 
50019 & tet Gem & 06 52 47.33887 & +33 57 40.5175 & -1.66 & -47.31 & 17.25 & 0.802 & 0.804 & 0.056 & 0.802 & 0.056 & C & 2 &  \\ 
73262 & del Hya & 08 37 39.36627 & +05 42 13.6057 & -70.19 & -7.9 & 20.34 & 0.464 & 0.465 & 0.032 & 0.465 & 0.032 & P & 2 &  \\ 
101501 & 61 UMa & 11 41 03.01636 & +34 12 05.8843 & -12.55 & -380.75 & 104.04 & $\ldots$ & $\ldots$ & $\ldots$ & 0.864 & 0.061 & C & 4 &  \\ 
219080 & 7 And & 23 12 33.00380 & +49 24 22.3455 & 90.23 & 95.56 & 40.67 & $\ldots$ & $\ldots$ & $\ldots$ & 0.665 & 0.047 & C & 4 &  \\ 
\enddata
\tablenotetext{C}{CHARA} 
\tablenotetext{N}{NPOI} 
\tablenotetext{P}{PTI}
\tablenotetext{1}{MIRC Calibrator database}
\tablenotetext{2}{\cite{LafrasseS.2010}}
\tablenotetext{3}{\cite{Maestro2013}}
\tablenotetext{4}{Value computed using SearchCal}
\end{deluxetable}

\end{landscape}

%
% tbl_ld_selection
%
\clearpage
\begin{deluxetable}{llllllll}
\tablecolumns{8}
\tablewidth{0pt}
\tabletypesize{\scriptsize}
\tablecaption{\label{tbl_ld_selection}
\textbf{Posterior odds ratios ($\Delta \log R$) relative to the uniform disk model}}
\tablehead{
\multicolumn{1}{l}{} & \multicolumn{7}{c}{Posterior odds ratio ($\Delta \log R$, see 
Section \ref{sec_bayesian_model_selection})} \\ 
\colhead{Epoch} & 
\colhead{\cite{ClaretA.2000a}} & 
\colhead{\cite{Fields2003}} & 
\colhead{Logarithmic} & 
\colhead{Power law $^\dagger$} & 
\colhead{Quadratic} & 
\colhead{Square root} & 
\colhead{Uniform disk} \\
}
\startdata
2011-09-18 & 9504 & 9518 & 9515 & 9506 & 9545 & 9506 & \\
2011-09-24 & 1954 & 1949 & 1959 & 1956 & 1971 & 1954 & \\
2011-10-10 & 2568 & 2615 & 2560 & 2561 & 2560 & 2558 & \\
2011-11-03 & 6151 & 5900 & 6165 & 6166 & 6227 & 6134 & \\

\enddata
\tablenotetext{\dagger}{Although there is slight evidence in favor 
of the quadratic limb darkening law, visual inspection of the visibility data 
shows no significant difference between it and the power-law limb darkening that 
we pragmatically adopted in this work.}
\end{deluxetable}

%
% tbl_parameters
%
\clearpage
\begin{deluxetable}{llll}
\tablecolumns{4}
\tablewidth{0pt}
\tabletypesize{\scriptsize}
\tablecaption{\label{tbl_parameters}
A summary of all parameters used in the modeling process}
\tablehead{
\colhead{Parameter} & 
\colhead{Range} & 
\colhead{Units} & 
\colhead{Description} \\
}
\startdata
\\
\sidehead{F-star parameters}                  
\UDD{}                    &  1-3              & mas        & Uniform disk diameter \\
\LDD{}                    &  1-3              & mas        & Limb darkened disk diameter \\
$\alpha_\text{LDD}$       &  0-1              &            & Power-law limb darkening coefficient \\
\sidehead{Geometric disk models}
$r_\text{in}$             &  0                & mas        & Inner radius \\
$r_\text{out}$            &  1-30             & mas        & Outer radius \\                 
$h_\text{disk}$           &  0-5              & mas        & Height \\                    
$\alpha$                  &  5-20             &            & Radial exponent \\
$\beta$                   &  0.1-5            &            & Height exponent \\  
\sidehead{Astrophysical density disk models}
$\kappa_0 \rho_0$         &  3,000-10,000      &            & Characteristic opacity \\    
$\alpha$                  &  5-20             &            & Radial exponent (Pascucci model) \\
$\beta$                   &  0.1-5            &            & Height exponent (Pascucci model) \\
$\gamma$                  &  0.001-20         &            & Radial exponent (Andrews model) \\
$h_c$                     &  0.001-20         & mas        & Disk scale height \\
$r_c$                     &  1-4              & mas        & Disk scale radius \\
$i_\text{disk}$           &  $\pm 10$         & deg        & Disk inclination   \\
$\Omega_\text{disk}$      &  $\pm 10$         & deg        & Disk position angle \\
\sidehead{Orbital parameters$^\dagger$}
$e$                       &  0.227 $\pm$ 0.01 &            & Eccentricy \\
$i$                       &  70-110           & deg        & Inclination \\
$\omega$                  &  39.2 $\pm$ 3.4   & deg        & Longitude of periastron \\
P                         &  9,896 $\pm$ 1.6  & day        & Period \\
$\Omega$                  &  90-145 and 270-325  & deg        & Position Angle \\
$\alpha_T$, $\alpha_1$, $\alpha_2$ & $13 < \alpha_T < 38$ & mas & Semi-major axis (total, F-star, disk) \\
T                         &  2,434,723 $\pm$ 80  & days       & Time of periastron \\
\enddata
\tablenotetext{\dagger}{$e$, $\omega$, $P$, and $T$ from \cite{Stefanik2010} }
\end{deluxetable}

%
% tbl_disk_model_selection
%
\begin{landscape}
\clearpage
\setlength{\tabcolsep}{0.05in}
\begin{deluxetable}{lllllllllllllllllll}
\tabletypesize{\scriptsize}
\tablecolumns{10}
\tablewidth{0pt}
\tablecaption{\label{tbl_disk_model_selection}
\textbf{Bayes factors relative to the cylinder model} and average reduced $\chi^2$ 
for the six disk models described in the Section \ref{sec_disk_models}.}
\tablehead{
% row 1
\multicolumn{2}{l}{} & 
\multicolumn{2}{c}{Orbit} & 
\multicolumn{10}{c}{Disk} &
\multicolumn{5}{c}{Fit Information} \\ 
\cmidrule(lr){3-4} \cmidrule(lr){5-14} \cmidrule(lr){15-19}
% row 2, headers
\colhead{Name} &
\colhead{Model$^{\dagger}$} & 
\colhead{$\alpha_T$} & 
\colhead{$i$} & 
\colhead{$r_\text{in}$} & 
\colhead{$r_\text{out}$} & 
\colhead{$h$} & 
\colhead{$\alpha$} & 
\colhead{$\beta$} & 
\colhead{$r_c$} & 
\colhead{$h_c$} & 
\colhead{$\kappa \rho$ $^{\ddagger}$} &  
\colhead{$\Omega_\text{disk}$} & 
\colhead{$i_\text{disk}$} &
\colhead{$\Delta \log R$} & 
\colhead{$\chi^2_r(\text{H})$} & 
\colhead{$\overline{\chi^2_r}(V^2)$} & 
\colhead{$\overline{\chi^2_r}(T_{3\text{A}})$} & 
\colhead{$\overline{\chi^2_r}(T_{3 \phi})$} \\
% row 3, units
\multicolumn{2}{l}{} & 
\multicolumn{1}{l}{(mas)} & 
\multicolumn{1}{l}{(deg)} & 
\multicolumn{1}{l}{(mas)} & 
\multicolumn{1}{l}{(mas)} & 
\multicolumn{1}{l}{(mas)} & 
\multicolumn{2}{l}{} & 
\multicolumn{1}{l}{(mas)} & 
\multicolumn{1}{l}{(mas)} & 
\multicolumn{1}{l}{} &
\multicolumn{1}{l}{(deg)} &
\multicolumn{1}{l}{(deg)} &
\multicolumn{4}{l}{} \\
}
\startdata
Cylinder & 1 & 21.7 & 88.4 & 0.0 & 4.85 & 0.54 & $\ldots$ & $\ldots$ & $\ldots$ & $\ldots$ & $\ldots$ & $\ldots$ & $\ldots$ & 0 & 260 & 10 & 12 & 48 \\ 
Ringed Disk & 2 & 27.0 & 88.7 & 0.0 & 7.20 & 0.71 & 0.13 & 1.64 & $\ldots$ & $\ldots$ & $\ldots$ & $\ldots$ & $\ldots$ & 63940 & 69 & 4.1 & 4.3 & 31 \\ 
RingedDisk (only $\beta$) & 3 & 27.5 & 88.8 & 0.0 & 6.39 & 0.79 & $\ldots$ & 0.02 & $\ldots$ & $\ldots$ & $\ldots$ & $\ldots$ & $\ldots$ & 59811 & 93 & 4.4 & 4.1 & 32 \\ 
RingedDisk (only $\alpha$) & 4 & 27.5 & 88.8 & 0.0 & 7.40 & 0.70 & 0.11 & $\ldots$ & $\ldots$ & $\ldots$ & $\ldots$ & $\ldots$ & $\ldots$ & 63792 & 67 & 4.1 & 4.4 & 31 \\ 
Pascucci Disk & 5 & 33.0 & 89.0 & 0.0 & $\ldots$ & $\ldots$ & 11.11 & 1.75 & 2.39 & 0.029 & 6496 & $\ldots$ & $\ldots$ & 67058 & 50 & 3.8 & 4.1 & 30 \\ 
Andrews Disk & 6 & 33.0 & 89.0 & 0.0 & $\ldots$ & $\ldots$ & 11.14 & 1.74 & 2.48 & 0.032 & 5161 & $\ldots$ & $\ldots$ & 67110 & 49 & 3.8 & 4.1 & 31 \\ 
Pascucci Disk w/ clearing & 7 & 32.6 & 88.9 & 3.8 & $\ldots$ & $\ldots$ & 10.94 & 0.75 & 2.32 & 0.079 & 6667 & $\ldots$ & $\ldots$ & 67254 & 51 & 3.8 & 4.0 & 31 \\ 
Tilted Pascucci Disk$^*$ & 8 & 31.2 & 88.9 & 0.0 & $\ldots$ & $\ldots$ & 13.33 & 3.69 & 2.77 & 0.007 & 6287 & -0.02 & 2.98 & 68806 & 50 & 3.7 & 3.7 & 28 \\ 
\enddata
\tablenotetext{\dagger}{See Section \ref{results_disk_selection} for model descriptions}
\tablenotetext{*}{Model 8 obtains the highest Bayes factor and is therefore
 adopted in this work.}
\end{deluxetable}

\end{landscape}

%
% tbl_bootstrap_results
%
\begin{landscape}
\clearpage
\setlength{\tabcolsep}{0.01in}
\begin{deluxetable}{lllllccccccccccccp{4cm}}
\tablecolumns{19}
\tablewidth{0pt}
%\rotate
\tabletypesize{\tiny}
\tablecaption{\label{tbl_bootstrap_results}
Bootstrapped nominal values and uncertainties for model 8 with a fixed tilt 
evaluated on a per-epoch basis subject to only the interferometric data.}
\tablehead{
% row 1, labels
\multicolumn{5}{l}{} &
\multicolumn{3}{c}{F-star} &
\multicolumn{5}{c}{Disk} &
\multicolumn{4}{c}{Statistical Information} &
\multicolumn{1}{l}{}
\\
% Horizontal rules under the above columns
\cmidrule(lr){6-8} 
\cmidrule(lr){9-13}
\cmidrule(lr){14-17}
% No \\ needed here. Just start the next row
\colhead{Data set } & 
\colhead{$N(V^2)$} & 
\colhead{$N(T_3)$} & 
\colhead{$N(UV)$} & 
\colhead{Effective JD} & 
\colhead{ \UDD{} } & 
\colhead{ \LDD{} } & 
\colhead{$\alpha_\text{LDD}$} & 
\colhead{$\kappa \rho$} & 
\colhead{$r_c$} & 
\colhead{$h_c$} & 
\colhead{$\alpha$} & 
\colhead{$\beta$ $^*$} & 
\colhead{$\chi^2_r$} &
\colhead{$\chi^2_r(V^2)$} &
\colhead{$\chi^2_r(T_{3A})$} &
\colhead{$\chi^2_r(T_{3\phi})$} &
\colhead{Notes}
\\
% row 2, units
\multicolumn{5}{l}{} & 
(mas) & 
(mas) &  
\multicolumn{2}{l}{} & 
(mas) & 
(mas) &
\multicolumn{4}{l}{} 
\\
% row 2, units
\multicolumn{7}{l}{} & 
$(\times 10^{-2})$ &
$(\times 10^{-2})$ &
\multicolumn{3}{l}{} 
\\
}
\startdata
1997-10-22-PTI      & 14 & \ldots & 16 & 2450744.0200 & $2.43 \pm 0.29$ & \ldots & \ldots & \ldots & \ldots & \ldots & \ldots & \ldots & 0.38 & 0.38 & \ldots & \ldots &  \\ 
1998-11-07-PTI      & 66 & \ldots & 69 & 2450761.9500 & $2.13 \pm 0.28$ & \ldots & \ldots & \ldots & \ldots & \ldots & \ldots & \ldots & 0.27 & 0.27 & \ldots & \ldots & Distribution not well constrained \\ 
1997-11-09-PTI      & 20 & \ldots & 21 & 2451124.9500 & $2.78 \pm 0.15$ & \ldots & \ldots & \ldots & \ldots & \ldots & \ldots & \ldots & 2.21 & 2.21 & \ldots & \ldots &  \\ 
1998-11-25-PTI      & 10 & \ldots & 11 & 2451142.9400 & $1.93 \pm 0.44$ & \ldots & \ldots & \ldots & \ldots & \ldots & \ldots & \ldots & 0.78 & 0.78 & \ldots & \ldots & Distribution well constrained. HD 30823 sole calibrator this night. Perhaps calibrator diameter over-estimated? \\ 
1998-11-26-PTI      & 5 & \ldots & 5 & 2451143.9300 & $2.06 \pm 0.53$ & \ldots & \ldots & \ldots & \ldots & \ldots & \ldots & \ldots & 0.53 & 0.53 & \ldots & \ldots & Distriubtion poorly constrained. Nominal value matches best-fit MultiNest estimate. \\ 
2006-02-NPOI        & 540 & 135 & 544 & 2453791.7434 & $2.09 \pm 0.06$ & \ldots & \ldots & \ldots & \ldots & \ldots & \ldots & \ldots & 2.35 & 2.65 & 2.00 & 1.49 &  \\ 
2007-03-NPOI        & 660 & 330 & 966 & 2454173.6316 & \ldots & $2.28_{-0.02}^{+0.07}$ & $0.42_{-0.05}^{+0.23}$ & \ldots & \ldots & \ldots & \ldots & \ldots & 2.28 & 0.90 & 1.67 & 5.67 &  \\ 
2007-10-19-PTI      & 100 & \ldots & 104 & 2454392.9900 & $2.13 \pm 0.13$ & \ldots & \ldots & \ldots & \ldots & \ldots & \ldots & \ldots & 0.59 & 0.59 & \ldots & \ldots &  \\ 
2007-10-20-PTI      & 81 & \ldots & 84 & 2454393.9700 & $2.08 \pm 0.14$ & \ldots & \ldots & \ldots & \ldots & \ldots & \ldots & \ldots & 0.18 & 0.18 & \ldots & \ldots &  \\ 
2007-10-21-PTI      & 40 & \ldots & 43 & 2454394.9800 & $2.16 \pm 0.24$ & \ldots & \ldots & \ldots & \ldots & \ldots & \ldots & \ldots & 1.35 & 1.35 & \ldots & \ldots &  \\ 
2007-11-27-PTI      & 20 & \ldots & 22 & 2454431.8500 & $2.55 \pm 0.32$ & \ldots & \ldots & \ldots & \ldots & \ldots & \ldots & \ldots & 2.16 & 2.16 & \ldots & \ldots & Distribution not constrained, highly skewed towards higher values. \\ 
2007-12-23-PTI      & 5 & \ldots & 5 & 2454457.7800 & $2.09 \pm 0.43$ & \ldots & \ldots & \ldots & \ldots & \ldots & \ldots & \ldots & 0.68 & 0.68 & \ldots & \ldots &  \\ 
2007-12-24-PTI      & 35 & \ldots & 39 & 2454458.7500 & $2.45 \pm 0.26$ & \ldots & \ldots & \ldots & \ldots & \ldots & \ldots & \ldots & 1.75 & 1.75 & \ldots & \ldots &  \\ 
2008-02-17-PTI      & 10 & \ldots & 10 & 2454513.6400 & $1.75 \pm 0.33$ & \ldots & \ldots & \ldots & \ldots & \ldots & \ldots & \ldots & 3.22 & 3.22 & \ldots & \ldots & Distribution is well constrained. Bad calibration? \\ 
2008-02-18-PTI      & 25 & \ldots & 27 & 2454514.6500 & $2.17 \pm 0.34$ & \ldots & \ldots & \ldots & \ldots & \ldots & \ldots & \ldots & 1.77 & 1.77 & \ldots & \ldots &  \\ 
2008-09-CHARA-MIRC  & 23 & 8 & 50 & 2454729.0153 & \ldots & $2.28_{-0.08}^{+0.08}$ & $0.69_{-0.20}^{+0.20}$ & \ldots & \ldots & \ldots & \ldots & \ldots & 1.91 & 2.46 & 1.94 & 0.29 & Errors limited by calibrator uncertainty \\ 
2008-10-17-PTI      & 65 & \ldots & 69 & 2454757.0000 & $2.37 \pm 0.20$ & \ldots & \ldots & \ldots & \ldots & \ldots & \ldots & \ldots & 5.71 & 5.71 & \ldots & \ldots &  \\ 
2008-10-26-PTI      & 80 & \ldots & 84 & 2454765.9800 & $2.01 \pm 0.15$ & \ldots & \ldots & \ldots & \ldots & \ldots & \ldots & \ldots & 0.88 & 0.88 & \ldots & \ldots &  \\ 
2008-11-CHARA-MIRC  & 138 & 76 & 268 & 2454778.5967 & \ldots & $2.22_{-0.06}^{+0.06}$ & $0.39_{-0.12}^{+0.12}$ & \ldots & \ldots & \ldots & \ldots & \ldots & 3.81 & 4.59 & 3.24 & 2.96 & Errors limited by calibrator uncertainty \\ 
\tablebreak
2008-11-08-PTI      & 10 & \ldots & 11 & 2454778.8700 & $2.17 \pm 0.40$ & \ldots & \ldots & \ldots & \ldots & \ldots & \ldots & \ldots & 0.08 & 0.08 & \ldots & \ldots &  \\ 
2008-11-09-PTI      & 20 & \ldots & 21 & 2454779.8700 & $2.34 \pm 0.49$ & \ldots & \ldots & \ldots & \ldots & \ldots & \ldots & \ldots & 2.18 & 2.18 & \ldots & \ldots & Distribution not constrained and highly skewed towards higher values. Bad calibration? \\ 
2008-11-16-PTI      & 10 & \ldots & 10 & 2454786.8700 & \ldots & \ldots & \ldots & \ldots & \ldots & \ldots & \ldots & \ldots & \ldots & \ldots & \ldots & \ldots &  \\ 
2008-11-22-PTI      & 50 & \ldots & 52 & 2454792.8200 & $2.30 \pm 0.23$ & \ldots & \ldots & \ldots & \ldots & \ldots & \ldots & \ldots & 1.32 & 1.32 & \ldots & \ldots &  \\ 
2008-12-CHARA-MIRC  & 38 & 8 & 65 & 2454810.8249 & \ldots & $2.36_{-0.06}^{+0.06}$ & $0.80_{-0.17}^{+0.17}$ & \ldots & \ldots & \ldots & \ldots & \ldots & 3.67 & 4.03 & 5.42 & 0.23 & Errors limited by calibrator uncertainty \\ 
2009-03-NPOI        & 840 & 420 & 1225 & 2454898.6326 & \ldots & $2.16_{-0.02}^{+0.07}$ & $0.37_{-0.04}^{+0.25}$ & \ldots & \ldots & \ldots & \ldots & \ldots & 1.85 & 1.65 & 1.29 & 2.80 &  \\ 
2009-11-CHARA-MIRC  & 1091 & 672 & 2575 & 2455138.9326 & \ldots & $2.29_{-0.03}^{+0.02}$ & $0.62_{-0.10}^{+0.08}$ & $6329_{-2038}^{+2038}$ & $1.79_{-0.02}^{+0.02}$ & $2.57_{-0.40}^{+0.40}$ & $9.19_{-0.06}^{+0.04}$ & $1.56_{-0.10}^{+0.12}$ & 4.01 & 2.64 & 0.00 & 8.57 &  \\ 
2009-12-CHARA-MIRC  & 730 & 392 & 1662 & 2455169.0375 & \ldots & $2.28_{-0.04}^{+0.04}$ & $1.00_{-0.12}^{+0.11}$ & $6842_{-2009}^{+2009}$ & $3.72_{-0.07}^{+0.07}$ & $12.18_{-1.00}^{+0.50}$ & $19.94_{-0.03}^{+0.01}$ & $0.10_{-0.02}^{+0.28}$ & 9.26 & 4.08 & 3.77 & 24.39 &  \\ 
2009-12-NPOI        & 290 & 29 & 293 & 2455185.6210 & \ldots & $2.05_{-0.04}^{+0.21}$ & $0.72_{-0.20}^{+0.28}$ & $7007_{-1847}^{+1847}$ & $3.23_{-0.03}^{+0.22}$ & $6.70_{-4.10}^{+8.50}$ & $16.72_{-0.06}^{+0.04}$ & $0.92_{-0.16}^{+0.21}$ & 2.14 & 2.20 & 1.89 & 1.73 & Visibities at short baselines are much higher than CHARA model would predict. \\  
2010-01\_AB-NPOI    & 3324 & 1376 & 4444 & 2455205.5082 & \ldots & $2.24_{-0.02}^{+0.03}$ & $0.76_{-0.07}^{+0.14}$ & $6211_{-1690}^{+1690}$ & $2.97_{-0.06}^{+0.08}$ & $4.10_{-0.30}^{+0.30}$ & $17.84_{-0.02}^{+0.02}$ & $2.43_{-0.02}^{+0.07}$ & 1.23 & 1.61 & 1.21 & 0.33 & Disk $r_c$ is clearly bimodal \\ 
2010-02-NPOI        & 810 & 265 & 814 & 2455243.1816 & \ldots & $2.37_{-0.10}^{+0.10}$ & $0.62_{-0.24}^{+0.24}$ & $6219_{-2000}^{+2000}$ & \ldots & $10.70_{-1.30}^{+1.30}$ & \ldots & $0.70_{-0.23}^{+0.23}$ & 1.42 & 1.47 & 1.81 & 0.87 &  \\ 
2010-02-CHARA-MIRC  & 96 & 64 & 236 & 2455245.7444 & \ldots & $2.01_{-0.04}^{+0.04}$ & $0.21_{-0.13}^{+0.13}$ & $3075_{-1600}^{+1600}$ & \ldots & $13.14_{-0.50}^{+0.50}$ & \ldots & $0.55_{-0.11}^{+0.11}$ & 6.88 & 1.64 & 1.18 & 20.45 & Interferometry + Photometry. Statistics from MultiNest distribution. \\ 
2010-04-NPOI        & 15 & 0 & 15 & 2455289.6026 & \ldots & $2.33_{-0.28}^{+0.28}$ & $0.56_{-0.24}^{+0.24}$ & $6500_{-2000}^{+2000}$ & \ldots & $11.50_{-7.70}^{+7.70}$ & \ldots & $0.88_{-1.05}^{+1.05}$ & 0.53 & 0.53 & 0.00 & 0.00 &  \\ 
2010-08-CHARA-MIRC  & 960 & 640 & 2164 & 2455430.5170 & \ldots & $2.33_{-0.04}^{+0.04}$ & $0.74_{-0.11}^{+0.09}$ & $9901_{-1334}^{+1334}$ & \ldots & $1.27_{-0.10}^{+0.10}$ & \ldots & $3.75_{-0.12}^{+0.07}$ & 10.74 & 3.97 & 3.90 & 27.74 &  \\ 
2010-09-CHARA-MIRC  & 1176 & 728 & 3020 & 2455464.4883 & \ldots & $2.37_{-0.02}^{+0.03}$ & $0.73_{-0.04}^{+0.06}$ & $5944_{-1888}^{+1888}$ & \ldots & $2.35_{-0.10}^{+0.20}$ & \ldots & $3.11_{-0.08}^{+0.10}$ & 3.38 & 1.96 & 1.98 & 7.07 &  \\ 
2010-10-CHARA-MIRC  & 288 & 152 & 732 & 2455496.4319 & \ldots & $2.26_{-0.03}^{+0.03}$ & $0.62_{-0.08}^{+0.08}$ & $6629_{-1973}^{+1973}$ & \ldots & $1.47_{-0.10}^{+0.80}$ & \ldots & $3.56_{-0.62}^{+0.18}$ & 1.77 & 1.31 & 0.77 & 3.67 &  \\ 
2010-11-CHARA-MIRC  & 288 & 192 & 763 & 2455505.4193 & \ldots & $2.18_{-0.01}^{+0.01}$ & $0.11_{-0.03}^{+0.05}$ & $3720_{-1817}^{+1817}$ & \ldots & $9.66_{-0.20}^{+0.20}$ & \ldots & $0.40_{-0.07}^{+0.11}$ & 4.64 & 2.93 & 1.16 & 10.71 &  \\ 
2010-12-CHARA-MIRC  & 191 & 112 & 475 & 2455543.7059 & \ldots & $2.22_{-0.03}^{+0.04}$ & $0.31_{-0.09}^{+0.10}$ & $5548_{-1985}^{+1985}$ & \ldots & $0.72_{-0.10}^{+0.10}$ & \ldots & $4.71_{-0.28}^{+0.28}$ & 5.72 & 2.29 & 1.36 & 15.92 &  \\ 
2011-01-CHARA-MIRC  & 310 & 182 & 860 & 2455580.2465 & \ldots & $2.18_{-0.02}^{+0.02}$ & $0.25_{-0.04}^{+0.04}$ & $3013_{-1838}^{+1838}$ & \ldots & $4.94_{-0.20}^{+1.10}$ & \ldots & $1.84_{-0.52}^{+0.16}$ & 5.01 & 2.57 & 1.36 & 12.83 &  \\ 
2011-04-CHARA-CLIMB & 41 & 14 & 45 & 2455655.0673 & \ldots & $2.33_{-0.06}^{+0.06}$ & $0.33_{-0.48}^{+0.48}$ & $6100_{-2000}^{+2000}$ & $3.48_{-0.30}^{+0.30}$ & $13.20_{-3.80}^{+3.80}$ & $18.55_{-1.96}^{+1.96}$ & $0.64_{-0.50}^{+0.50}$ & 3.23 & 4.06 & 1.04 & 2.97 &  \\ 
2011-09-18-CHARA-MIRC & 201 & 240 & 756 & 2455823.0305 & \ldots & $2.25_{-0.04}^{+0.02}$ & $0.62_{-0.10}^{+0.06}$ & \ldots & \ldots & \ldots & \ldots & \ldots & 5.06 & 3.55 & 3.02 & 8.37 &  \\ 
2011-09-24-CHARA-MIRC & 120 & 160 & 394 & 2455829.0277 & \ldots & $2.17_{-0.03}^{+0.03}$ & $0.36_{-0.07}^{+0.06}$ & \ldots & \ldots & \ldots & \ldots & \ldots & 5.40 & 3.32 & 1.67 & 10.69 &  \\ 
2011-10-10-CHARA-MIRC & 400 & 480 & 1412 & 2455844.9422 & \ldots & $2.12_{-0.05}^{+0.04}$ & $0.34_{-0.12}^{+0.11}$ & \ldots & \ldots & \ldots & \ldots & \ldots & 16.97 & 5.49 & 6.82 & 36.70 &  \\ 
2011-11-03-CHARA-MIRC & 831 & 1119 & 2677 & 2455868.8509 & \ldots & $2.25_{-0.07}^{+0.03}$ & $0.57_{-0.17}^{+0.08}$ & \ldots & \ldots & \ldots & \ldots & \ldots & 12.14 & 5.64 & 7.21 & 21.90 &  \\ 
\enddata
\tablenotetext{*}{The changes seen in height power, $\beta$, hint that there 
may be some asymmetric structure in the disk.}
\end{deluxetable}

\end{landscape}

%
% tbl_aggregate_stats
%
\clearpage
\begin{deluxetable}{llllll}
\tablecolumns{6}
\tablewidth{0pt}
\tabletypesize{\scriptsize}
\tablecaption{\label{tbl_aggregate_stats}
Aggregate statistics for all interferometric data with uncertainties determined 
from the maximum of the upper/lower averaged bootstrapped uncertainties or the 
standard deviation of the nominal values.}
\tablehead{
\colhead{Quantity} &
\colhead{Units} & 
\colhead{NPOI (V)} & 
\colhead{MIRC (H)} & 
\colhead{CLIMB (K)} & 
\colhead{PTI (K)} \\ 
}
\startdata
Quantity &  & NPOI (V) & MIRC (H) & CLIMB (K) & PTI (K) \\ 
\UDD{} & (mas) & 2.09 $\pm$ 0.06 & 2.10 $\pm$ 0.15 & $\ldots$ & 2.22 $\pm$ 0.53 \\ 
\LDD{} & (mas) & 2.21 $\pm$ 0.28 & 2.22 $\pm$ 0.09 & 2.33 $\pm$ 0.06 & $\ldots$ \\ 
$\alpha_\text{LDD}$ &  & 0.47 $\pm$ 0.28 & 0.50 $\pm$ 0.26 & 0.33 $\pm$ 0.48 & $\ldots$ \\ 
$\Omega_\text{disk}$ & (deg) & $\ldots$ & $1.30 \pm 0.67$ & $\ldots$ & $\ldots$ \\ 
$i_\text{disk}$ & (deg) & $\ldots$ & $-0.51 \pm 1.03$ & $\ldots$ & $\ldots$ \\ 
$\kappa \rho ^\ddagger$ &  & 6676 $\pm$ 2000 & 5667 $\pm$ 2188 & 6100 $\pm$ 2000 & $\ldots$ \\ 
$r_c$ & (mas) & 3.10 $\pm$ 0.22 & 2.76 $\pm$ 1.36 & 3.48 $\pm$ 0.30 & $\ldots$ \\ 
$h_c$ & (mas) & 0.07 $\pm$ 0.09 & 0.05 $\pm$ 0.05 & 0.13 $\pm$ 0.04 & $\ldots$ \\ 
$\alpha$ &  & 17.28 $\pm$ 0.79 & 14.56 $\pm$ 7.60 & 18.55 $\pm$ 1.96 & $\ldots$ \\ 
$\beta$ &  & 1.23 $\pm$ 1.05 & 2.18 $\pm$ 1.67 & 0.64 $\pm$ 0.50 & $\ldots$ \\ 
\enddata
\tablecomments{Orbital values of
$\Omega = 297.60 \pm 0.06$ (deg), $i = 88.89 \pm 0.03$ (deg), and 
$\alpha_T = 31.2 \pm 0.9$ (mas) were used in these models.}
\tablenotetext{\ddagger}{Not well constrained}
\end{deluxetable}

%
% tbl_physical_values
%
\clearpage
\setlength{\tabcolsep}{0.05in}
\begin{deluxetable}{lllllllll}
\tablecolumns{9}
\tablewidth{0pt}
\tabletypesize{\scriptsize}
\tablecaption{\label{tbl_physical_values}
Representative$^*$ linear equivalent of our results if the system were at various 
distances in literature. }
\tablehead{
% row 1, upper label
\multicolumn{3}{l}{} &
\multicolumn{5}{c}{Nominal distance estimates (pc)} & 
\multicolumn{1}{l}{} \\ 
% row 2, column headings
\colhead{Quantity} &
\multicolumn{2}{c}{This work} &
\colhead{600$^\text{a,b}$} &
\colhead{653$^\text{c}$} &
\colhead{737$^\text{d}$} &
\colhead{1000$^\text{e}$} &
\colhead{1500$^\text{f}$} &
\colhead{Linear units}  \\ 
}
\startdata
$\Omega$ & $297 \pm 3$ & (deg) &  &  &  &  &  &  \\ 
$i$ & $89 \pm 1$ & (deg) &  &  &  &  &  &  \\ 
$\alpha_T = \alpha_1 + \alpha_2$ & $31 \pm 3$ & (\mas{}) & $18.72 \pm 1.80$ & $20.37 \pm 1.96$ & $22.99 \pm 2.21$ & $31.20 \pm 3.00$ & $46.80 \pm 4.50$ &  (AU) \\ 
F-star Radius & $1.11 \pm 0.05$ & (\mas{}) & $143.25 \pm 5.81$ & $155.90 \pm 6.32$ & $175.96 \pm 7.13$ & $238.75 \pm 9.68$ & $358.13 \pm 14.52$ & (\RSolar{}) \\ 
F-star LDD coeff & $0.50 \pm 0.26$ & &  &  &  &  &  &  \\ 
Disk scale Height ($h_c$) & $1.038 \pm 0.139$ & (\mas{}) & $0.03 \pm 0.03$ & $0.03 \pm 0.03$ & $0.04 \pm 0.04$ & $0.05 \pm 0.05$ & $0.07 \pm 0.07$ & (AU) \\ 
Disk scale radius ($r_c$) & $7.416 \pm 0.276$ & (\mas{}) & $1.66 \pm 0.82$ & $1.80 \pm 0.89$ & $2.03 \pm 1.00$ & $2.76 \pm 1.36$ & $4.14 \pm 2.04$ & (AU) \\ 
\enddata
\tablecomments{These values average over all interferometric epochs. 
Therefore, these estimates are biased and are skewed by the outliers in 
Table \ref{tbl_bootstrap_results}.}
\tablenotetext{*}{We caution the reader that this aggregation of data 
is performed in a wavelength and model-agnostic fashion. Thus, any asymmetries 
in the system, which we argue exist, have biased the values quoted here. These 
values are supplied to ease the creation of a radiative transfer model including 
dust physics. We do not advocate that these values be quoted elsewhere.}
\tablenotetext{a}{\cite{VandeKamp1978}}
\tablenotetext{b}{\cite{Heintz1994}}
\tablenotetext{c}{\cite{VanLeeuwenFloor2008}}
\tablenotetext{d}{\cite{Kloppenborg2012}}
\tablenotetext{e}{\cite{Strand1959}} 
\tablenotetext{f}{\cite{Guinan2012}}
\end{deluxetable}

% Appendix
\appendix
\setcounter{table}{0}
\renewcommand{\thetable}{A \arabic{table}}

%
% tbl_multinest_eclipse
%
\begin{landscape}
\clearpage
\setlength{\tabcolsep}{0.02in}
\begin{deluxetable}{ccccccccccrrrrr}
\tablecolumns{13}
\tablewidth{0pt}
%\rotate
\tabletypesize{\tiny}
\tablecaption{\label{tbl_eclipse_multinest}
Best-fit values and statistical information for single-epoch MultiNest
minimizations involving three variants$^*$ of the tilted Pascucci disk model (model
8) \textbf{The posterior odds ratio ($\Delta \log R$) is relative to the Pascucci zero
tilt disk model.}
}
\tablehead{
% row 1, multi-column labels
\multicolumn{1}{l}{} &
\multicolumn{2}{c}{F-star} & 
\multicolumn{7}{c}{Disk} & 
\multicolumn{5}{c}{Statistical Information} 
\\
% Horizontal rules under the above columns
\cmidrule(lr){2-3} 
\cmidrule(lr){4-10} 
\cmidrule(lr){11-15}
% row 2, labels
\multicolumn{1}{l}{} &
\colhead{\LDD{}} & 
\colhead{$\alpha_\text{LDD}$} & 
\colhead{$i_\text{disk}$} &
\colhead{$\Omega_\text{disk}$} & 
\colhead{$\alpha$} & 
\colhead{$r_c$} & 
\colhead{$\beta$} & 
\colhead{$h_c$} & 
\colhead{$\kappa \rho$} & 
\colhead{$\Delta \log R$} & 
\colhead{$\chi^2_r$} & 
\colhead{$\chi^2_r(V^2)$} & 
\colhead{$\chi^2_r(T_{3\text{A}})$} & 
\colhead{$\chi^2_r(T_{3 \phi})$} \\
% row 3, units
\multicolumn{1}{l}{} &
(mas) &  
\multicolumn{1}{l}{} &
(deg) & 
(deg) &  
\multicolumn{1}{l}{} &
(mas) &  
\multicolumn{1}{l}{} &
(mas) &
\multicolumn{5}{l}{} &
\\ \hline
}
\startdata
Model: Pascucci zero tilt &  &  &  &  &  &  &  &  &  &  &  &  &  &  \\ 
2009-11 & 2.30 & 0.63 & $\ldots$ & $\ldots$ & 9.19 & 1.87 & 1.67 & 0.023 & 4708  & $\ldots$ & 4.92 & 4.23 & 2.62 & 8.34 \\ 
2009-12 & 2.28 & 1.00 & $\ldots$ & $\ldots$ & 19.95 & 3.71 & 0.10 & 0.118 & 7046 & $\ldots$ & 9.39 & 4.00 & 3.78 & 25.02 \\ 
2010-02 & 2.43 & 0.98 & $\ldots$ & $\ldots$ & $\ldots$ & $\ldots$ & 0.75 & 0.173 & 3046 & $\ldots$ & 3.47 & 2.83 & 3.26 & 4.66 \\ 
2010-08 & 2.33 & 0.74 & $\ldots$ & $\ldots$ & $\ldots$ & $\ldots$ & 3.78 & 0.011 & 4349 & $\ldots$ & 10.25 & 4.00 & 3.69 & 26.18 \\ 
2010-09 & 2.37 & 0.72 & $\ldots$ & $\ldots$ & $\ldots$ & $\ldots$ & 3.00 & 0.021 & 3336 & $\ldots$ & 3.24 & 1.98 & 2.05 & 6.48 \\ 
2010-10 & 2.29 & 0.66 & $\ldots$ & $\ldots$ & $\ldots$ & $\ldots$ & 4.20 & 0.007 & 5569 & $\ldots$ & 1.78 & 1.33 & 0.91 & 3.49 \\ 
2010-11 & 2.18 & 0.10 & $\ldots$ & $\ldots$ & $\ldots$ & $\ldots$ & 0.36 & 0.093 & 3106 & $\ldots$ & 4.80 & 2.91 & 1.14 & 11.29 \\ 
2010-12 & 2.29 & 0.41 & $\ldots$ & $\ldots$ & $\ldots$ & $\ldots$ & 4.83 & 0.005 & 3726 & $\ldots$ & 4.94 & 2.92 & 1.49 & 11.84 \\ 
2011-01 & 2.13 & 0.18 & $\ldots$ & $\ldots$ & $\ldots$ & $\ldots$ & 3.05 & 0.015 & 3030 & $\ldots$ & 5.73 & 3.31 & 1.86 & 13.71 \\ 
\hline
Average & 2.29 $\pm$ 0.09 & 0.60 $\pm$ 0.32 & 0 & 0 & 14.57 $\pm$ 7.61 & 2.79 $\pm$ 1.30 & 2.41 $\pm$ 1.75 & 0.052 $\pm$ 0.061 & 4215 $\pm$ 1373 & $\ldots$ & 5.39 & 3.06 & 2.31 & 12.34 \\ 
 &  &  &  &  &  &  &  &  &  &  &  &  &  &  \\ 
Model: Pascucci fixed tilt &  &  &  &  &  &  &  &  &  &  &  &  &  &  \\ 
2009-11 & 2.29 & 0.62 & $\ldots$ & $\ldots$ & 9.19 & 1.79 & 1.56 & 0.026 & 6329  & -37  & 4.91 & 3.98 & 2.64 & 8.69 \\ 
2009-12 & 2.28 & 1.00 & $\ldots$ & $\ldots$ & 19.94 & 3.72 & 0.10 & 0.122 & 6842 &  119 & 9.22 & 3.93 & 3.63 & 24.66 \\ 
2010-02 & 2.43 & 0.99 & $\ldots$ & $\ldots$ & $\ldots$ & $\ldots$ & 0.72 & 0.179 & -5   & 296 & 3.53 & 2.92 & 3.39 & 4.59 \\ 
2010-08 & 2.33 & 0.74 & $\ldots$ & $\ldots$ & $\ldots$ & $\ldots$ & 3.35 & 0.014 & -493 & -2999 & 10.65 & 3.93 & 3.82 & 27.57 \\ 
2010-09 & 2.39 & 0.76 & $\ldots$ & $\ldots$ & $\ldots$ & $\ldots$ & 2.81 & 0.028 & -129 & 5802 & 3.37 & 2.02 & 1.95 & 6.98 \\ 
2010-10 & 2.29 & 0.66 & $\ldots$ & $\ldots$ & $\ldots$ & $\ldots$ & 3.66 & 0.011 & -8   & 1235 & 1.70 & 1.34 & 0.88 & 3.18 \\ 
2010-11 & 2.18 & 0.10 & $\ldots$ & $\ldots$ & $\ldots$ & $\ldots$ & 0.37 & 0.097 &  29  & 849 & 4.69 & 2.91 & 1.15 & 10.91 \\ 
2010-12 & 2.29 & 0.39 & $\ldots$ & $\ldots$ & $\ldots$ & $\ldots$ & 4.80 & 0.006 & -56  & 394 & 5.22 & 2.97 & 1.56 & 12.73 \\ 
2011-01 & 2.13 & 0.17 & $\ldots$ & $\ldots$ & $\ldots$ & $\ldots$ & 2.27 & 0.030 &  81  & 661 & 5.50 & 3.02 & 1.68 & 13.53 \\ 
\hline
Average & 2.29 $\pm$ 0.09 & 0.60 $\pm$ 0.32 & 1.3 & -0.02 & 14.56 $\pm$ 7.60 & 2.76 $\pm$ 1.36 & 2.18 $\pm$ 1.62 & 0.057 $\pm$ 0.061 & 4948 $\pm$ 2245 & -55 & 5.42 & 3.00 & 2.30 & 12.54 \\ 
 &  &  &  &  &  &  &  &  &  &  &  &  &  &  \\ 
\tablebreak
Model: Pascucci free tilt &  &  &  &  &  &  &  &  &  &  &  &  &  &  \\ 
2009-11 & 2.30 & 0.62 & 1.102 & 0.16 & 8.93 & 1.74 & 1.55 & 0.025 & 6018   & 349  & 4.65 & 4.37 & 2.52 & 7.25 \\ 
2009-12 & 2.27 & 1.00 & 1.148 & -0.15 & 19.96 & 3.73 & 0.10 & 0.120 & 6417 & 164  & 9.14 & 3.65 & 3.29 & 25.22 \\ 
2010-02 & 2.43 & 0.95 & 1.088 & -2.72 & $\ldots$ & $\ldots$ & 1.10 & 0.176 & 46   & 347 & 3.06 & 2.37 & 2.63 & 4.52 \\ 
2010-08 & 2.33 & 0.78 & 2.330 & 0.70 & $\ldots$ & $\ldots$ & 4.75 & 0.005  & 1298 & -1208 & 16.17 & 7.69 & 7.96 & 37.09 \\ 
2010-09 & 2.39 & 0.78 & 1.153 & 0.51 & $\ldots$ & $\ldots$ & 2.70 & 0.027  & 422  & 6353 & 2.92 & 1.71 & 1.67 & 6.11 \\ 
2010-10 & 2.32 & 0.71 & 0.353 & -0.29 & $\ldots$ & $\ldots$ & 4.92 & 0.003 & 13   & 1256 & 4.43 & 2.11 & 1.19 & 12.06 \\ 
2010-11 & 2.24 & 0.29 & 6.879 & -0.95 & $\ldots$ & $\ldots$ & 0.24 & 0.000 & 691  & 1511 & 5.22 & 2.93 & 1.37 & 12.52 \\ 
2010-12 & 2.23 & 0.38 & 2.324 & -0.96 & $\ldots$ & $\ldots$ & 4.97 & 0.000 & 330  & 780 & 114.24 & 22.38 & 14.86 & 370.25 \\ 
2011-01 & 2.20 & 0.23 & 1.108 & -0.87 & $\ldots$ & $\ldots$ & 2.08 & 0.037 & 1391 & 1971 & 1.62 & 1.25 & 0.87 & 3.01 \\ 
\hline
Average & 2.30 $\pm$ 0.08 & 0.64 $\pm$ 0.28 & 1.943 $\pm$ 1.955$^\dagger$ & -0.51 $\pm$ 1.03 & 14.45 $\pm$ 7.80 & 2.73 $\pm$ 1.40 & 2.49 $\pm$ 1.97 & 0.044 $\pm$ 0.062 & 4223 $\pm$ 1319 & 522 & 17.94 & 5.39 & 4.04 & 53.12 \\ 
\enddata
\tablenotetext{*}{The variations either assume the disk has zero tilt, a fixed 
tilt, or a per-epoch tilt with respect to the orbital plane. The averaged 
inclination of $1.33 \pm 0.67$ degrees agrees well with the multi-epoch 
minimizations. The variations in scale height appear to be real.}
\tablenotetext{\dagger}{Excluding the 2010-11 result, this becomes $1.33 \pm 0.67$
in agreement with our fixed-tilt model.}
\end{deluxetable}

\end{landscape}

\end{document}